\documentclass[12pt,sort&compress]{article}

\usepackage[top=2.5cm, bottom=2.5cm, left=2.5cm, right=2.5cm]{geometry}

\usepackage[utf8]{inputenc}
\usepackage[french,english]{babel}
\usepackage[T1]{fontenc}
\usepackage{lmodern} 
\usepackage{color}
\usepackage{indentfirst}
\usepackage{lineno}
\modulolinenumbers[0]
\usepackage{soul}

\usepackage{appendix}

\usepackage{graphicx,color}
\usepackage{xcolor} 
\usepackage{subfigure}
\usepackage{epstopdf}

\usepackage{multicol}
\usepackage{multirow}
\usepackage{colortbl}
\usepackage{tabularx}

\let\oldtabular=\tabular
\def\tabular{\footnotesize\oldtabular} 
\usepackage[labelsep=period,%
			textfont=it,
            justification=raggedright]{caption}
\definecolor{darkgreen}{rgb}{0,0.7,0}
\definecolor{gris}{rgb}{0.5,0.5,0.5}
\definecolor{darkblue}{rgb}{0.0,0.0,0.5}
\definecolor{oneblue}{rgb}{0,0.0,0.75}
\definecolor{dkblue}{rgb}{0,0,0.7}
\definecolor{grey}{rgb}{0.4,0.4,0.4}
\definecolor{lightgrey}{rgb}{0.6,0.6,0.6}
\definecolor{dkgreen}{rgb}{0,0.52,0.14}

\usepackage{amsmath, amsfonts, amssymb, amstext, amsthm, dsfont,mathrsfs}
\usepackage[ugly]{nicefrac}

\renewcommand{\b}{\mathrm{b \,}}

\newcommand*\od[2]{\frac{\mathrm{d} #1}{\mathrm{d} #2}}
\newcommand*\pd[2]{\frac{\partial #1}{\partial #2}}

\newcommand{\eqdef}{\mathop{\stackrel{\,\mathrm{def}}{:=}\,}}

\newcommand{\const}{\mathrm{const}}
\newcommand{\Mat}{\mathrm{Mat}\,}

\newcommand{\half}{\frac{1}{2}} 

\newcommand*\e[1]{\cdot 10^{\,#1}}

\newcommand*\egal{\ = \ }
\newcommand*\plus{\ + \ }
\newcommand*\moins{\ - \ }

\RequirePackage[sort&compress, semicolon, round, authoryear]{natbib}
\setcitestyle{aysep={}}

\usepackage[colorlinks,
            urlcolor=oneblue,
            linkcolor=oneblue,
            citecolor=oneblue,
            bookmarksopen=false,
            pagebackref=true]{hyperref}
            
\usepackage{url}

\newcommand{\A}{\mathcal{A}}

\newcommand{\CN}{\textsc{Crank}--\textsc{Nicolson}}
\newcommand{\dx}{\Delta x}
\newcommand{\dt}{\Delta t}
\newcommand{\Eu}{\textsc{Euler}}
\newcommand{\Fo}{Fo}
\newcommand{\M}{\mathcal{M}}
\newcommand{\Pv}{P_{\,v}}
\newcommand{\R}{\mathds{R}}
\newcommand{\RK}{\textsc{Runge}--\textsc{Kutta}}

\newcommand{\ts}{t^{\,\star}}
\newcommand{\xs}{x^{\,\star}}

\newcommand*\unit[1]{ \ [\,\mathsf{ #1 }\,]}
\newcommand*\RC[1]{R$#1$C}

\begin{document}

\title{On the comparison of three numerical methods applied to building simulation: finite-differences, RC circuit approximation and a spectral method.}
\author{Julien Berger\textsuperscript{a}$^{\ast}$ \& Suelen Gasparin\textsuperscript{b,c} \& Denys Dutykh\textsuperscript{b} \& Nathan Mendes\textsuperscript{c} \vspace{0.5cm} \\
}
\date{\small \today}

\maketitle

\begin{center}
\small
\textsuperscript{a} Univ. Grenoble Alpes, Univ. Savoie Mont Blanc, CNRS, LOCIE,
73000 Chambéry, France  \\
\textsuperscript{b} Univ. Grenoble Alpes, Univ. Savoie Mont Blanc, CNRS, LAMA,	
73000 Chambéry, France \\
\textsuperscript{c} Thermal Systems Laboratory, Mechanical Engineering Graduate Program, \\
Pontifical Catholic University of Paran\'a, Rua Imaculada Conceição, 1155, CEP : 80215-901,
Curitiba - Paran\'a, Brazil\\
$^{\ast}$\emph{Corresponding author. E-mail address: julien.berger@univ-smb.fr}\\
\end{center}

\begin{abstract}

Predictions of physical phenomena in buildings are carried out by using physical models formulated as a mathematical problem and solved by means of numerical methods, aiming at evaluating, for instance, the building thermal or hygrothermal performance by calculating distributions and fluxes of heat and moisture transfer. Therefore, the choice of the numerical method is crucial since it is a compromise among (i) the solution accuracy, (ii) the computational cost to obtain the solution and (iii) the complexity of the method implementation. An \emph{efficient} numerical method enables to compute an accurate solution with a minimum computational run time (CPU). On that account, this article brings an investigation on the performance of three numerical methods. The first one is the standard and widely used finite-difference approach, while the second one is the so-called RC approach, which is a particular method brought to the building physics area by means of an analogy of electric circuits. The third numerical method is the spectral one, which has been recently proposed to solve nonlinear diffusive problems in building physics. The three methods are evaluated in terms of accuracy on the assessment of the dependent variable (temperature or vapor pressure) or of density of fluxes for three different cases: i) heat diffusion through a concrete slab, ii) moisture diffusion through an aerated concrete slab and iii) heat diffusion using measured temperatures as boundary conditions. Results highlight the spectral approach as the most accurate method. The RC based model with a few number of resistances does not provide accurate results for temperature and vapor pressure distributions neither to flux densities nor conduction loads. 


\end{abstract}
\textit{keywords}: {\small Heat Transfer; Moisture Transfer; Numerical Methods; Finite Differences; Thermal Circuit Model; Spectral Method.}

\section{Introduction}

As the building sector represents almost $33$\% of the world global energy consumption, current environmental issues lead to focus on energy efficiency of building envelopes \cite{IEA_2015}. Within this context, several tools have been developed  since the $1970$'s for the accurate assessment of building energy performance. Many of them have been reported in the frame of the International Energy Agency Annex $41$ published by \textsc{Woloszyn} and \textsc{Rode} in \cite{woloszyn_2008} and more recently in \textsc{Mendes} \textit{et al.} \cite{Mendes_2017}.

Among all the physical phenomena involved in building energy efficiency, energy losses associated to heat and moisture transfer through the building envelope are of major importance. They represent an important part of building energy consumption and moisture may considerably impact on conduction loads and on the size of HVAC systems \cite{Mendes_2003}, besides promoting severe disorders when reaching high levels \cite{berger_2015}. Thus, it is of primary importance to have numerical models enabling to accurately represent the physical phenomena for the evaluation of heat losses and gains through building envelopes. 

The numerical models are elaborated from the main governing equations representing the heat and/or moisture transfer in building porous materials detailed for instance in \cite{Mendes_2017}. The use of analytical solutions is often limited due to the nonlinearity of the material properties and to the non-periodicity of the boundary conditions. Thus, most of models referenced in literature are based on numerical approaches.

Consequently, the main challenge arises in elaborating efficient numerical models to perform the simulation. The word \emph{efficiency} can designate several features. One decisive aspect is the accuracy of the computed solution. It is of capital importance for the design of energy efficient buildings. To predict reliable energy consumption, one must be certain that the numerical errors of the model are negligible. When comparing the predictions to experimental observations, such as performed in \cite{Yang_2015}, researchers often assume that the numerical errors are always lower than the uncertainties present in the measurements, in the inputs parameters and in the mathematical model that described the physical phenomena. Another important feature is the computational run time of the numerical model to compute the solution. Even with the increase of computer power in the recent decades, it is still a crucial issue. The numerical model needs to save the computational efforts to ease the work of building designers and engineers. It is also relevant in the research context of sensitivity analysis or parameter estimation problem, where a large number of direct model computations is required.

Surprisingly, despite the widespread use of models in research and practice, the efficiency of the numerical models have received little attention in the literature. Models are implemented in software such as EnergyPlus \cite{Crawley_2001}, ESP-r \cite{Clarke_2013}, BSim \cite{Rode_2003}, etc. The governing equations are well detailed but the description of the numerical methods is generally brief and no discussion on their efficiency are provided. Moreover, the validity of the assumption that the numerical error can be negligible has never been verified. Thus, this article proposes to overcome this issue by presenting a detailed evaluation of the efficiency of the three numerical methods.

The first one, is the standard finite-differences based approach, which is probably the most-used method to compute the solution of the diffusion problem. Different variations of this approach have been reported in the literature such as the implicit \Eu ~in \cite{mendes_2005,steeman_2009}, the explicit \Eu ~in \cite{tariku_2010,kalagasidis_2007} or the \CN ~in \cite{vangenuchten_1982}. The second method is the so-called RC approach. This method was first used during the second world war \cite{Kirkpatrick_1943} where analogous electric networks were built to solve the solution of transient heat-flow problems. Since there were no computer devices, it was an ingenuous way to rapidly simulate the solution of the problem. Interesting details can be found in \cite{Lawson_1953,Robertson_1958} with an investigation of the error devices as a function of the number of (physical) thermal resistances. Although the appearance of digital computers started in the fifties and the rapid and progressive hardware evolution since the seventies, this RC method is still used in many algorithms to solve the partial differential equation of heat transfer as for instance in \cite{Fraisse_2002,Roels_2017,Naveros_2015}.
The third method is more advanced spectral method which was recently applied for the solution of diffusion problems through porous building elements \cite{Gasparin_2017a,Gasparin_2017b}.

The manuscript is organized as follows. The physical problem of heat and moisture transfer in building porous materials is recalled in Section~\ref{sec:physical_problem}. The three numerical methods are described in Section~\ref{sec:num_methods}. Then, three case studies are analysed. The first one, in Section~\ref
{sec:linear_heat_diff}, considers a linear heat transfer in a concrete wall. Then, in Section~\ref{sec:NL_moisture_diff}, a nonlinear case of moisture diffusion is investigated. In Section~\ref{sec:real_case}, the efficiencies of the numerical methods are evaluated considering real measured temperatures as boundary conditions. Some conclusion and final remarks are outlined in Section~\ref{sec:conclusion}.

\section{Physical problem and mathematical formulation}
\label{sec:physical_problem}

\subsection{Physical phenomenon of heat transfer}

The physical problem involves heat conduction in a wall of thickness $L$ composed of a single material in which the thermal conductivity is denoted as $k$, the density as $\rho$ and the specific heat as $c\,$. The problem can be formulated by the \textsc{Fourier} (or heat) equation, for $x \ \in \ \bigl[\,0 \,,\, L \,\bigr]$ and $t \ \in \ \bigl[\, 0 \,,\, \tau \,\bigr]\,$:
\begin{align}
\label{eq:heat1d_dim}
\rho \, c \, \pd{T}{t} \moins k \, \pd{^{\,2}\,T}{x^{\,2}} \egal 0 \,,
\end{align}
where $T(\,x\,,\,t\,)$ is the temperature within the wall at the distance $x \ \in \ \bigl[\, 0 \,,\, L \,\bigr]$ and time $t \ \geqslant \ 0 \,$. 

For the sake of simplicity\footnote{Different approaches are reported in literature to represent the \textsc{Neumann} or \textsc{Robin} boundary conditions within the RC model framework. Thus, to limit the possible sources of error, the numerical investigation was performed considering \textsc{Dirichlet} boundary conditions. }, \textsc{Dirichlet} boundary conditions are assumed at the extremity of the wall:
\begin{align*}
& T \egal T_{\,L} (\,t\,) \,, && x \egal 0 \,, \\
& T \egal T_{\,R} (\,t\,)\,, && x \egal L \,.
\end{align*}

At the initial state, the temperature of the wall is assumed to be uniform:
\begin{align*}
& T \egal T_{\,0} \,, \qquad t \egal 0 \,, \qquad \forall \ x \ \in \ \bigl[\, 0 \,,\, L \,\bigr] \,.
\end{align*}

One of the interesting outputs in the building physics framework is the heat flux density at $x_{\,0}  \ \in \ \bigl[\, 0 \,,\, L \,\bigr] $, defined as:
\begin{align}
\label{eq:heatflux}
q\, (\,t\,) &\ \eqdef \ - \, k \, \pd{T}{x}\biggl|_{\,x \egal x_{\,0}} \,,
\end{align}
Particularly, we denote as $q_{\,R}$ the heat flux density computed at $x \egal L$:
\begin{align*}
q_{\,R}\, (\,t\,) &\ \eqdef \ - \, k \, \pd{T}{x}\biggl|_{\,x\egal L} \,,
\end{align*}
The conduction loads represent the heat fluxes at the building envelope internal surface and, in terms of the energy density, it can be evaluated as: 
\begin{align}
\label{eq:heat_load}
E \ \eqdef \  \int_{t_{\,1}}^{t_{\,2}} q(\,\tau\,) \, \mathrm{d}\tau \,,
\end{align}
which can be evaluated, for instance, over daily or monthly periods.

\subsection{Physical phenomenon of moisture transfer}

The moisture transfer occurs under isothermal conditions in a wall of thickness $L\,$, with a single material of permeability $\kappa$ and moisture capacity $\xi$, both depending on the vapor pressure. The formulation of the problem, for $x \ \in \ \bigl[\,0 \,,\, L \,\bigr]$ and $t \ \in \ \bigl[\, 0 \,,\, \tau \,\bigr]\,$, yields to:
\begin{align}
\label{eq:moisture_1d_dim}
\xi \, (\Pv) \, \pd{\Pv}{t} \moins \pd{}{x} \, \biggl(\, \kappa\, (\Pv) \, \pd{\,\Pv}{x} \, \biggr) \egal 0 \,,
\end{align}
where $\Pv$ is the vapor pressure within the wall. 

The boundary conditions at the extremity of the wall are: 
\begin{align*}
& \Pv \egal P_{\,v,\,L} \, (\,t\,) \,, && x \egal 0 \,, \\
& \Pv \egal P_{\,v,\,R} \, (\,t\,)\,, && x \egal L \,.
\end{align*}

A uniform vapor pressure is assumed as initial condition:
\begin{align*}
& \Pv \egal P_{\,v,\,0} \,, \qquad t \egal 0 \,, \qquad \forall \ x \ \in \ \bigl[\, 0 \,,\, L \,\bigr] \,.
\end{align*}

The vapor flux density is similarly computed according to: 
\begin{align*}
g(\,t\,) &\ \eqdef \ - \, \kappa \, \pd{\Pv}{x}\biggl|_{\,x\egal x_{\,0}} \,.
\end{align*}

\subsection{Dimensionless formulation}

\begin{figure}
\centering
\includegraphics[width=.95\textwidth]{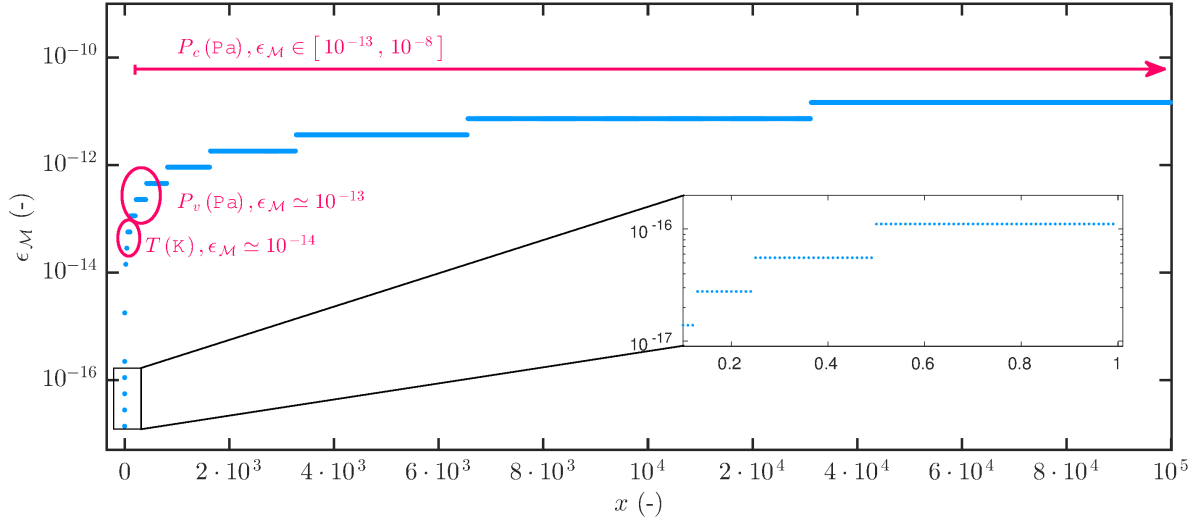}
\caption{Accuracy of the floating points in \texttt{Matlab\texttrademark} environment with analysis of the physical potential commonly used in the physical model of heat or mass transfer. }
\label{fig:float}
\end{figure}

While performing a mathematical and numerical analysis of a given practical problem, it is of capital importance to obtain a unitless formulation of governing equations, due to a number of good reasons. First of all, it enables to determine important scaling parameters such as the \textsc{Biot} and \textsc{Fourier} numbers. Henceforth, solving one dimensionless problem is equivalent to solve a whole class of dimensional problems sharing the same scaling parameters. Then, dimensionless equations allow to estimate the relative magnitude of various terms, and thus, eventually to simplify the problem using asymptotic methods \cite{Nayfeh_2000}. Finally, the floating point arithmetics is designed such as the rounding errors are minimal if computer manipulates the numbers of the same magnitude \cite{Kahan_1979}. Moreover, the floating point numbers have the highest density within the interval $(\, 0,\,1 \,)$ and their density decays exponentially when we move further away from zero. Figure~\ref{fig:float} shows the accuracy of the floating points in \texttt{Matlab\texttrademark} environment. It is generated using the $\mathrm{eps}(\,x\,$) function in the \texttt{Matlab} environment. For a numerical model written using dimensionless equations, the accuracy of the floating points scales with $\mathcal{O}(\,10^{\,-17}\,)\,$. The potential commonly used in the physical model of heat and mass transfer is usually the temperature $T$ in $\unit{K}$, the vapor pressure $\Pv$ or the capillary pressure $P_{\,c}$ both pressures in $\unit{Pa}\,$. According to Figure~\ref{fig:float}, if the numerical model is written using the temperature or the vapor pressure with their physical dimension, the accuracy of the floating points scales between $10^{\,-14}$ and $10^{\,-13}\,$, respectively. The range of the capillary pressure is between $10^{\,3} \unit{Pa}$ and $10^{\,8} \unit{Pa}\,$. Therefore, the accuracy of the floating point can loose up to $8$ orders compared to a dimensionless numerical model. So, it is always better to manipulate numerically the quantities of the order of $\mathcal{O}(1)$ to avoid severe round-off errors and to likely improve the conditioning of the problem in hands.

\subsubsection{Heat transfer}

In this way, according to \cite{Incropera_2007}, we define the following dimensionless quantities for the temperature: 
\begin{align*}
& u \egal \frac{T}{T_{\,0}} \,, 
&& u_{\,R} \egal \frac{T_{\,R}}{T_{\,0}} \,,
&& u_{\,L} \egal \frac{T_{\,L}}{T_{\,0}} \,.
\end{align*}
The time and space domains are also modified through a unitless formulation: 
\begin{align*}
& \xs \egal \frac{x}{L} \,, 
&& \ts \egal \frac{t}{t^{\,\circ}} \,,
\end{align*}
where $t^{\,\circ}$ is a characteristic time. The \textsc{Fourier} dimensionless number is defined, characterizing the importance of the heat transfer through the material: 
\begin{align*}
& \Fo_{\,T} \egal \frac{k \, t^{\,\circ}}{\rho \, c \, L^{\,2}} \,.
\end{align*}

Therefore, the unitless system of the differential equation of heat transfer is formulated as: 
\begin{align}
\label{eq:heat1d}
\pd{u}{\ts} \egal \, \Fo_{\,T} \, \pd{^{\,2} u}{x^{\,\star\, 2}} \,,
\end{align}
together with the boundary conditions:
\begin{align*}
& u \egal u_{\,R}  \,, && x \egal 0 \,, \\
& u \egal u_{\,L}  \,, && x \egal 1 \,, \\
\end{align*}
and the initial condition:
\begin{align*}
u \egal 1 \,, && t \egal 0 \,.
\end{align*}

\subsubsection{Mass transfer}

In a very similar way, we define the following dimensionless quantities related to the vapor pressure field \cite{Luikov_1966}: 
\begin{align*}
& v \egal \frac{\Pv}{P_{\,v,\,0}} \,, 
&& v_{\,R} \egal \frac{\Pv}{P_{\,v,\,0}} \,,
&& v_{\,L} \egal \frac{\Pv}{P_{\,v,\,0}} \,. 
\end{align*}
The unitless formulation of the time and space domains are:
\begin{align*}
& \xs \egal \frac{x}{L} \,, 
&& \ts \egal \frac{t}{t^{\,\circ}} \,.
\end{align*}
The vapor pressure dependent moisture properties are transformed according to: 
\begin{align*}
& \kappa^{\,\star} \egal \frac{\kappa}{\kappa^{\,\circ}} \,,
&& \xi^{\,\star} \egal \frac{\xi}{\xi^{\,\circ}} \,,
\end{align*}
where $\kappa^{\,\circ}$ and $\xi^{\,\circ}$ are reference property values. 
The \textsc{Fourier} dimensionless number is defined, characterizing here the importance of the moisture transfer through the material: 
\begin{align*}
\Fo_{\,m} \egal \frac{\kappa^{\,\circ} \, t^{\,\circ}}{\xi^{\,\circ} \, L^{\,2}} \,.
\end{align*}
The \textsc{Fourier} number quantifies the first order of the diffusion transfer, while the dimensionless parameters $\kappa^{\,\star}$ and $\xi^{\,\star}$ define the distortion or the nonlinearity of the phenomenon.

The unitless system of differential equation for moisture transfer is:
\begin{align}
\label{eq:mass1d}
\xi^{\,\star}\,(\,v\,) \, \pd{v}{\ts} \egal \, \Fo_{\,m} \, \pd{}{x} \, \biggl(\, \kappa^{\,\star}\,(\,v\,) \, \pd{ v}{x^{\,\star \, 2} } \, \biggr) \,,
\end{align}
with the boundary conditions:
\begin{align*}
& v \egal v_{\,R}  \,, && x \egal 0 \,, \\
& v \egal v_{\,L}  \,, && x \egal 1 \,, \\
\end{align*}
and the initial condition:
\begin{align*}
v \egal 1 \,, && t \egal 0 \,.
\end{align*}

\section{Numerical methods}
\label{sec:num_methods}

In order to describe the numerical schemes, let's first consider a uniform discretisation for simplicity of the interval $\Omega_{\,x} \ \rightsquigarrow\ \Omega_{\,h}\,$:
\begin{equation*}
  \Omega_{\,h}\ =\ \bigcup_{j\,=\,1}^{N} [\,x_{\,j},\;x_{\,j+1}\,]\,, \qquad
  x_{\,j+1}\ -\ x_{\,j}\ \equiv\ \Delta x\,, \qquad \forall j\ \in\ \bigl\{\,1,\,\ldots,\,N\,\bigr\}\,.
\end{equation*}
For the RC model, the time layers are uniformly spaced as well $t^{\,n}\ =\ m\,\Delta t\,$, $\Delta t\ =\ \const\ >\ 0\,$, $m\ =\ 0,\,1,\,2,\,\ldots, \, N_{\,t}$. The values of the function $u(x,\,t)$ in discrete nodes will be denoted by $u_{\,j}^{\,m}\ \eqdef\ u\,(x_{\,j},\,t^{\,m}\,)\,$. 

For the sake of simplicity, the standard finite-differences scheme and spectral approach will be described for the linear dimensionless heat diffusion equation~\eqref{eq:heat1d}. To our knowledge, the RC model is always described in the literature considering the physical dimension of the equation. In this way, the approach will be presented considering the heat diffusion equation~\eqref{eq:heat1d_dim}. 

\subsection{The standard finite-differences method}

The standard semi--discrete scheme based on central finite-differences can be written as:
\begin{align}\label{eq:exp}
& \od{u_{\,j}}{t} \ =\ \Fo\;\frac{u_{\,j-1} \moins 2\,u_{\,j} \plus u_{\,j+1}}{\Delta x^{\,2}}\,, \qquad j\ =\ 1,\,\ldots,\,N-1\,, \qquad n\ \geqslant\ 0\,,
\end{align}
whose starting value is directly obtained from the initial condition:
\begin{align*}
  u_{\,j}\, (\,0\,) \ =\ 1 \,.
\end{align*}
Many approaches can be used for the temporal discretisation of Eq.~\eqref{eq:exp}. Here the algorithm is implemented in \texttt{Matlab\texttrademark} environment using the function \texttt{ode45} providing an efficient explicit \textsc{Runge}--\textsc{Kutta} scheme. In whole figures, this approach will be referenced as FDM. 

\subsection{The RC model}

\begin{figure}
\centering
\subfigure[a][\label{fig:schemaRC_stencil}]{\includegraphics[width=.45\textwidth]{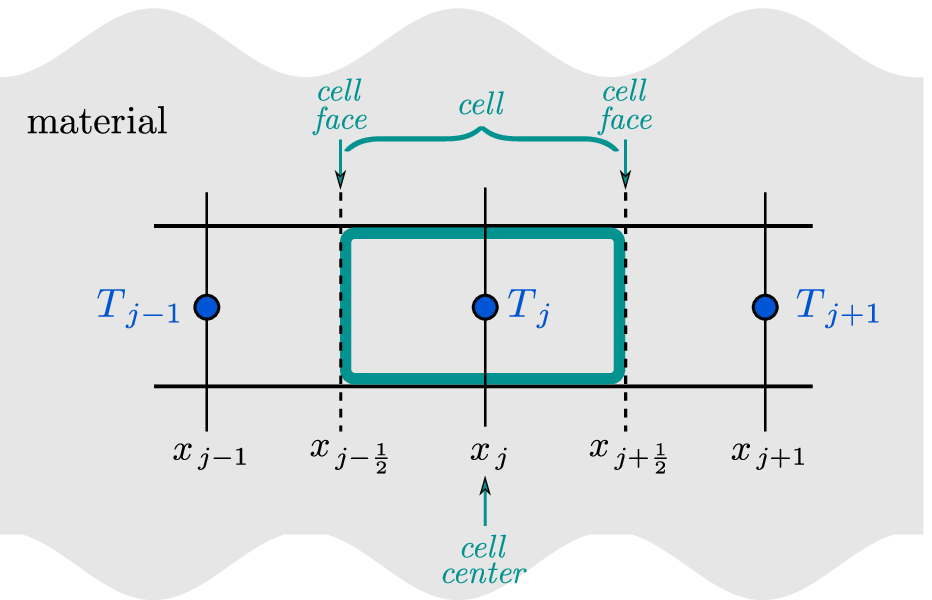}} \hspace{0.5cm}
\subfigure[b][\label{fig:schemaRC}]{\includegraphics[width=.45\textwidth]{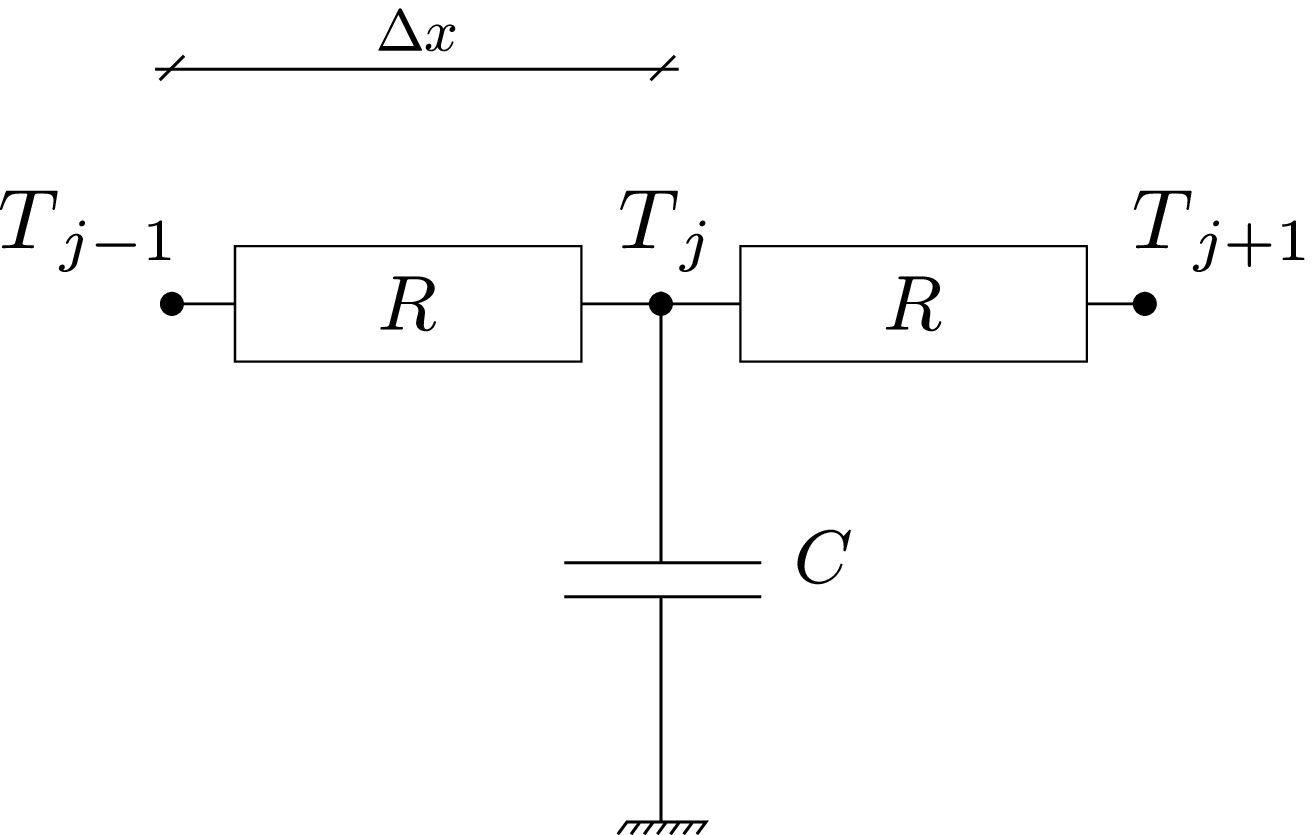}} 
\caption{ Illustration of the RC model: \emph{(a)} stencil and \emph{(b)} electrical analogy of the heat conduction transfer. }
\end{figure}

Both sides of the heat equation~\eqref{eq:heat1d_dim} is integrated over $x$ for the cell illustrated in Figure~\ref{fig:schemaRC_stencil}:
\begin{align}
\label{eq:heat1d_integrated}
\int_{x_{j-1/2}}^{\,x_{\,j+1/2}} \rho \, c \, \pd{T}{t}  \, \mathrm{d}x 
\egal \int_{x_{\,j-1/2}}^{\,x_{j+1/2}}\  k \pd{^{\,2} T}{x^{\,2}} \, \mathrm{d}x \,.
\end{align}
The average temperature of the cell is defined as 
\begin{align*}
T_{\,j} \ \eqdef \ \frac{1}{\dx} \, 
\displaystyle \int_{x_{\,j-1/2}}^{\,x_{j+1/2}} \ T(\,x\,,\,t\,)  \, \mathrm{d}x .
\end{align*}
Thus, Eq.~\eqref{eq:heat1d_integrated} becomes: 
\begin{align*}
\dx \, \rho \, c \, T_{\,j} \egal q_{\,j+ 1/2} \moins q_{\,j + 1/2} \,.
\end{align*}
From the electric analogy of the heat conduction  \cite[Chap. 10]{Davies_2004}, the heat flux is approximated by:
\begin{align*}
q_{\,j+1/2} \egal \frac{1}{R} \; \bigl(\, T_{\,j+1} \moins T_{\,j} \,\bigr) \,,
\end{align*}
where $R$ is the thermal resistance defined as:
\begin{align*}
& R \ \eqdef \ \frac{\dx}{k} \,.
\end{align*}
Finally, for each node $j$, the temperature is computed using: 
\begin{align}
\label{eq:schema_RC}
C \, \Delta x \, \od{T_{\,j}}{t} 
\egal  \frac{1}{R} \; \bigl(\, T_{\,j+1} \moins T_{\,j} \,\bigr)
\moins \frac{1}{R} \; \bigl(\, T_{\,j} \moins T_{\,j-1} \,\bigr) \,,
\end{align}
$C$ being the thermal capacity defined as:
\begin{align*}
& C \ \eqdef \ \rho \, c \,.
\end{align*}
The electric analogy of the heat conduction equation is illustrated in Figure~\ref{fig:schemaRC}.
It can be noted that Eq.~\eqref{eq:schema_RC} corresponds to the central finite-differences discretisation of the second space derivative. The RC model states that the temperature can be computed using a user-defined number $r \, \in \bigl\{\, 1 \,,\,\ldots\,,\, N \,\bigr\}$ of resistances. Usually, such a model is denoted as $R_{\,r}\,C\,$ with $r$ of the order of the unity $r \, \simeq \, \mathcal{O}(\,1\,)\,$. This hypothesis corresponds to compute the temperature using $r \moins 1$ points of discretisation. It can also be seen as a low fidelity model to represent the physical phenomena of heat or moisture transfer in building porous material. 

In this work, an \Eu ~explicit approach is associated with the semi-discrete scheme~\eqref{eq:schema_RC}, in agreement with \cite{Biddulph_2014}. In this case, it is important to note that the explicit scheme is only conditionally stable under the following \textsc{Courant}--\textsc{Friedrichs}--\textsc{Lewy}-type condition \cite{Courant_1928}:
\begin{align}\label{eq:cfl}
  \Delta t\ \leqslant\ \frac{1}{2\,\Fo}\;\Delta x^{\,2} \,.
\end{align}
The algorithm is implemented in the \texttt{Matlab\texttrademark} environment. Interested readers are invited to consult \cite{Davies_2004,Fraisse_2002} for more details on this approach and \cite{Deconinck_2016,Reynders_2014,Jimenez_2009} for examples of applications in building physics.

\subsection{The advanced spectral method}
\label{sec:spectral_method}

The spectral method has different approach, other than the central differences and thus, the RC ones. It assumes that the unknown $u\,(\,x,\,t\,)$ from Eq.~\eqref{eq:heat1d} can be approximately represented as a finite sum \cite{Mendes_2017}:
\begin{equation}
\label{eq:series_ap}
u\, (\,x,\, t\,)\ \approx\ u_{\, n}\, (\,x,\, t\,) \egal \sum_{i\, =\, 0}^{n} \, a_{\,i}\, (\,t\,)\, \mathsf{T}_{\,i}\, (\,x\,) \,.
\end{equation}
\
Here, $\{\mathsf{T}_{\,i}\, (\,x\,)\}_{\,i\, =\, 0}^{\,n}$ is a set of basis functions that remains constant in time. In this study, the \textsc{Chebyshev} polynomials are used as the basis functions since they are optimal in $\mathcal{L}_{\infty}$ approximation norm \cite{Gautschi_2004}. The functions $\{a_{\,i}\, (\,t\,)\}_{\,i\, =\, 0}^{\,n}$ are the corresponding time-dependent spectral coefficients. The parameter $n$ represents the number of degrees of freedom of the solution, also denoted as the order of the solution with $n \,\simeq\,\mathcal{O}(\, 10\,)\, \,$. The main advantage of the spectral method is that $n\, \ll\, p$, where $p$ is the number of degrees of freedom needed to solve problem~\eqref{eq:heat1d} by means of conventional methods such as finite-differences, finite-volume or finite-element methods. For these reasons, the spectral method is also denoted as the spectral-Reduced Order Method (spectral-ROM) \cite{Gasparin_2017a,Gasparin_2017b}. 

The derivatives are written as follows:
\begin{subequations}\label{eq:derivatives}
\begin{align}
\pd{u_{\,n}}{x} &\egal \sum_{i\, =\, 0}^n \, a_{\,i}\,(\,t\,)\, \pd{\mathsf{T}_{\,i}}{x}\,(\,x\,)\egal \sum_{i\, =\, 0}^n \tilde{a}_{\,i}\,(\,t\,)\, \mathsf{T}_{\,i}\,(\,x\,)\,,\label{eq:derivative1}\\
\pd{^{\,2} u_{\,n}}{x^{\,2}} &\egal \sum_{i\, =\, 0}^n \, a_{\,i}\,(\,t\,)\, \pd{^{\,2} \mathsf{T}_{\,i}}{x^{\,2}}\,(\,x\,)\egal \sum_{i\, =\, 0}^n \Tilde{\Tilde{a}}_{\,i}\,(\,t\,)\, \mathsf{T}_{\,i}\,(\,x\,) \,, \label{eq:derivative2}\\
\pd{u_{\,n}}{t} &\egal \sum_{i\, =\, 0}^n \, \dot{a}_{\,i}\,(\,t\,)\, \mathsf{T}_{\,i}\,(\,x\,)\,,\label{eq:derivative3} 
\end{align}
\end{subequations}
where the dot denotes $\dot{a}_{\,i}\, (\,t\,) \ \eqdef \ \dfrac{\mathrm{d}a\, (\,t\,) }{\mathrm{d}t} $ according to \textsc{Newton} notation. Using the properties of the \textsc{Chebyshev} polynomials, the space derivatives are re-expanded in the same \textsc{Chebyshev} basis function. 
The connection is explicitly given from the recurrence relation of the \textsc{Chebyshev} polynomial derivatives  \cite{peyret_2002}:
\begin{align*}
& \tilde{a}_{\,i} \egal \dfrac{2}{c_{\,i}} \sum_{\substack{p\, =\,i\,+\,1 \\ p\,+\,i\; \text{odd}}}^{\,n} \, p \, a_{\,p}\, , & 
i \egal 0,\ldots,n-1, 
\\[3pt]
& \Tilde{a}_{\,n} \ \equiv \ 0 \,, \\[3pt]
& \Tilde{\Tilde{a}}_{\,i} \egal \dfrac{1}{c_{\,i}} \sum_{\substack{p\, =\, i\,+\,2 \\ p\,+\,i\; \text{even}}}^{\,n}\, p\,\Bigl(\,p^{\,2} \moins i^{\,2}\,\Bigr)\, a_{\,p}\, , & 
i \egal 0,\ldots,n-2, 
\\[3pt]
& \Tilde{\Tilde{a}}_{\,n-1} \ \equiv \  \Tilde{\Tilde{a}}_{\,n} \ \equiv \ 0 \,,
\end{align*}
with,
\begin{align*}
c_{\,i} \egal \left\lbrace 
\begin{matrix}
2 \,,& \text{if} & i \egal 0\,,\\
1 \,,& \text{if} & i \ >\ 0\,.
\end{matrix} \right.
\end{align*}

Using the expression of the derivatives provided by Eqs.~\eqref{eq:derivative2} and \eqref{eq:derivative3}, the residual of the diffusion equation~\eqref{eq:heat1d} is: 
\begin{align}
\label{eq:heat_eq_residual}
R \,(\,x\,,\,t \,) \egal \sum_{i\, =\, 0}^n \, \Bigl[\, \dot{a}_{\,i}\, (\,t\,) \moins \Fo \ \Tilde{\Tilde{a}}_{\,i}\, (\,t\,)\, \Bigr]\, \mathsf{T}_{\,i}\, (\,x\,)\,,
\end{align}
which is considered a misfit of the approximate solution. The purpose is to minimize the residual:
\begin{align*}
\bigl|\,\bigl|\, R(\,x\,,\,t \,) \, \bigr|\,\bigr| \longrightarrow \min \,,
\end{align*}
which is realised via the the \textsc{Tau}--\textsc{Galerkin} method, which requires Eq.~\eqref{eq:heat_eq_residual} to be orthogonal to the \textsc{Chebyshev} basis functions $\langle\,R\,,\mathsf{T}_{\,i} \,\rangle \,=\, 0$. Here, the the scalar product is defined by :
\begin{align*}
\langle \, f \,,\, g \, \rangle \egal \int_{-1}^{\,1} \, \dfrac{f\,(\,x\,) \, g\,(\,x\,)}{\sqrt{1 \moins x^{\,2}}} \; \mathrm{d}x \,.
\end{align*}
Thus, it leads to the following relations among spectral coefficients: 
\begin{align*}
\dot{a}_{\,i}\, (\,t\,) \moins \nu \, \Tilde{\Tilde{a}}_{\,i}\, (\,t\,) \egal 0\,, & & 
i \egal 0, \,1, \,\ldots,\, n-2 \,.
\end{align*}

Finally, after the projection and expansion of the residual, the original partial differential equation~\eqref{eq:heat1d} is reduced to a system of ordinary differential equations plus two algebraic expressions enabling to compute the time dependent coefficients $\{a_{\,i}\, (\,t\,) \}\,$. For linear problems, the system of ordinary differential equations is explicitly built:
\begin{align*}
& \begin{cases} 
& \dot{a} \egal \A \, a \plus \b\, (\,t\,) \,, \\
& a\,(\,0\,) \egal a_{\,0} \,,
\end{cases}
\end{align*}
where $ \A \in \Mat_{n\times n}(\R )\,$, with constant coefficients, $\b(\,t\,) \in \R^{\,n}$ is a vector resulting from boundary conditions and $a_{\,0}$ is the vector of initial coefficients. Initial values of the coefficients $\{a_{\,i}\,(t\, =\, 0)\}$ are calculated by the \textsc{Galerkin} projection of the initial condition \cite{canuto_spectral_2006}:
\begin{align}
\label{eq:cond_initial_sp}
a_{\,0\,,\,i} \ \equiv \ a_{\,i}\,(\, 0\,) \egal \dfrac{2}{\pi\, c_{\,i}}\, \int_{-1}^{\,1}\, \dfrac{u_{\, 0}\,(\,x\,)\, \mathsf{T}_{\,i}\,(\,x\,)}{\sqrt{1 \moins x^{\,2}}}\, \mathrm{d}x\,, & &
i \egal 0, \,1, \,\ldots,\, n \,,
\end{align}
where $u_{\, 0}\,(\,x\,)$, is the dimensionless initial condition. Interested readers may refer to \cite{Gasparin_2017a,Gasparin_2017b} for further details on the spectral method.

\subsection{Extension of the methods for nonlinear problems}

The extension of the three methods for nonlinear problem is now detailed. For the standard semi--scheme based on central finite-differences for  Eq.~\eqref{eq:mass1d}, it is formulated as:
\begin{align*}
\xi^{\,\star} (\,v_{\,j}\,) \, \od{v_{\,j}}{\ts} 
\egal \, \Fo \, \frac{1}{\dx^{\,2}} \,
\Biggl(\, \kappa^{\,\star} \bigl(\,v_{\,j+\half}\,\bigr) \, v_{\,j+1} 
& \\ \moins 
& \Bigl(\, \kappa^{\,\star} \bigl(\,v_{\,j+\half}\,\bigr) \plus \kappa^{\,\star} \bigl(\,v_{\,j-\half}\,\bigr) \,\Bigr) \, v_{\,j} 
 \plus
\kappa^{\,\star} \bigl(\,v_{\,j-\half}\,\bigr) \, v_{\,j-1}  \,\Biggr) \,,
\end{align*}
where 
\begin{align*}
\kappa^{\,\star} \bigl(\,v_{\,j+\half}\,\bigr) \egal \kappa^{\,\star} 
\Biggl(\, \frac{1}{2} \, \Bigl(\, v_{\,j} \plus v_{\,j+1} \,\Bigr) \, \Biggr) \,.
\end{align*}

For the RC model, the extension for nonlinear problem in its physical dimension Eq.~\eqref{eq:moisture_1d_dim} is given by:
\begin{align*}
C_{\,j} \, \dx \, \od{P_{\,v\,,\,j}}{t} 
\egal 
\frac{P_{\,v\,,\,j+1}}{R_{\,j+1} \plus R_{\,j}}
\moins \biggl(\, \frac{1}{R_{\,j+1} \plus R_{\,j}} \plus \frac{1}{R_{\,j-1} \plus R_{\,j}} \,\biggr) \; P_{\,v\,,\,j} 
\plus \frac{P_{\,v\,,\,j-1}}{R_{\,j-1} \plus R_{\,j}} \,,
\end{align*}
where $R_{\,j}$ is the vapor resistance and $C_{\,j}$ being the moisture capacity, both defined respectively as:
\begin{align*}
& R_{\,j} \ \eqdef \ \frac{\dx}{2 \, \kappa \bigl(\,P_{\,v \,,\, j}\,\bigr) } \,, && C_{\,j} \ \eqdef \ \xi \bigl(\,P_{\,v \,,\, j}\,\bigr) \,. &&
\end{align*}

For the spectral method, Eq.~\eqref{eq:mass1d} is rearranged as follows:
\begin{align}
\label{eq:mass1d_Spectral}
\pd{v}{t} &\egal \nu \, (\, v \, ) \, \pd{^{\,2}v}{x^{\,2}} \plus \lambda \, (\, v \, )\, \pd{v}{x} \,,
\end{align}
where,
\begin{align*}
& \nu \, (\, v \, ) \eqdef \dfrac{\kappa^{\,\star}\, (\, v \,)}{\xi^{\,\star}\, (\, v \,) } \,, &&
\lambda \, (\, v \, ) \eqdef \dfrac{1}{\xi^{\,\star}\, (\, v \,) }  \cdot  \dfrac{\mathrm{d} \bigl(\, \kappa^{\,\star}\, (\, v \,)\, \bigr)}{\mathrm{d} v}\,.
\end{align*}

As described in Section~\ref{sec:spectral_method}, the unknown $v\,(\,x\,,t\,)$ is approximated by the finite sum \eqref{eq:series_ap} with \textsc{Chebyshev} polynomials as basis functions. The derivatives are written as in the linear case, by Eqs.~\eqref{eq:derivative1}, \eqref{eq:derivative2} and \eqref{eq:derivative3}. Substituting them into Eq.~\eqref{eq:mass1d_Spectral}, we get: 
\begin{align}
\label{eq:mass1d_Spectral2}
\sum_{i\, = \, 0}^{n} \,  \dot{a}_{\,i}\, (\,t\,)\, \mathsf{T}_{\,i}\, (\,x\,) \egal \nu \, \Biggl(\, \sum_{i\, = \, 0}^{n} \, a_{\,i} \,(\,t\,)\, \mathsf{T}_{\,i}\,(\,x\,) \, \Biggr) \, \sum_{i\, =\, 0}^{n} \, \Tilde{\Tilde{a}}_{\,i}\, (\,t\,)\, \mathsf{T}_{\,i}\, (\,x\,) \plus \nonumber \\ 
\lambda \, \Biggl( \, \sum_{i\, = \, 0}^{n} \, a_{\,i}\, (\,t\,)\, \mathsf{T}_{\,i}\, (\,x\,)\,  \Biggr) \sum_{i\, =\, 0}^{n}\, \tilde{a}_{\,i}\,(\,t\,)\, \mathsf{T}_{\,i}\,(\,x\,) \,.
\end{align}
Then, the \textsc{Tau}--\textsc{Galerkin} method is used to minimize the residual of the equation. The integrals of the nonlinear coefficients $\nu \,(\, v \, )$ and $\lambda \, (\, v \, )$ are computed using the \textsc{Chebyshev--Gau}\ss{} quadrature. At the end, it results in a system of Differential-Algebraic Equations (DAEs) with the following form:
\begin{align*}
\M \, \dot{a}_{\,n}\,(\,t\,) \egal \A \, a_{\,n}\,(\,t\,) \plus \b\,(\,t\,) \,,
\end{align*}
where, $\M$ is a diagonal and singular matrix containing the coefficients of the \textsc{Chebyshev} weighted orthogonal system, $\b\,(\,t\,)$ is a vector containing the boundary conditions and, $\A \cdot a_{\,n}\,(\,t\,)$ is composed by the right member of Eq.~\eqref{eq:mass1d_Spectral2} projected on the \textsc{Chebyshev} basis functions. The initial condition is given by Eq.~\eqref{eq:cond_initial_sp} and the DAE system is solved by \texttt{ode}15s or \texttt{ode}23t from \texttt{Matlab\texttrademark}.

\subsection{Methods implementation and metrics of their efficiency}

All the numerical algorithms for the classic and advanced schemes are written using dimensionless variables, while for the RC model, the physical dimensional variables are used. The numerical solutions are computed using an adaptive time step $\dt$ using \texttt{Matlab\texttrademark} function \texttt{ode45} \cite{Shampine_1997} with an absolute and relative tolerances set to $10^{\,-4}$. The efficiencies of the method are evaluated in terms of three criteria: (i) the global error of the numerical solution, (ii) the significant digits of the solution and (iii) the computational run time to compute the solution.

To evaluate the error, a reference solution $u^{\, \mathrm{ref}}\, (\,x \,, t \,)$ is computed using a numerical pseudo--spectral approach obtained with the \texttt{Matlab\texttrademark} open source toolbox \texttt{Chebfun} \cite{Chebfun_2014}. Using the function \texttt{pde23t}, it permits to compute a numerical solution of a partial derivative equation using the \textsc{Chebyshev} functions. This useful package enables to compute reference solutions for one dimensional space-time problems. This tools is chosen since it can deal with more complex problems than analytical solution. Indeed, it can consider nonlinear coefficients or \textsc{Robin}-type time-dependent boundary conditions, which are more related to building physics application. The error between the solution, obtained by the numerical methods described above, and the reference one is computed as a function of $x$ by the following formula:
\begin{align*}
\varepsilon_{\,2}\, (\, x\,)\ &\ \eqdef \ \sqrt{\,\frac{1}{N_{\,t}} \, \sum_{j\, =\, 1}^{N_{\,t}} \, \left( \, u_{\, j}\, (\,x \,, t \,) \moins u_{\, j}^{\mathrm{\, ref}}\, (\,x \,, t \,) \, \right)^{\,2}}\,,
\end{align*}
where $N_{\,t}$ is the number of temporal steps. 
The global uniform error $\L_{\, \infty}$ is given by the maximum value of $\varepsilon_{\,2}\, (\, x\,)\,$: 
\begin{align*}
\varepsilon_{\, \infty}\ &\ \eqdef \ \sup_{x \ \in \ \bigl[\, 0 \,,\, L \,\bigr]} \, \varepsilon_{\,2}\, (\, x\,) \,.
\end{align*}

The significant correct digits of the solution are evaluated according to \cite{Soderling_2003}:
\begin{align*}
\mathrm{scd}\, (\, u\,) &\ \eqdef \moins 
\log_{\,10} \, \biggl|\biggl|\, 
\frac{u\, (\,x \,, \tau \,) \moins u^{\, \mathrm{ref}}\, (\,x \,, \tau \,)}{u^{\, \mathrm{ref}}\, (\,x \,, \tau \,)} 
\, \biggr|\biggr|_{\,\infty} \,.
\end{align*}

As the RC approach computes directly the fields in their physical dimension, a scaling transformation is performed to compute the errors. The last criteria is the computational (CPU) run time required by the numerical model to compute the solution. It is measured using the \texttt{Matlab\texttrademark} environment with a computer equipped with \texttt{Intel} i$7$ CPU and $32$ GB of RAM. We define the ratio:
\begin{align*}
R_{\,\mathrm{cpu}} \ \eqdef \ \frac{t_{\,\mathrm{cpu}}}{\tau} \,,
\end{align*}
where $t_{\,\mathrm{cpu}} \ \unit{s}$ is the CPU time and $\tau$ is the final physical time of the simulation.

\section{Numerical investigations: linear heat diffusion}
\label{sec:linear_heat_diff}

Since the three methods have been presented, a first linear case of heat diffusion is considered to evaluate their efficiency, considering the following input values for a concrete slab:
\begin{align*}
& k \egal 2.0 \unit{W/(m.K)} \,, 
&& \rho \egal 1000 \unit{kg/m^{\,3}} \,,
&& c \egal 2000 \unit{J/(kg.K)} \,, \\
& L \egal 0.1 \unit{m} \,, 
&& \tau \egal 24 \unit{h} \,, 
&& T_{\,0} \egal 20 \unit{^\circ C} \,.
\end{align*}
The boundary conditions are defined as sinusoidal variations:
\begin{align*}
T_{\,L} & \egal T_{\,0} \plus 10\, \sin \biggl(\, \frac{2 \, \pi }{24 \cdot 3600} \, t \,\biggr) \,, && 
T_{\,R} \egal T_{\,0} \plus 4\, \sin \biggl(\, \frac{2 \, \pi }{3 \cdot 3600} \, t \,\biggr) \,.
\end{align*}

\begin{figure}
\centering
\subfigure[a][\label{fig_AN1:RC_T24h_fx}]{\includegraphics[width=.45\textwidth]{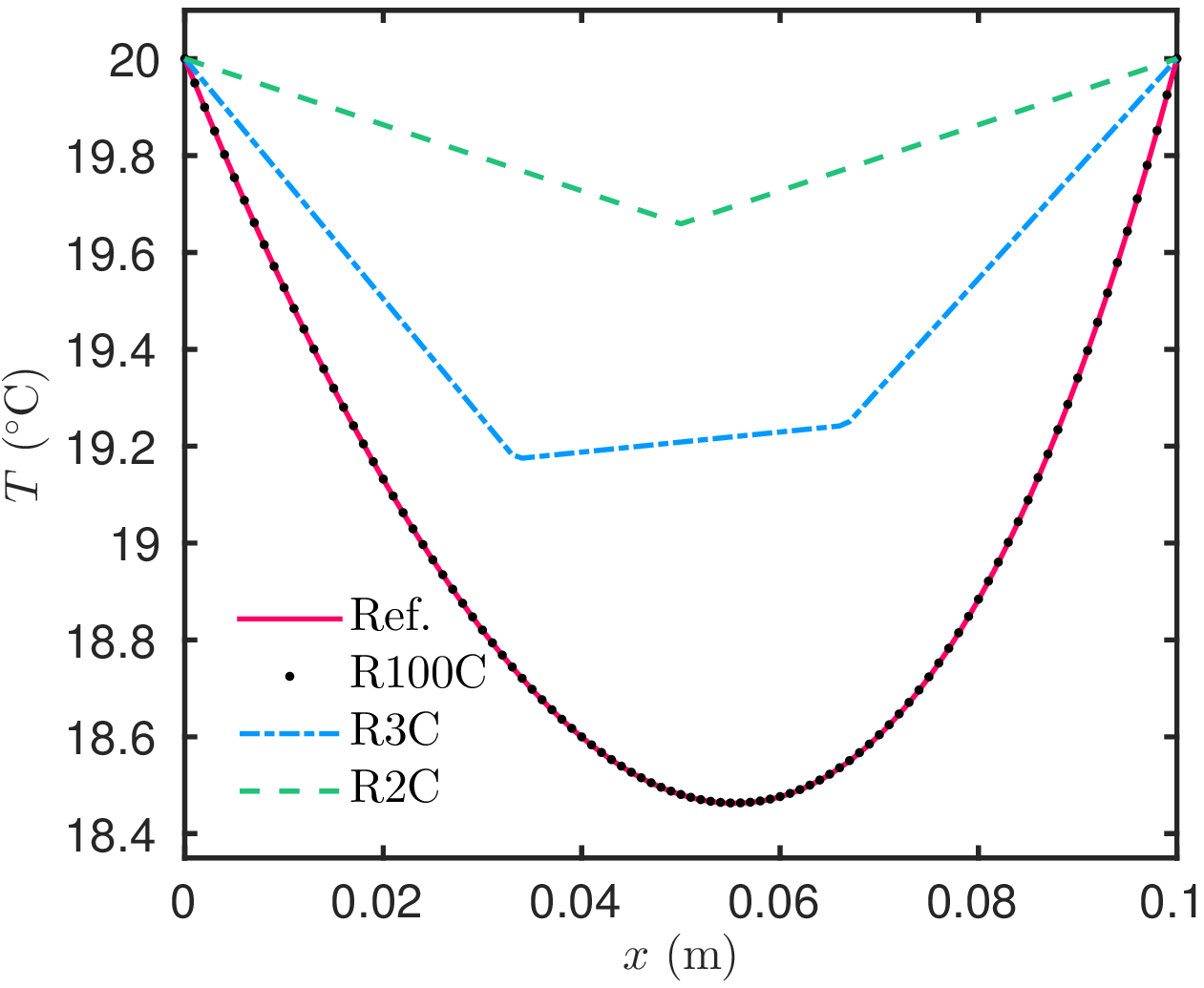}}
\subfigure[a][\label{fig_AN1:SP_EU_T24h_fx}]{\includegraphics[width=.45\textwidth]{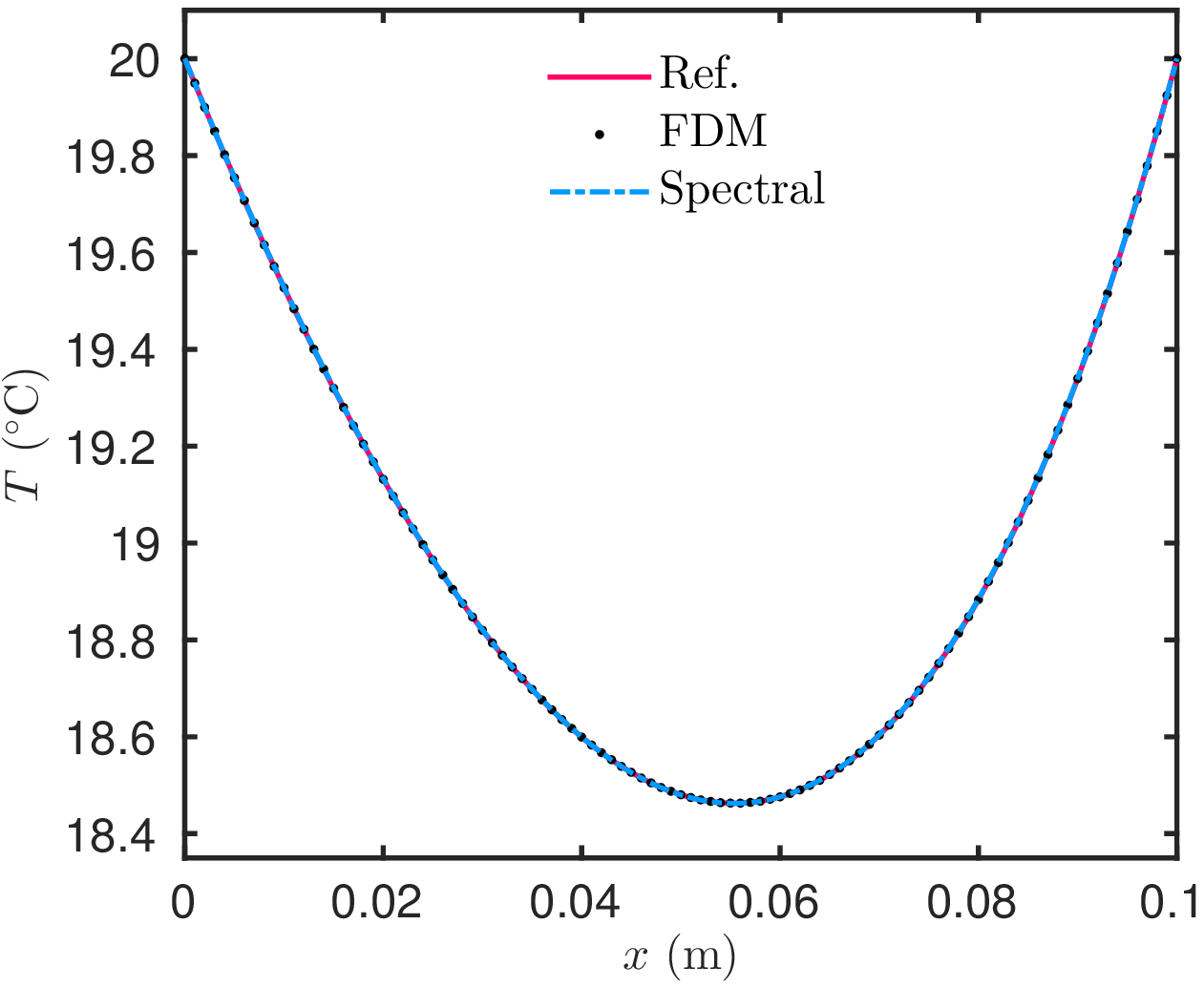}}
\caption{Temperature profiles at $t \egal 24 \unit{h}$.}
\label{fig_AN1:T24h_fx}
\end{figure}

\begin{figure}
\centering
\subfigure[a][\label{fig_AN1:RC_Tmid_ft}]{\includegraphics[width=.45\textwidth]{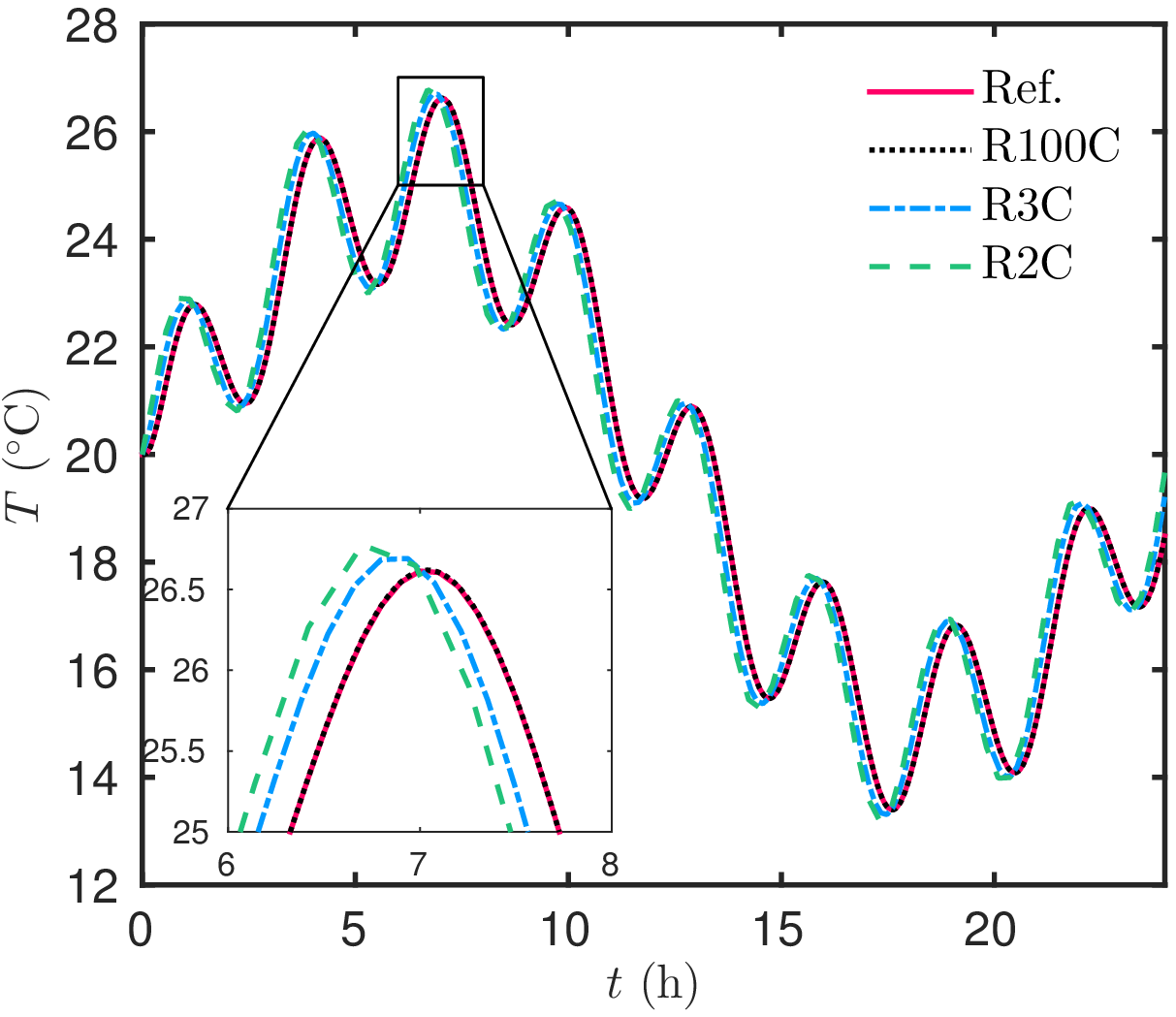}} 
\subfigure[b][\label{fig_AN1:SP_EU_Tmid_ft}]{\includegraphics[width=.45\textwidth]{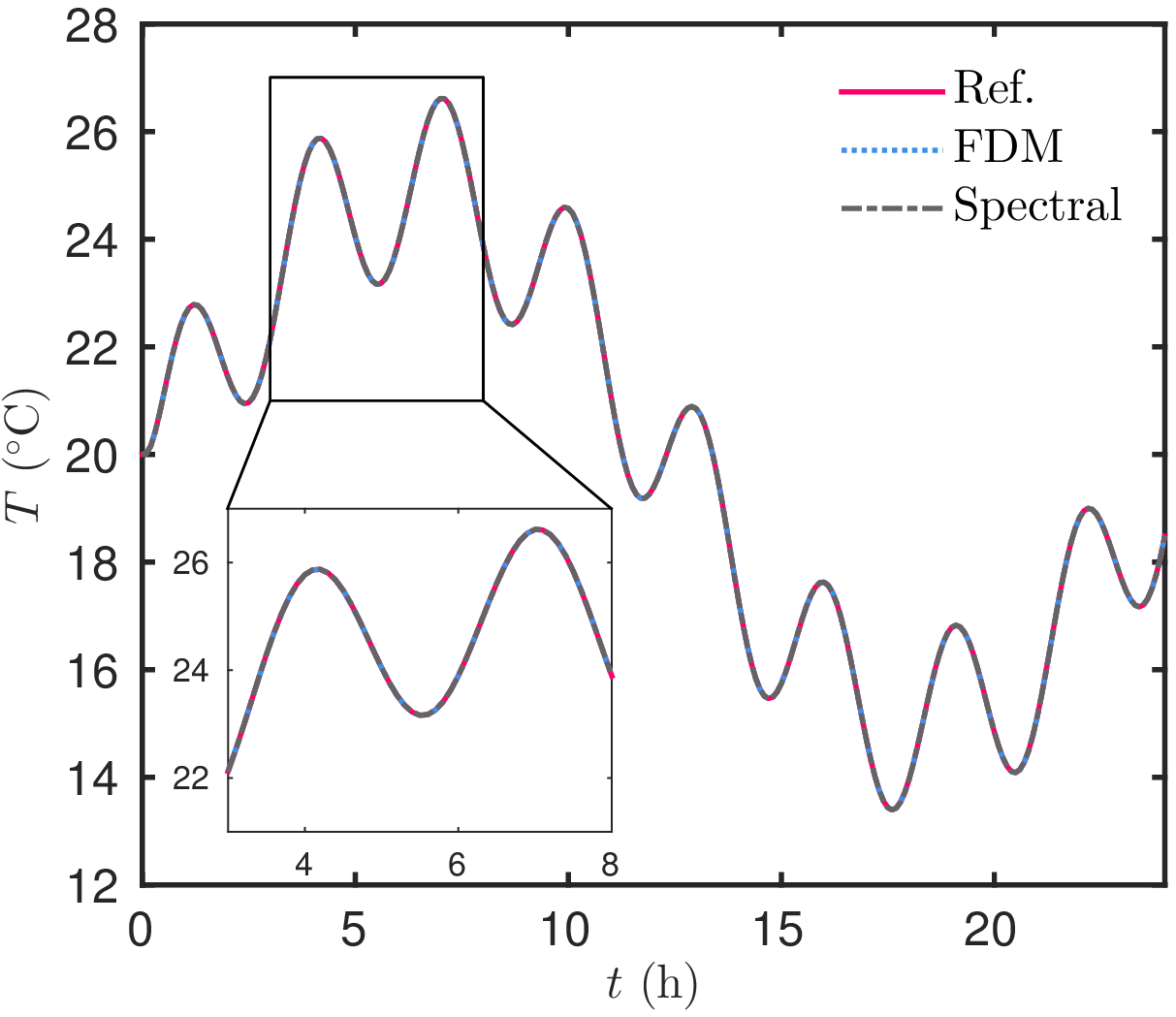}} 
\subfigure[c][\label{fig_AN1:RC_Emid_ft}]{\includegraphics[width=.45\textwidth]{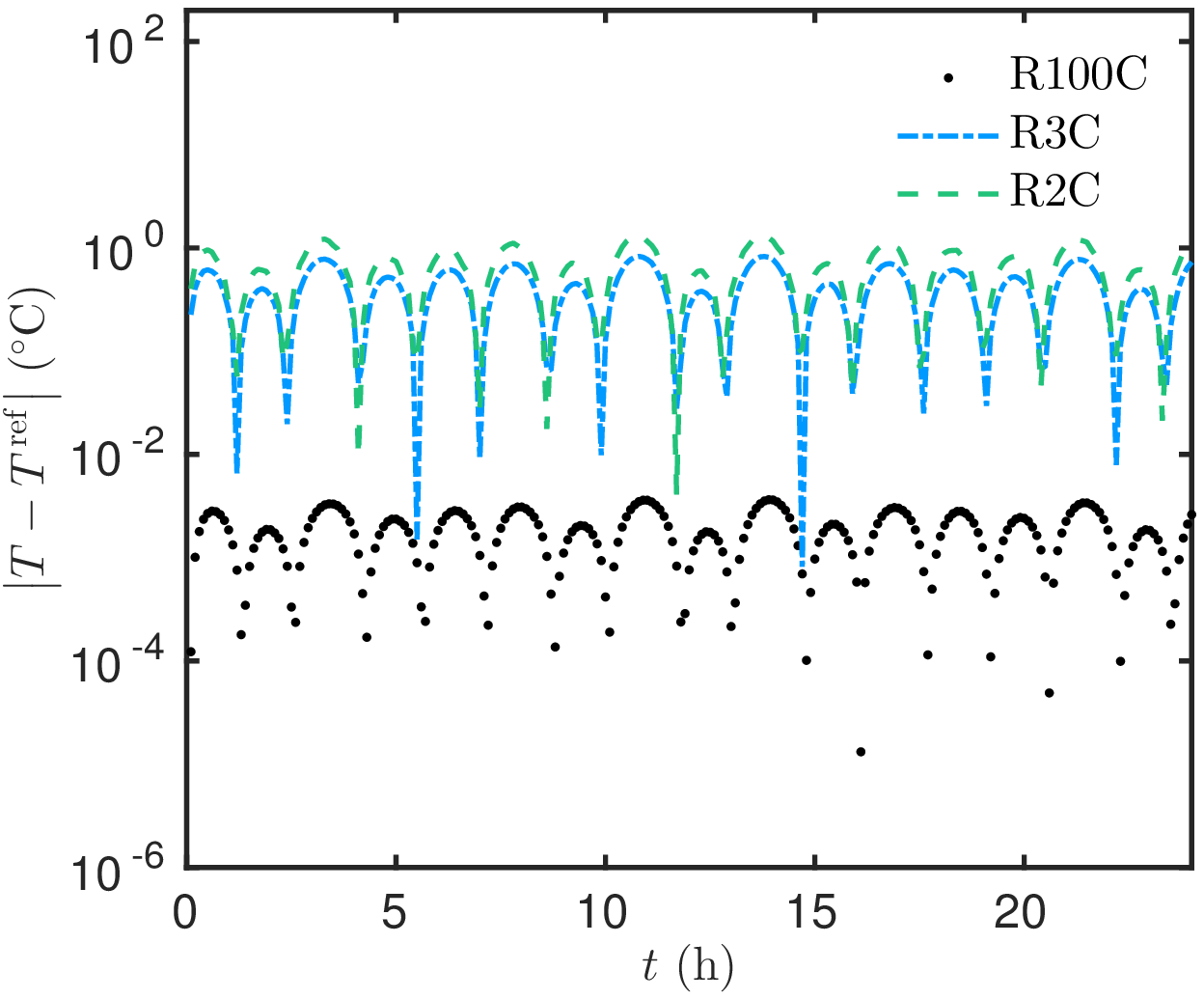}} 
\subfigure[d][\label{fig_AN1:SP_EU_Emid_ft}]{\includegraphics[width=.45\textwidth]{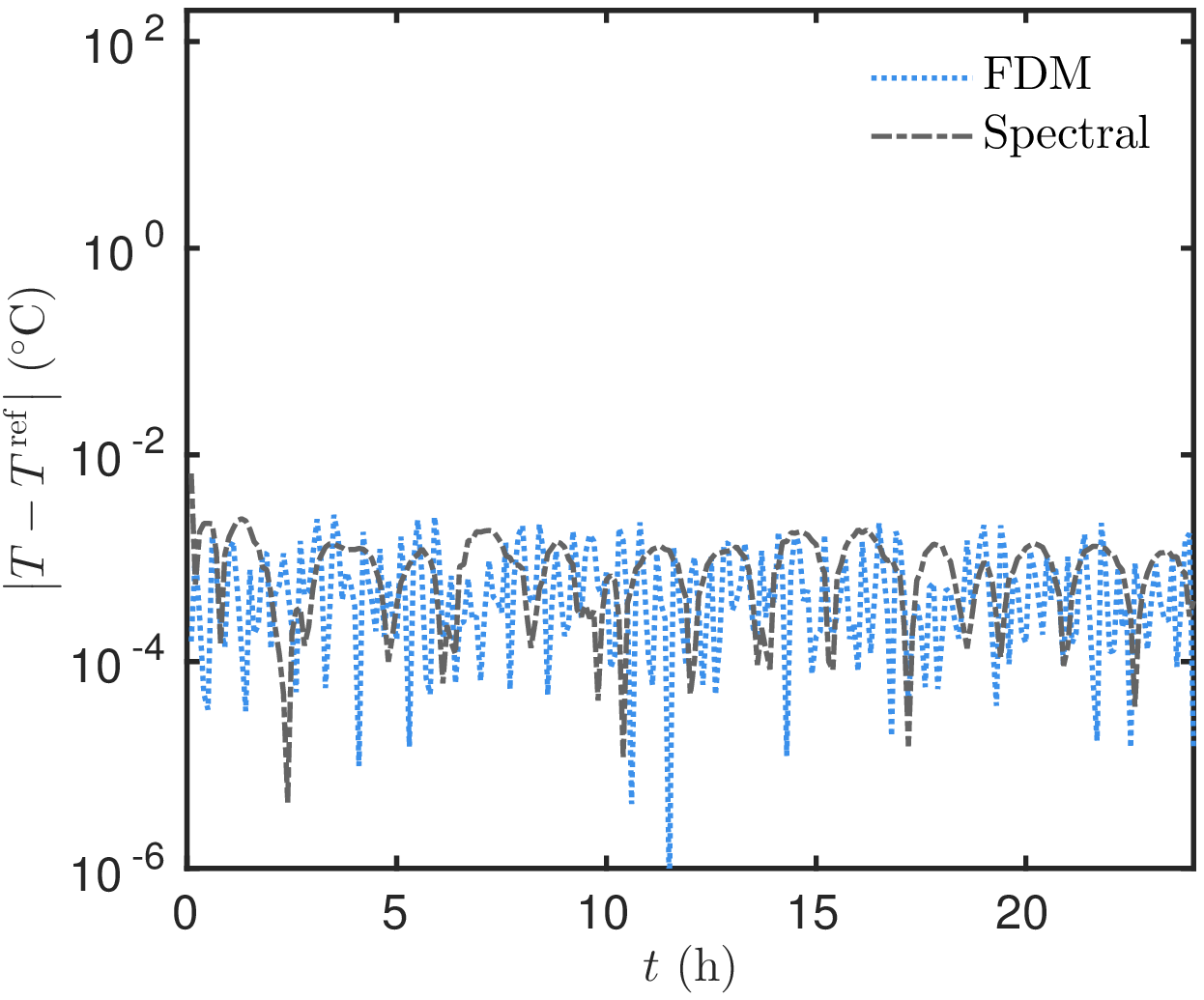}} 
\caption{\emph{(a,b)} Temperature evolution at $x \egal 0.05 \unit{cm}$. \emph{(c,d)} Temperature difference with respect to the reference solution.}
\end{figure}

The temperature is computed using three RC model approaches with the thermal resistances $r \in \bigl\{\,2 \,,\, 3 \,,\, 100 \,\bigr\} \,$. In addition, the problem is solved using the standard finite-differences method and spectral approach with $N \egal 6$ modes. A spatial discretisation step $\dx \egal 10^{\,-3} \ \mathsf{m}$ is considered for both approaches. The temperature profiles at the last time of the simulation are shown in Figures~\ref{fig_AN1:T24h_fx}~\textit{(a,b)}. It can be noted that the approaches with two or three resistances cannot compute an accurate temperature profile. A perfect agreement is observed between the reference and the solution computed using the \RC{100}, the standard finite-differences and the spectral approaches. The time evolution of the temperature in the middle of the wall is shown in Figures~\ref{fig_AN1:RC_Tmid_ft} and \ref{fig_AN1:SP_EU_Tmid_ft}. Apparently, it seems that each approach enables to represent the temperature evolution. However, as shown in Figure~\ref{fig_AN1:RC_Emid_ft}, the difference with the reference solution can reach $0.2 \unit{^\circ C}$ for the approach with two resistances. For the spectral approach, the difference is of the order $\mathcal{O}(\,10^{\,-3}\,)\,$.

\begin{figure}
\centering
\subfigure[a][\label{fig_AN1:RC_qR_ft}]{\includegraphics[width=.45\textwidth]{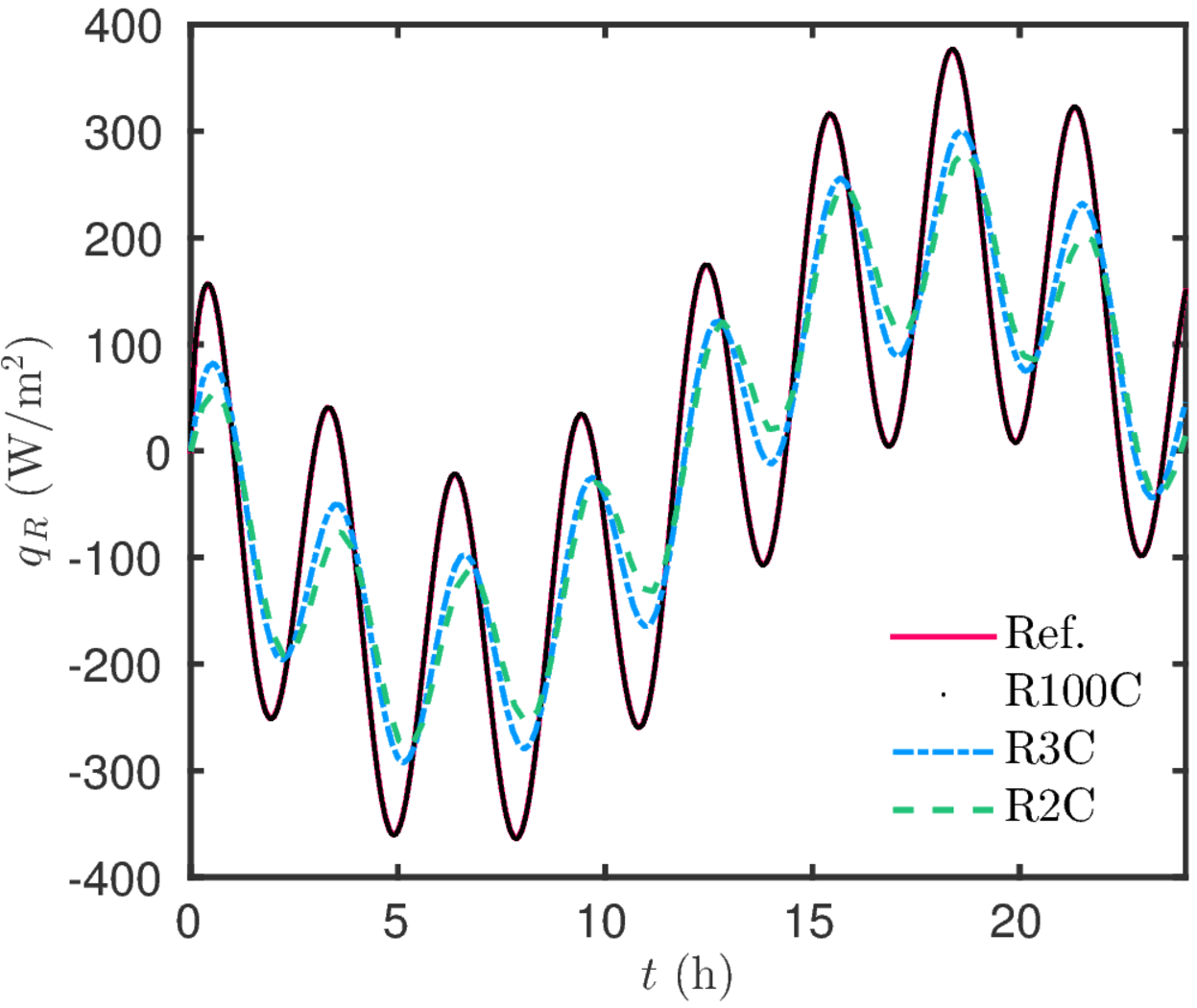}} 
\subfigure[b][\label{fig_AN1:SP_EU_qR_ft}]{\includegraphics[width=.45\textwidth]{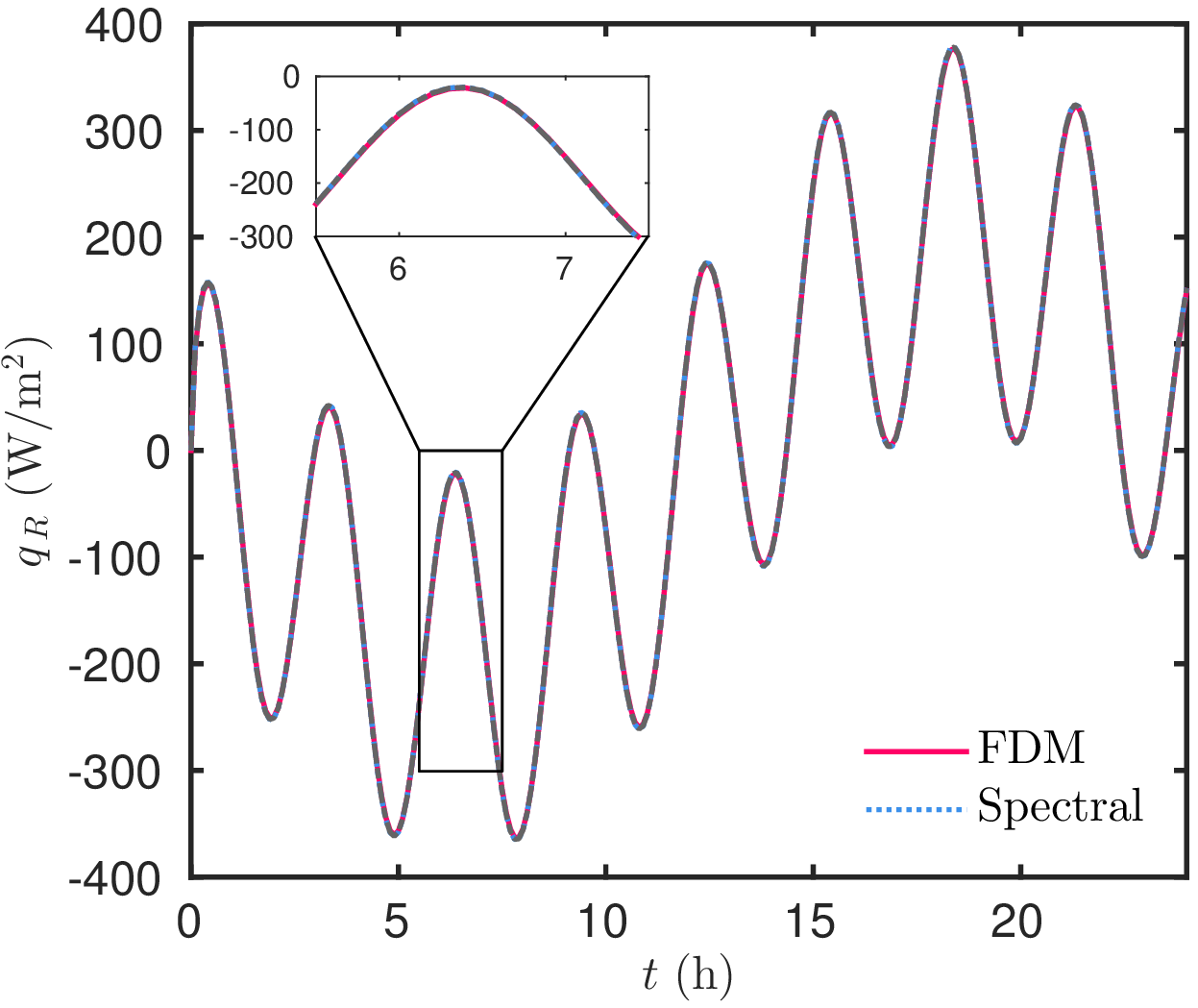}} 
\subfigure[a][\label{fig_AN1:RC_EqR_ft}]{\includegraphics[width=.45\textwidth]{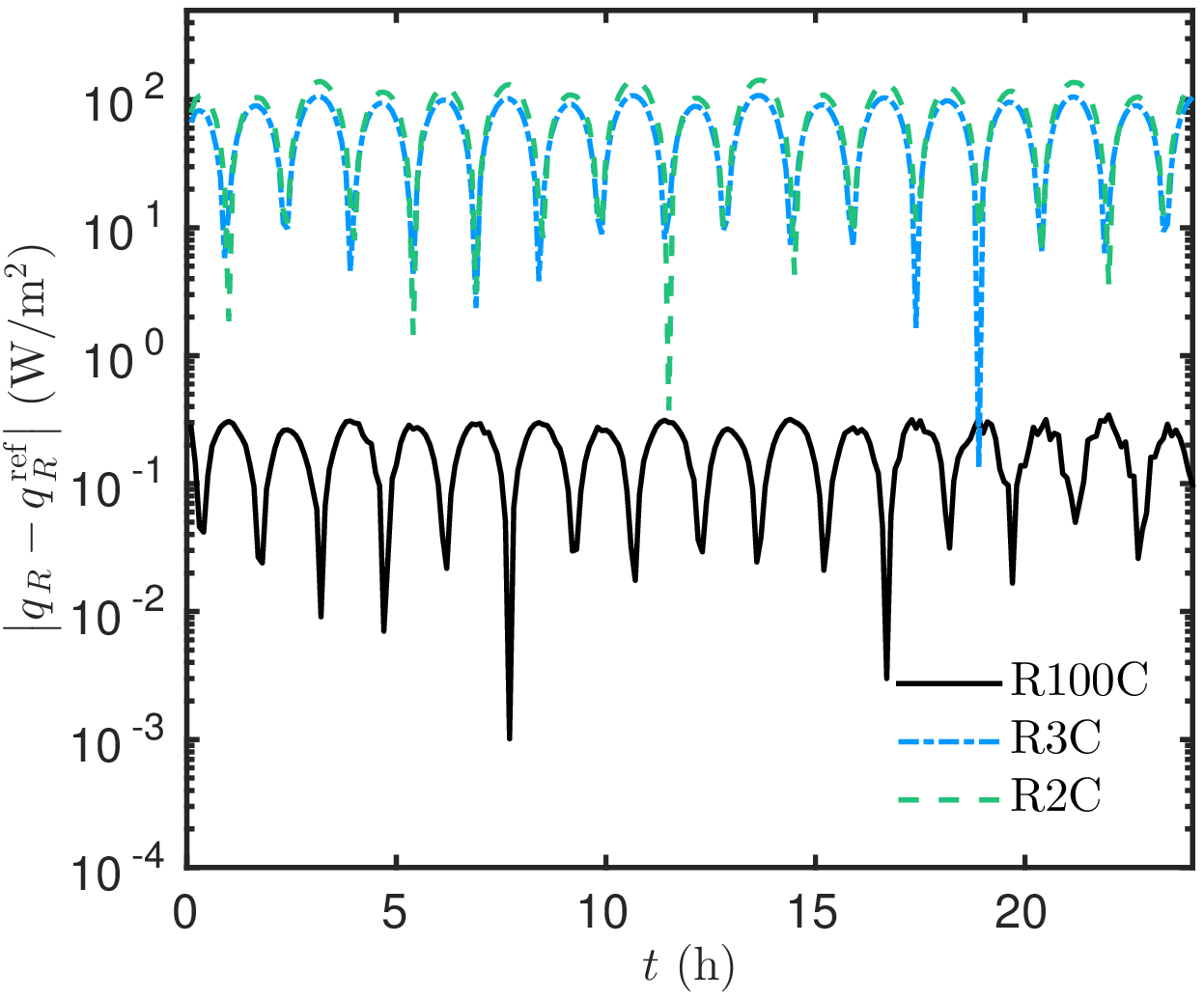}} 
\subfigure[b][\label{fig_AN1:SP_EU_EqR_ft}]{\includegraphics[width=.45\textwidth]{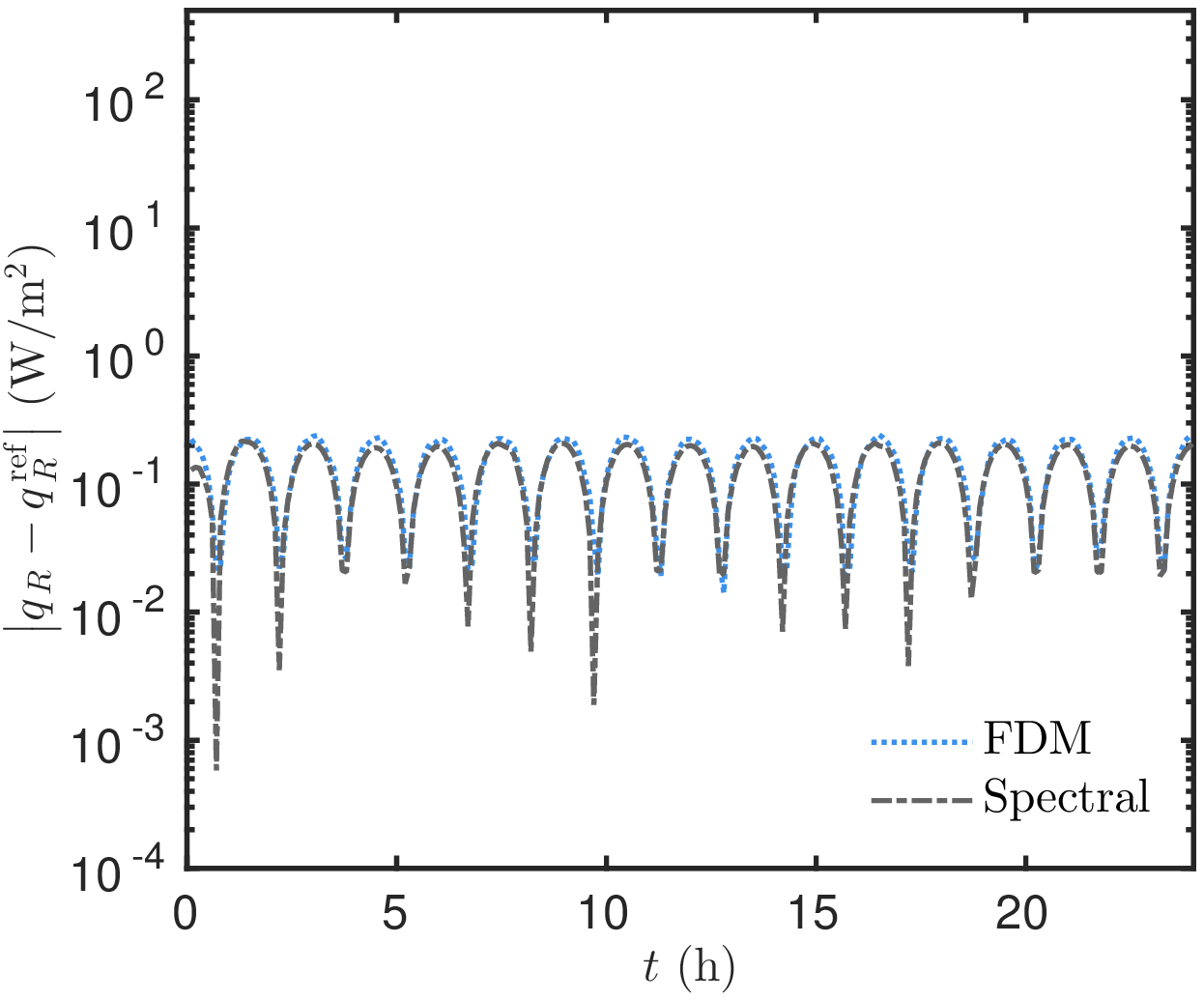}} 
\caption{\emph{(a,b)} Evolution of the heat flux density at the right boundary. \emph{(c,d)} Heat flux density difference to the reference solution.}
\end{figure}

\begin{figure}
\centering
\subfigure[a][\label{fig_AN1:e_fx}]{\includegraphics[width=.45\textwidth]{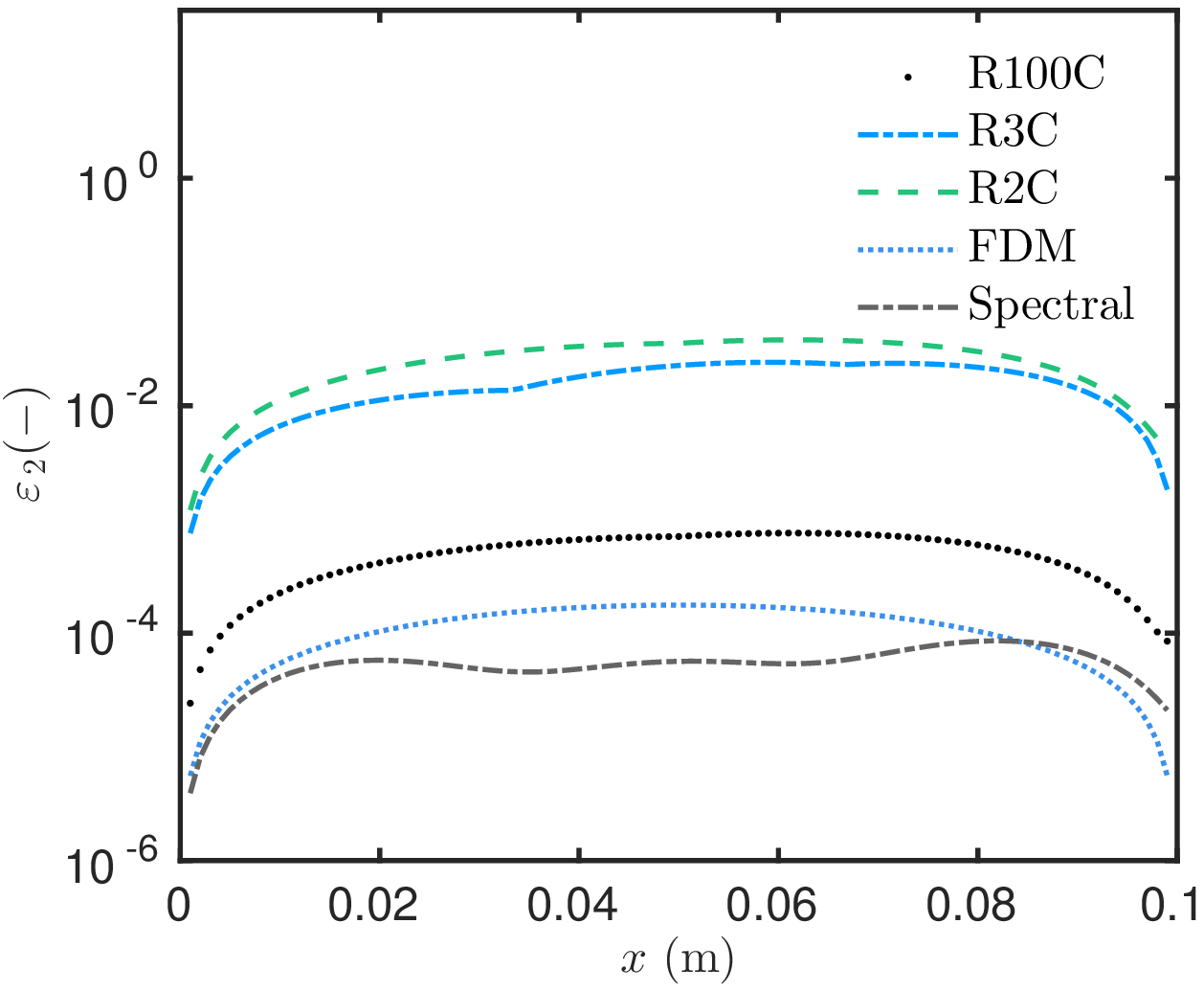}} 
\subfigure[a][\label{fig_AN1:eq_fx}]{\includegraphics[width=.45\textwidth]{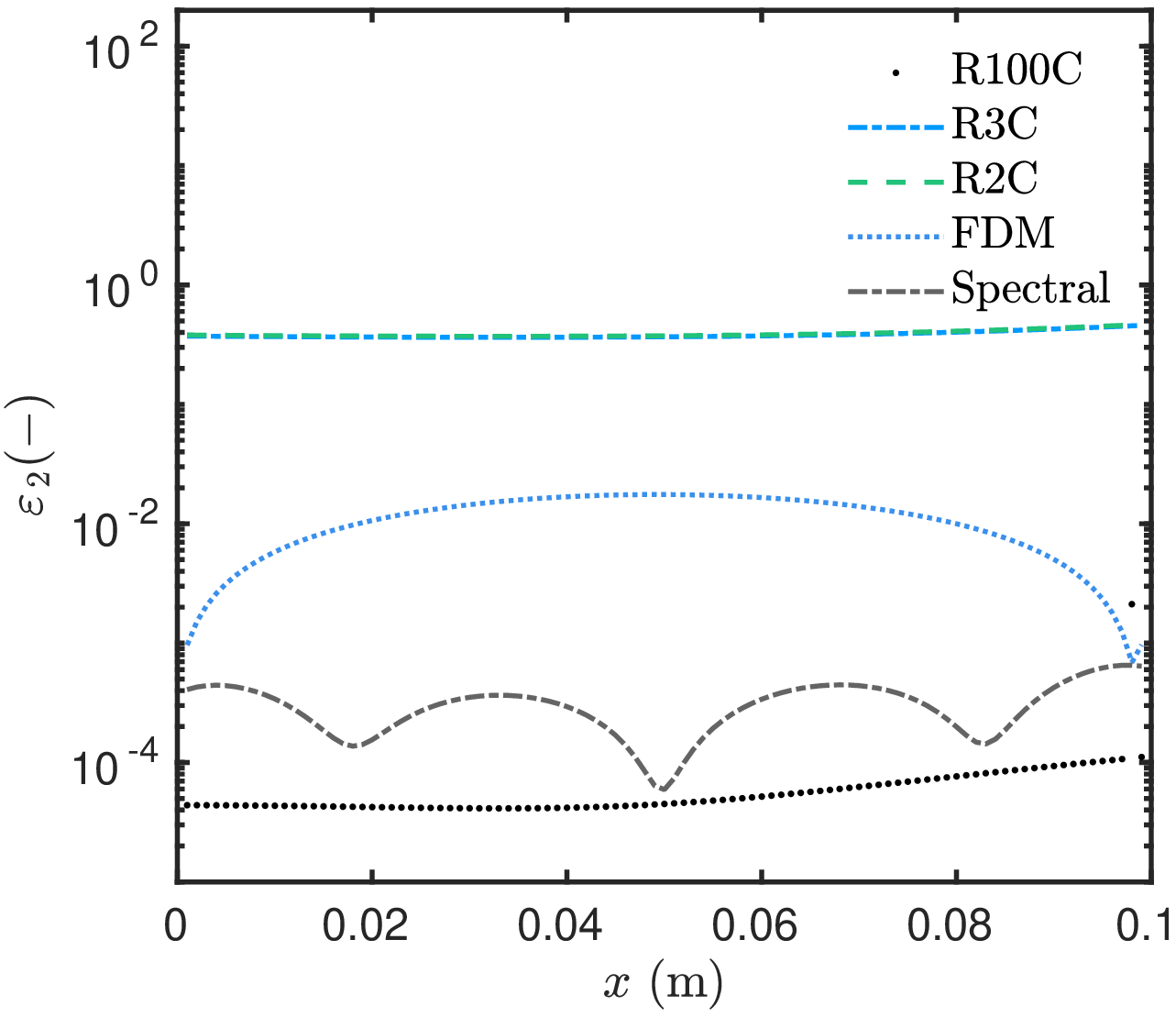}} 
\caption{Variation of the error $\varepsilon_{\,2}\,(\,x\,)$ on the temperature \emph{(a)} and on the heat flux density \emph{(b)}.}
\end{figure}

One can argue that the differences, between the reference temperature and the one computed with the RC approach using two or three resistances, are acceptable. However, the discrepancy increases drastically for the heat flux density -- which is directly dependent on the temperature derivative at the boundary --, going up to $100 \, \%$ as highlighted in Figure~\ref{fig_AN1:RC_qR_ft}. For the RC model with $100$ resistances, the heat flux density is computed with a very satisfying accuracy. As expected, similar results are observed for the standard finite-differences approach. The heat flux density is however more accurate when computed with the spectral approach. 
Figures~\ref{fig_AN1:e_fx} and \ref{fig_AN1:eq_fx} shows the error $\varepsilon_{\,2}$, computed using the dimensionless fields. It confirms that the temperature, computed with an RC model approach with only two or three resistances, lacks of accuracy to represent the physical phenomenon of heat diffusion. It should be noted that the error of the spectral method scales with $\mathcal{O}(\,10^{\,-5}\,)$ with only $N \egal 6$ modes. 
For the flux, the best accuracy reaches $\mathcal{O}(\,10^{\,-2}\,)\,$, which is obtained with the spectral approach. The error of the RC model with two or three resistances is completely unacceptable. It can be noted that the accuracy is more sensible to the heat flux density. This can also be understood by analyzing the propagation of the numerical errors. According to the definition in Eq.~\eqref{eq:heatflux}, the finite-difference approximation of the heat flux density is given by 
\begin{align*}
q_{\,j} \ \simeq \ k\, \frac{T_{\,j} \moins T_{\,j-1}}{\dx} \,.
\end{align*}
If we consider that the temperature is computed with a numerical perturbation $T \simeq T \plus \delta \,T$, with $\bigl|\bigl|\, \delta \,T \,\bigr|\bigr| \ \ll \ 1\,$, the approximation of the heat flux is:
\begin{align*}
q_{\,j} \egal k\, \frac{T_{\,j} \moins T_{\,j-1}}{\dx} \plus \frac{\delta \,T_{\,j} \moins \delta \,T_{\,j-1}}{\dx} \,.
\end{align*}
Since the perturbations are uncorrelated, $\delta \,T_{\,j} \moins \delta \,T_{\,j-1} \simeq 2 \, \delta T$ and the approximation of the heat flux becomes
\begin{align*}
q_{\,j} \egal k\, \frac{T_{\,j} \moins T_{\,j-1}}{\dx} \plus \frac{2 \, \delta T}{\dx} \,.
\end{align*}
Since $\dx \ \ll \ 1 \,$, the error on the heat flux density is higher than the one on the temperature. Considering the numerical values read in Figure~\ref{fig_AN1:e_fx}, the numerical perturbation scales with $ \delta T \egal \mathcal{O}(\,10^{\,-4}\,)$ for the standard finite-difference approach. With the spatial discretisation $\dx \egal 10^{\,-2}\,$, the term $\displaystyle \frac{2 \, \delta T}{\dx}$ is of the order $\mathcal{O}(\,10^{\,-2}\,)\,$. Thus, the error on the flux cannot be lower than this value. This value is in consistent with the results observed in Figure~\ref{fig_AN1:eq_fx}.

In terms of digits accuracy, the results are reported in Table~\ref{tab_AN1:efficiency_num_models}. It is noticed the RC approach with $100$ resistances computes the field with more than twice digits accuracy than the \RC{2} one. The spectral method presents the highest accuracy. For the computational time, as expected, the numerical models with $2$ and $3$ resistances are the fastest approach. Indeed they require a very few computations at each time step ($3$ and $4\,$, respectively). However, the speed of computation decreases with the accuracy of the solution. The computational effort of the standard finite-difference approach scales with the RC one for $100$ resistances. A good compromise between speed of computation and accuracy is the Spectral approach.

\begin{figure}
\centering\includegraphics[width=.45\textwidth]{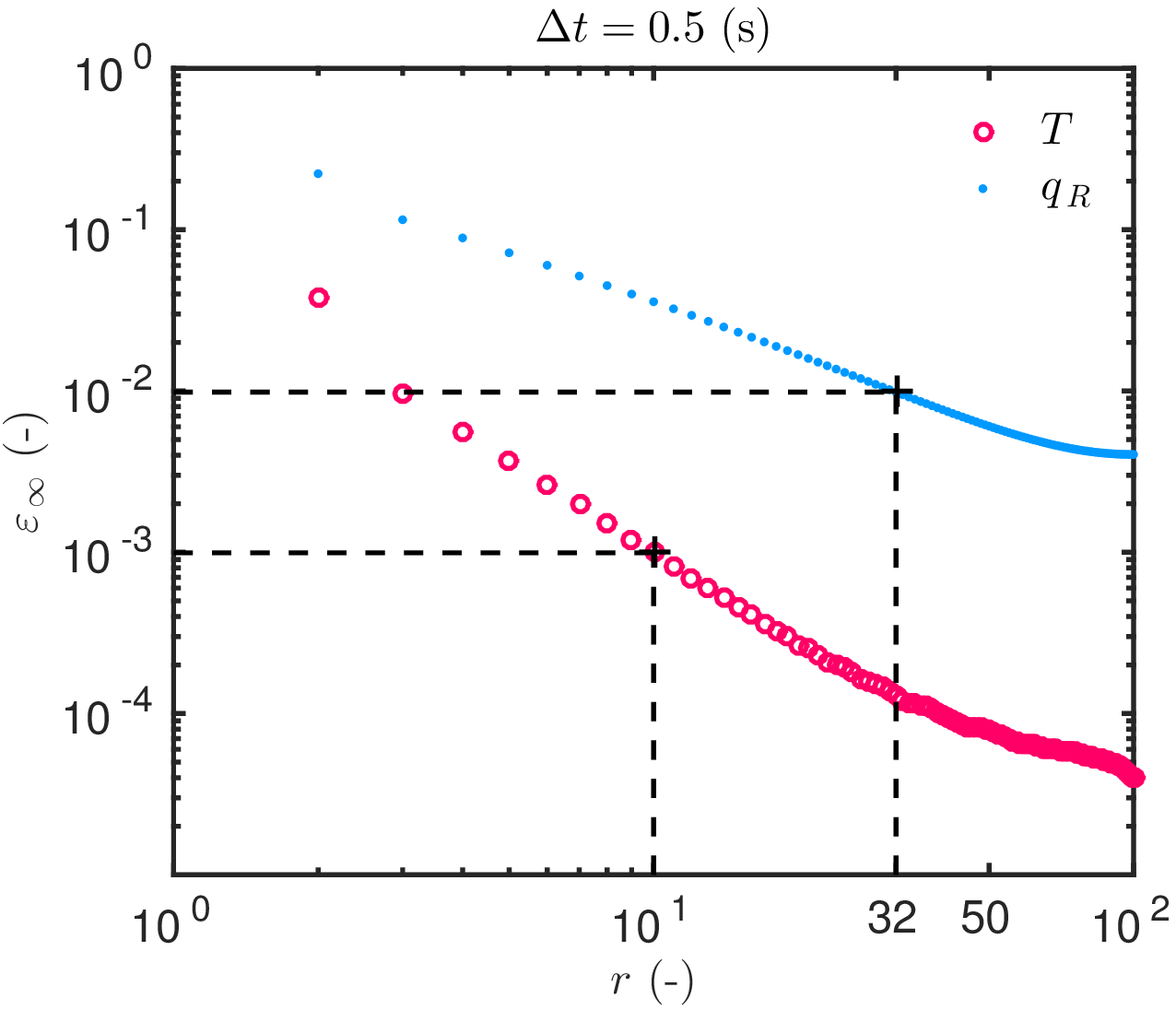}
\caption{Variation of the error $\varepsilon_{\,\infty}$ with the number of resistances $r$.}
\label{fig_AN1:err_fN}
\end{figure}

These results highlight that the RC model with a small number of resistance cannot provide an accurate solution. A natural question rises: how many resistances are required to accurately compute the temperature within the wall? The answer to this question strongly depends on the numerical values of the application. For this case study, the error with the reference solution has been computed as a function of the resistance number $r$, as shown in Figure~\ref{fig_AN1:err_fN}. If the field of interest is the temperature, it can be noted that a number $r \egal 10$ of resistances is sufficient. The error remains stable at $\mathcal{O}(\,10^{\,-4}\,)$ corresponding to the tolerance set in the \texttt{ode45} solver of the \texttt{Matlab\texttrademark} environment. Although, if one is interested in the heat flux density, the minimum number is $r \egal 30\,$  to reach an error lower than $10^{\,-2}\,$. 

\begin{table}
\centering
\caption{Efficiency of the numerical models for linear heat diffusion.}
\label{tab_AN1:efficiency_num_models}
\begin{tabular}{c |c c c| c}
\hline
\hline
\textit{Numerical Model} 
& $\mathrm{scd} \ \unit{-}$ 
& $\varepsilon_{\,\infty} \ \unit{-}$ for $T$
& $\varepsilon_{\,\infty} \ \unit{-}$ for $q$
& $R_{\,\mathrm{cpu}} \ \unit{ms/h}$\\ \hline \hline 
\RC{2}  
& $1.5$ 
& $0.02$
& $0.47$
&  $0.03$  \\ 	
\RC{3}  
& $1.8$ 
& $0.02$
& $0.45$ 
& $0.06$  \\ 
\RC{100}  
& $3.6$ 
& $8 \e{-5}$ 
& $6 \e{-3}$ 
&  $12$  \\ 
FDM  
& $4.2$ 
& $7 \e{-5}$ 
& $1.7 \e{-3}$ 
& $15$  \\ 
Spectral 
& $4.3$ 
& $4 \e{-5}$ 
& $6 \e{-4}$ 
& $1.23$  \\ 
\hline
\hline
\end{tabular}
\end{table}

\section{Numerical investigations: nonlinear moisture diffusion}
\label{sec:NL_moisture_diff}

The previous section considered a linear model of diffusion. It is important to evaluate the performance of the methods for nonlinear problems since the accuracy of the solution can be deteriorated. For this, a nonlinear moisture diffusion problem is considered. The length of the wall is set to $L \egal 0.1 \unit{m}$ and the simulation horizon to $t \egal 72 \unit{h} \,$. The material properties are inspired from \cite{Bednar_2005} with a constant moisture capacity and a vapor pressure dependent moisture permeability:
\begin{align*}
& \xi_{\,m} \egal 1.88 \e{-2} \unit{s^{\,2}/m^{\,2}} \,,
&& \kappa \egal 6.72 \e{-13} \ \cdot \Pv \plus 3 \e{-10} \unit{s}\,.
\end{align*}
Sinusoidal variations are imposed as boundary conditions:
\begin{align*}
P_{\,v,\,L}\,(\,t\,) & \egal  \frac{1}{P_{\,\mathrm{sat}}\,(\,T^{\,\circ}\,)} \, \biggl[\, 0.5  \plus 0.4 \, \sin \biggl(\, \frac{2 \, \pi }{12 \cdot 3600} \, t \,\biggr) \, \biggr] \unit{Pa}  \,, \\
P_{\,v,\,\,R}\,(\,t\,) & \egal \frac{1}{P_{\,\mathrm{sat}}\,(\,T^{\,\circ}\,)} \, \biggl[\, 0.5 \plus 0.1 \, \sin \biggl(\, \frac{2 \, \pi }{6 \cdot 3600} \, t \,\biggr) \, \biggr] \unit{Pa}  \,,
\end{align*}
where $P_{\,\mathrm{sat}}\,(\,T^{\,\circ}\,)$ is the saturation pressure at $T^{\,\circ} \egal 25 \unit{^{\,\circ}C}\,$. These conditions correspond to variations around the relative humidity $0.5$ to the dry and almost-saturated states.

\begin{figure}
\centering
\subfigure[a][\label{fig_AN1:RC_P24h_fx}]{\includegraphics[width=.45\textwidth]{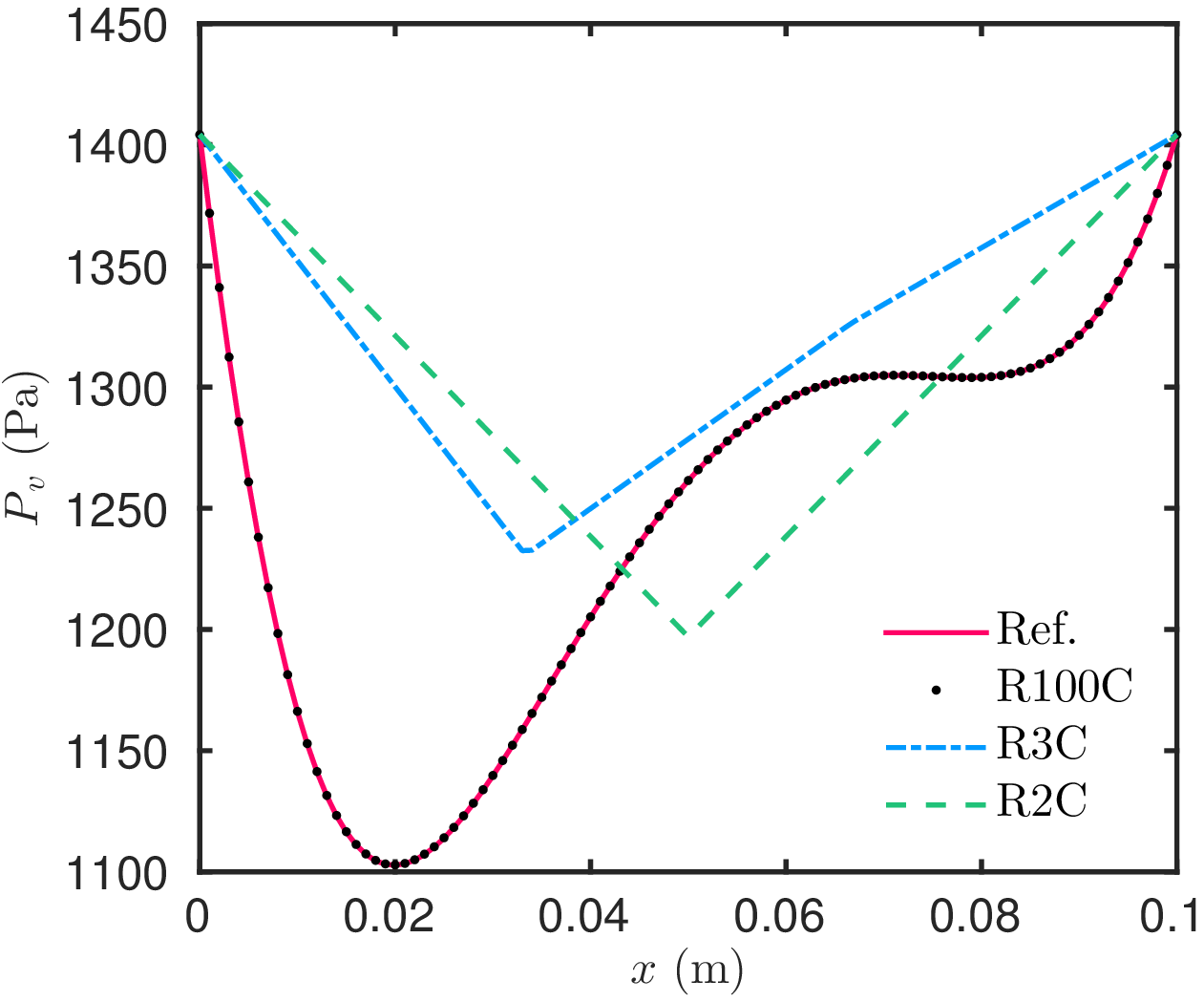}}
\subfigure[a][\label{fig_AN1:SP_EU_P24h_fx}]{\includegraphics[width=.45\textwidth]{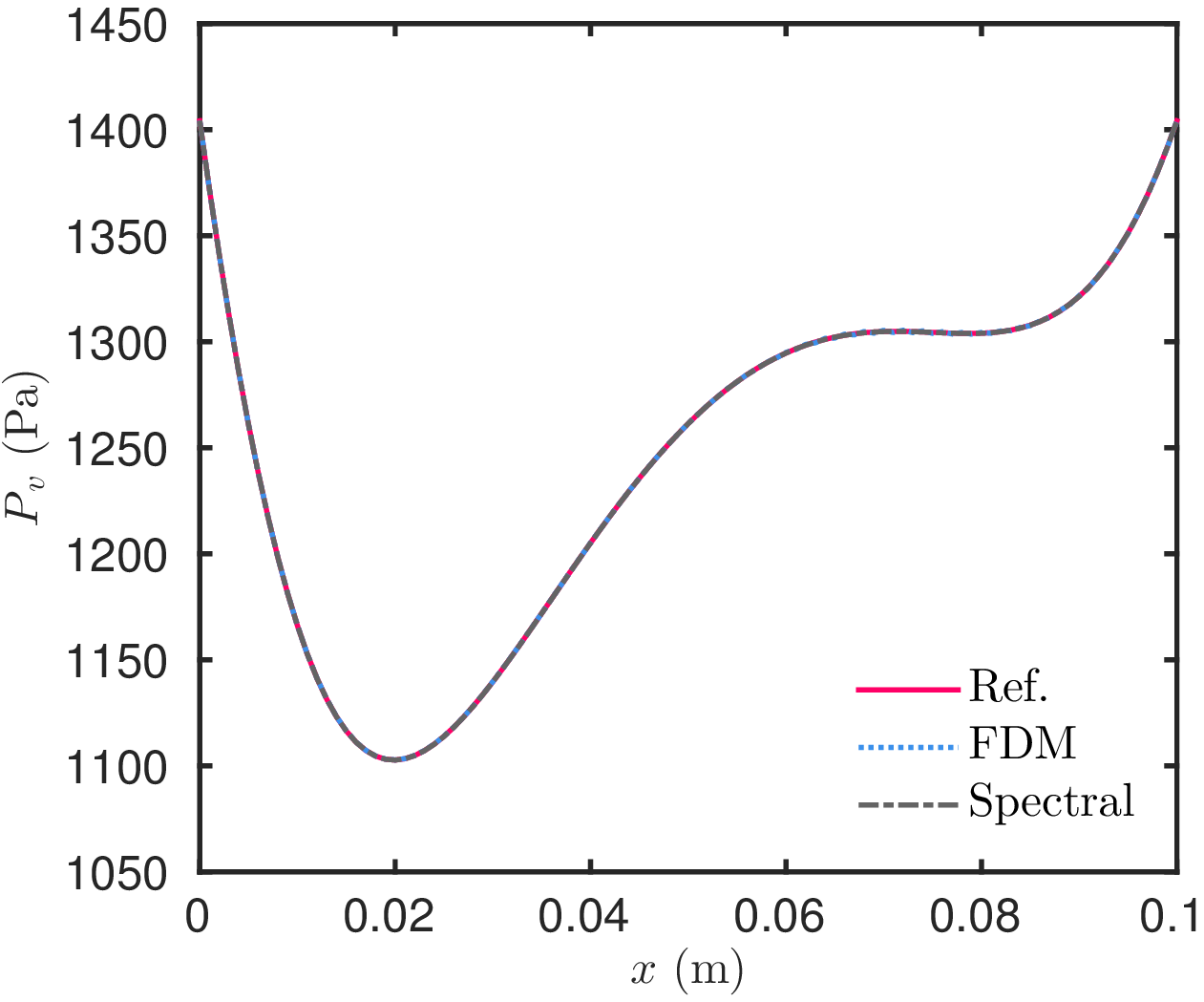}}
\caption{Vapor pressure profiles at $t \egal 72 \unit{h}$.}
\label{fig_AN2:P24h_fx}
\end{figure}

\begin{figure}
\centering
\subfigure[a][\label{fig_AN2:RC_Pmid_ft}]{\includegraphics[width=.45\textwidth]{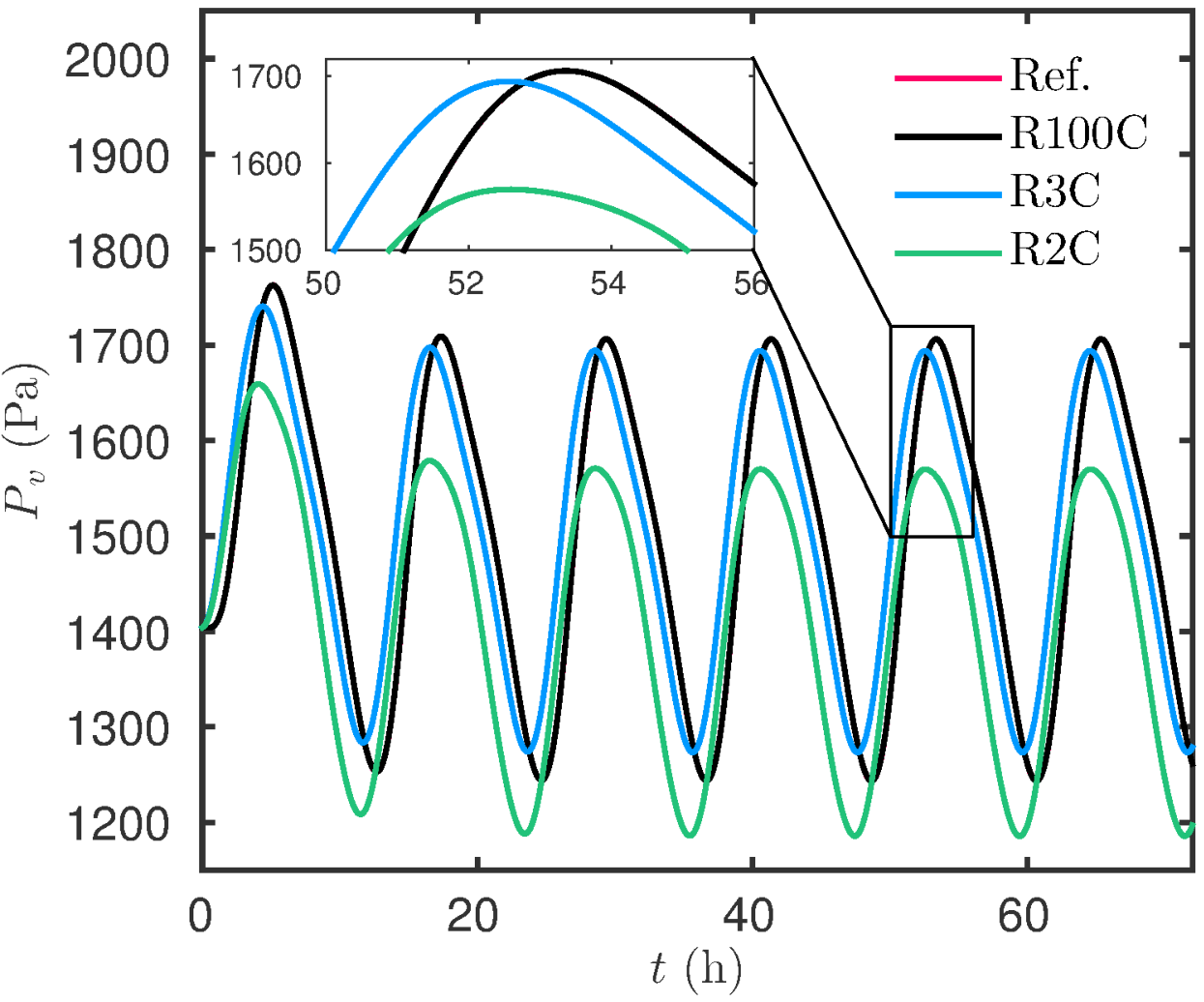}} 
\subfigure[b][\label{fig_AN2:SP_EU_Pmid_ft}]{\includegraphics[width=.45\textwidth]{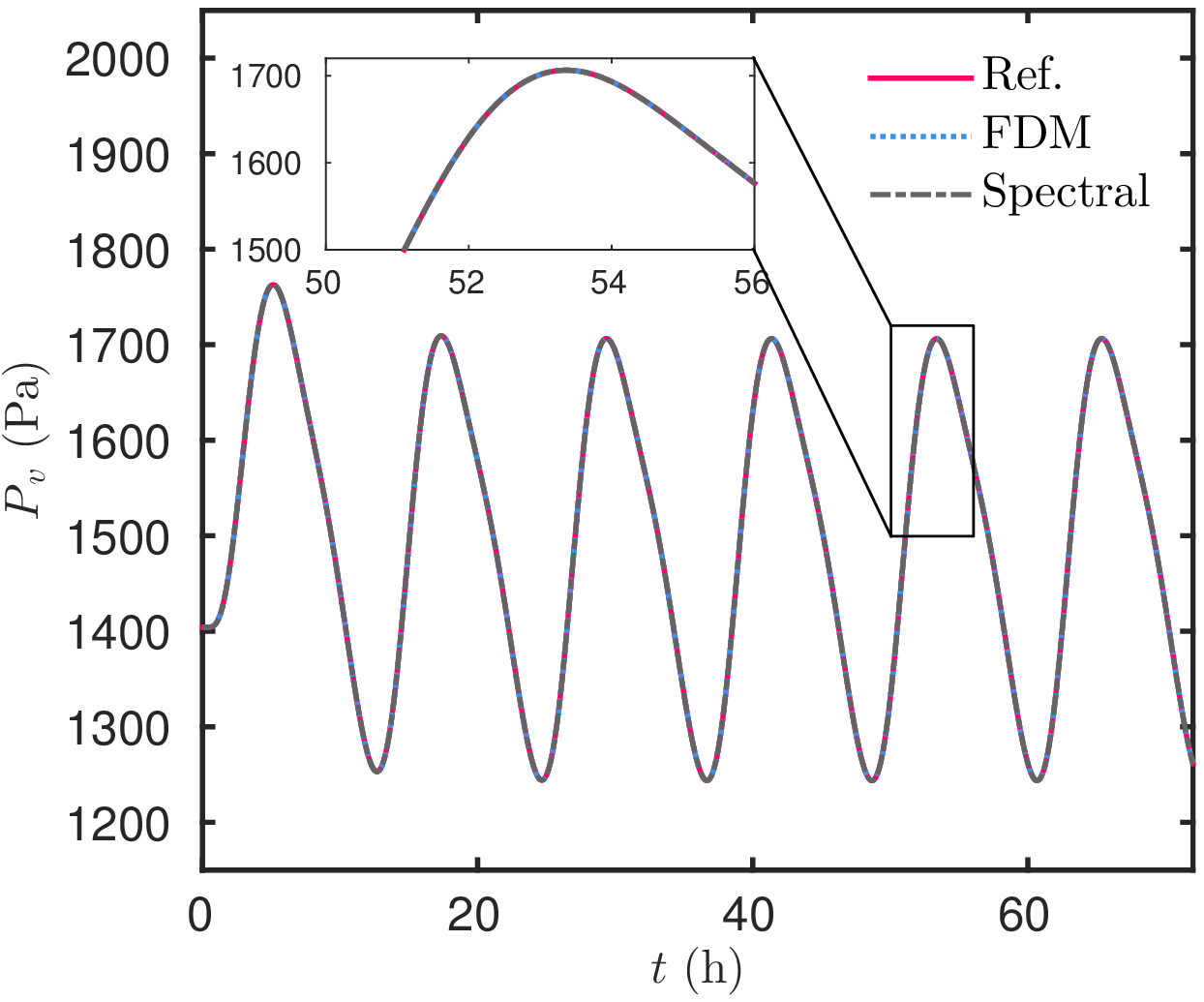}} 
\subfigure[c][\label{fig_AN2:RC_Emid_ft}]{\includegraphics[width=.45\textwidth]{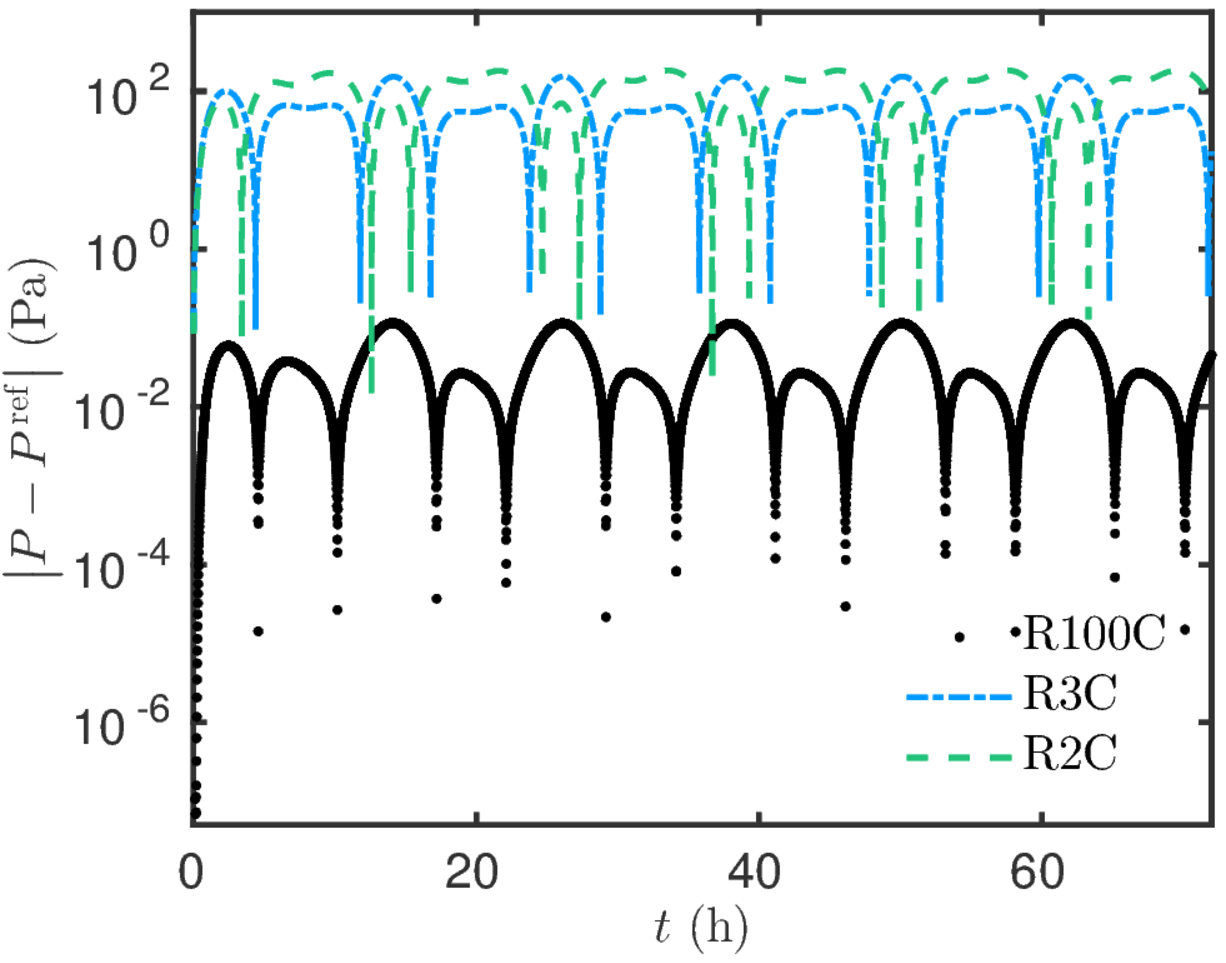}} 
\subfigure[d][\label{fig_AN2:SP_EU_Emid_ft}]{\includegraphics[width=.45\textwidth]{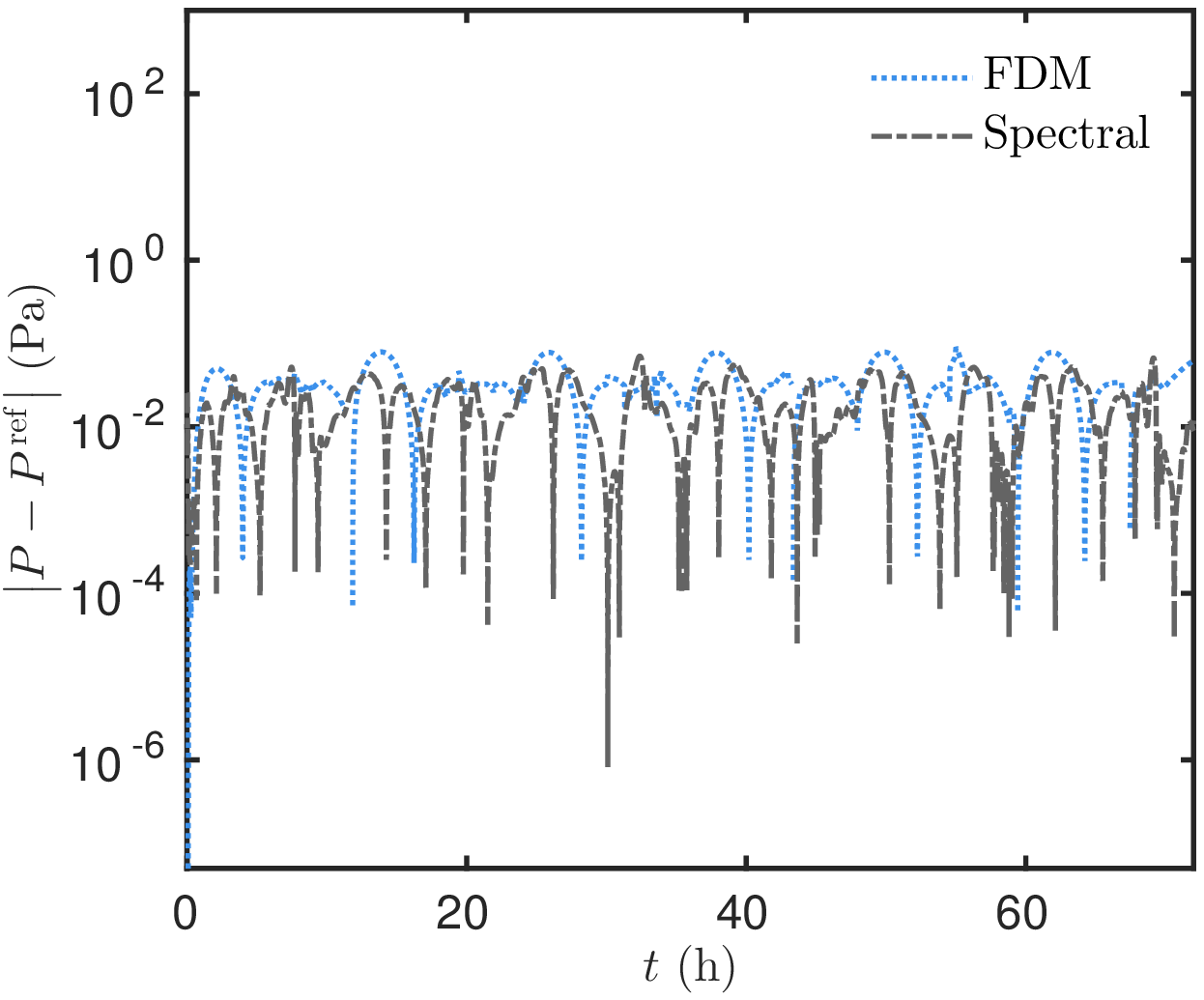}} 
\caption{\emph{(a,b)} Vapor pressure evolution at $x \egal 0.05 \unit{cm}$. \emph{(c,d)} Vapor pressure difference to the reference solution.}
\end{figure}

\begin{figure}
\centering
\subfigure[\label{fig_AN2:RC_gR_ft}]{\includegraphics[width=.45\textwidth]{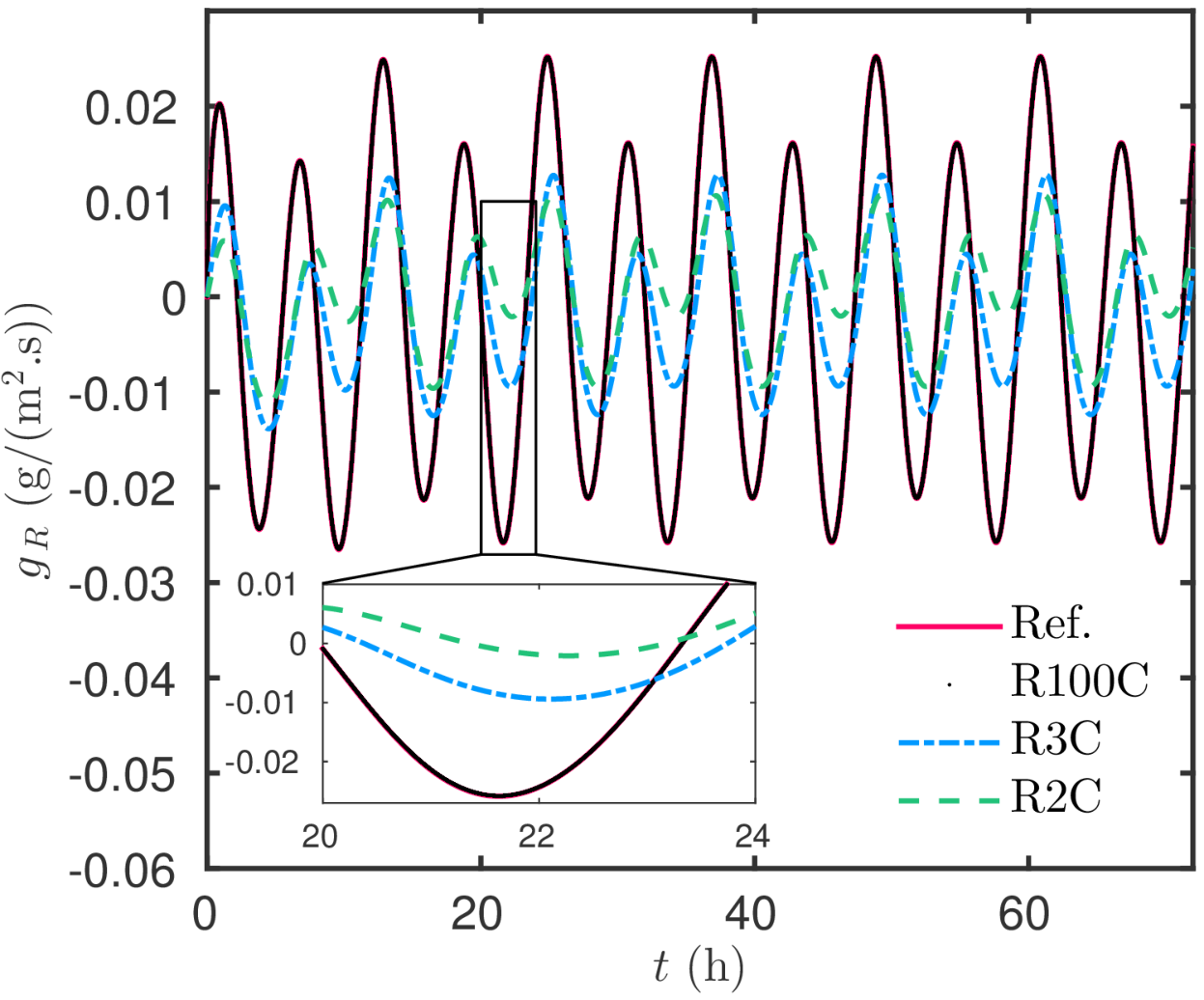}} 
\subfigure[\label{fig_AN2:SP_EU_gR_ft}]{\includegraphics[width=.45\textwidth]{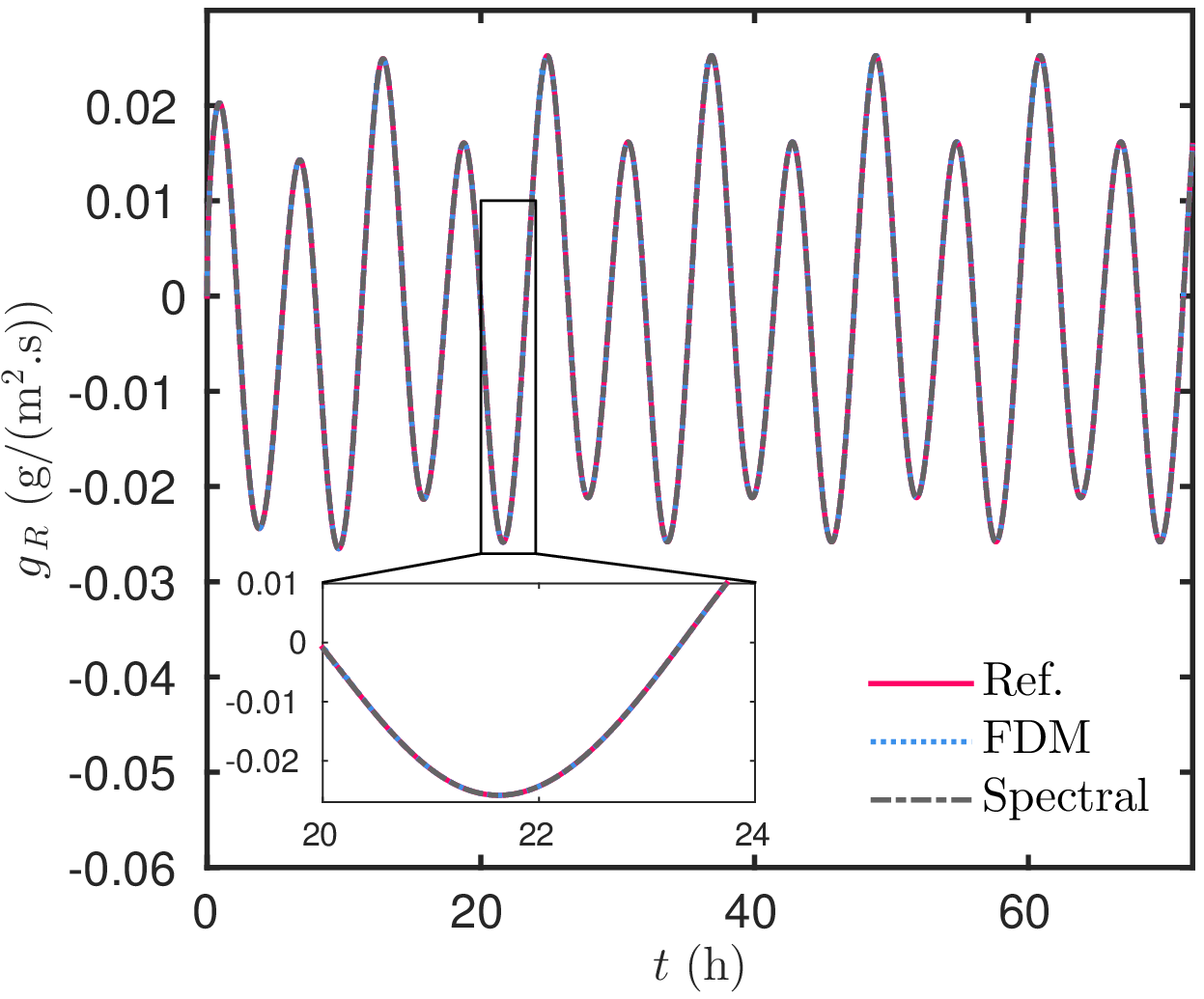}} 
\caption{\emph{(a,b)} Evolution of the moisture flux density at the right boundary.}
\end{figure}

\begin{figure}
\centering
\subfigure[\label{fig_AN2:RC_EagR_ft}]{\includegraphics[width=.45\textwidth]{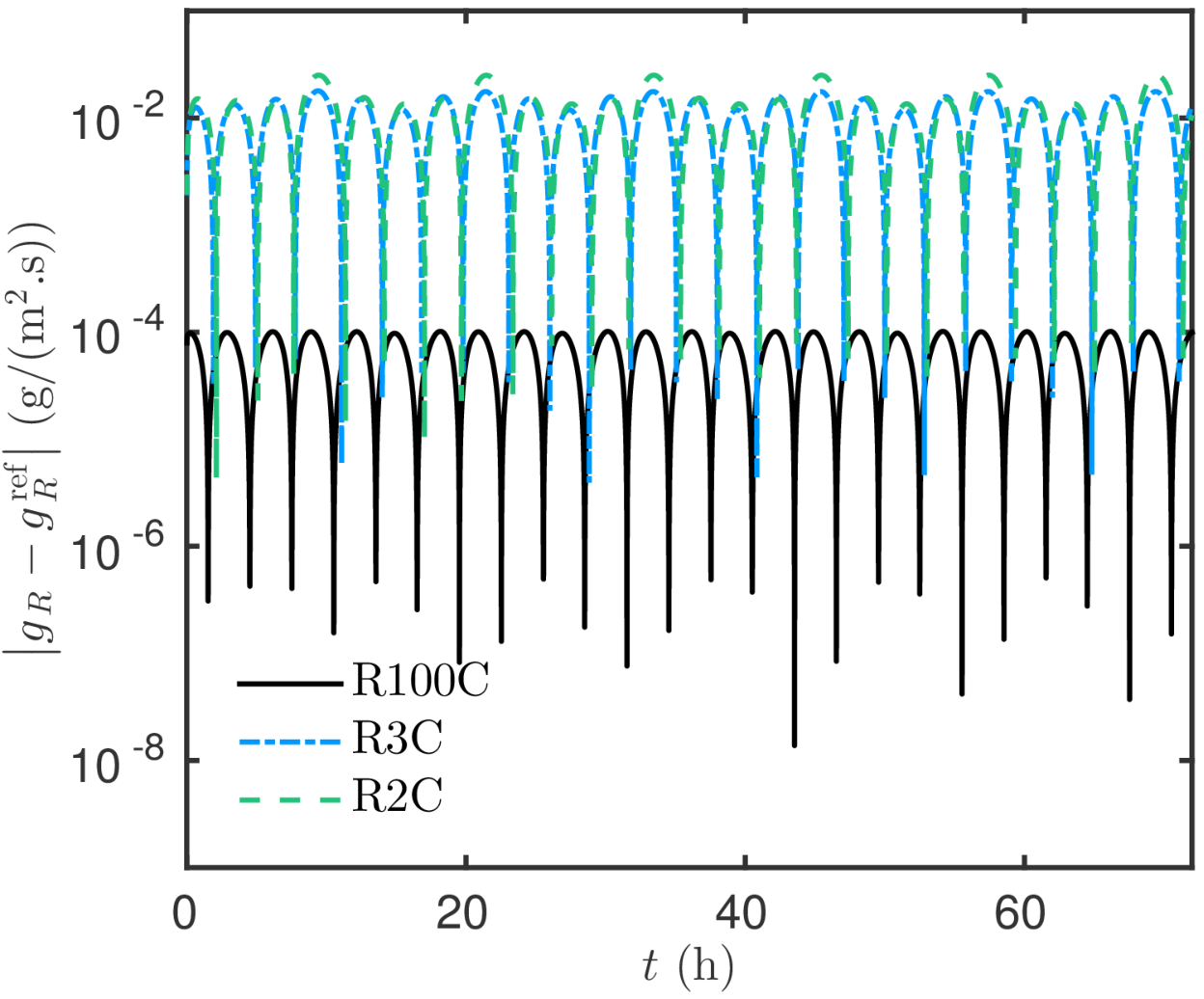}} 
\subfigure[\label{fig_AN2:SP_EU_EagR_ft}]{\includegraphics[width=.45\textwidth]{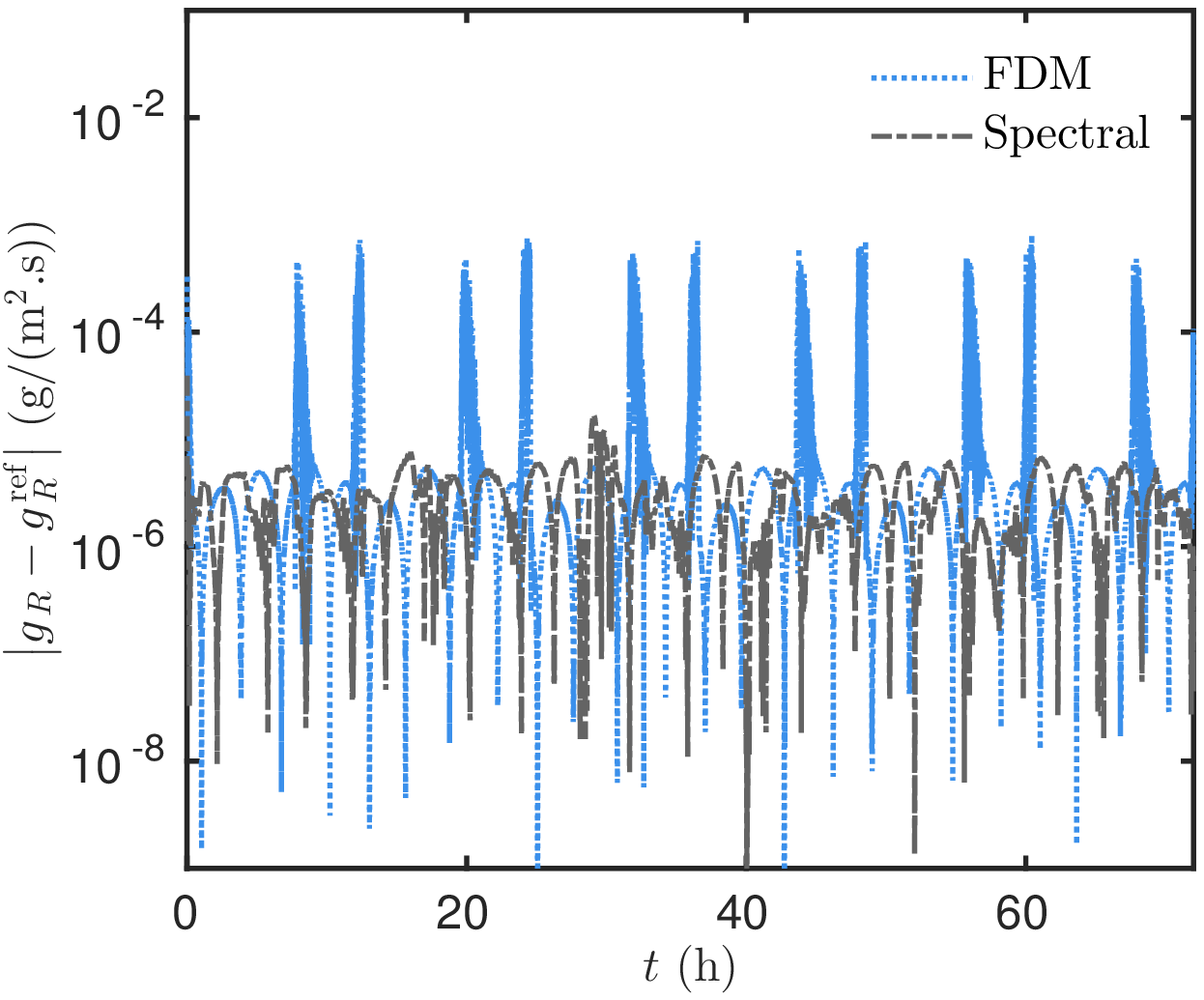}} 
\subfigure[\label{fig_AN2:RC_EgR_ft}]{\includegraphics[width=.45\textwidth]{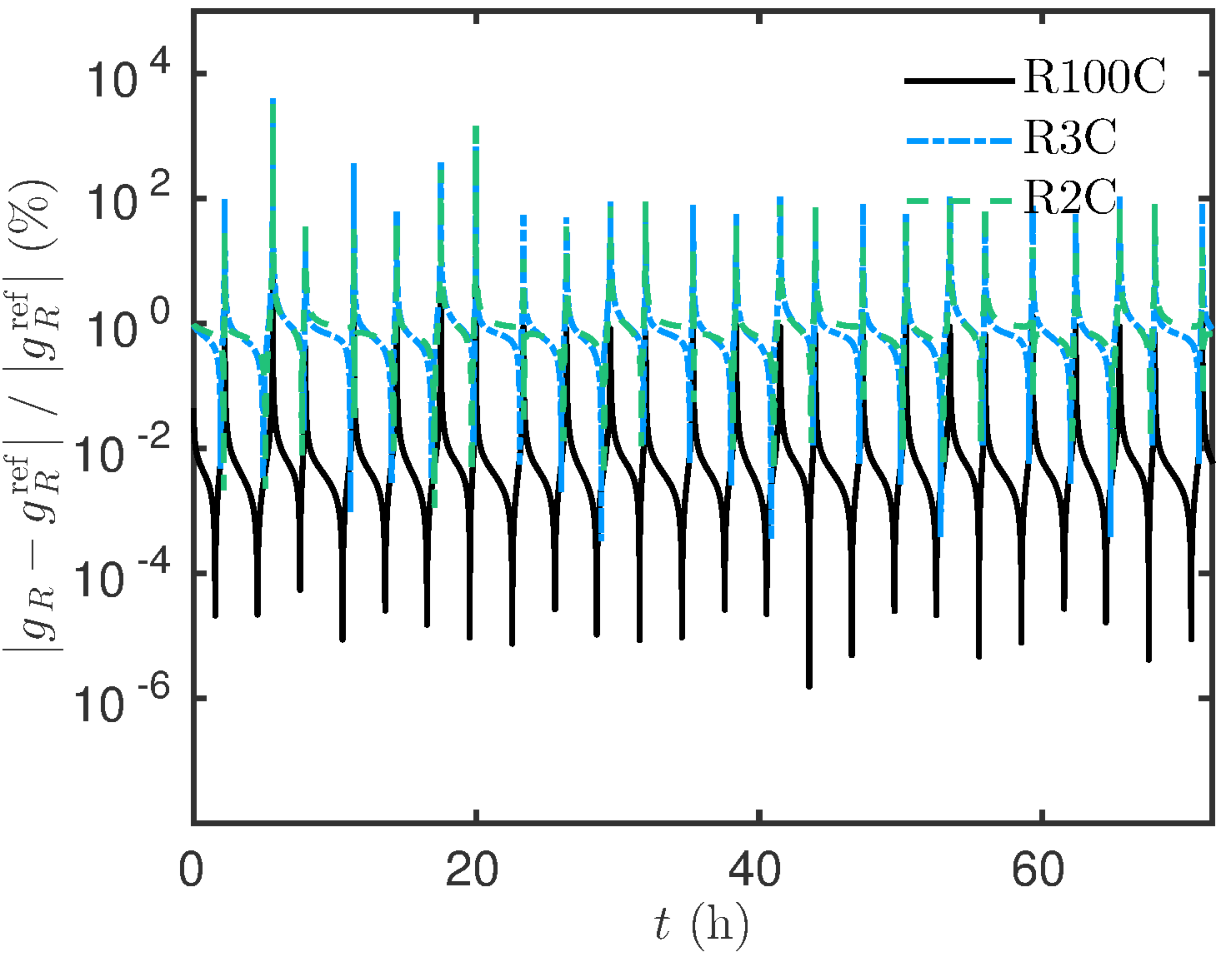}} 
\subfigure[\label{fig_AN2:SP_EU_EgR_ft}]{\includegraphics[width=.45\textwidth]{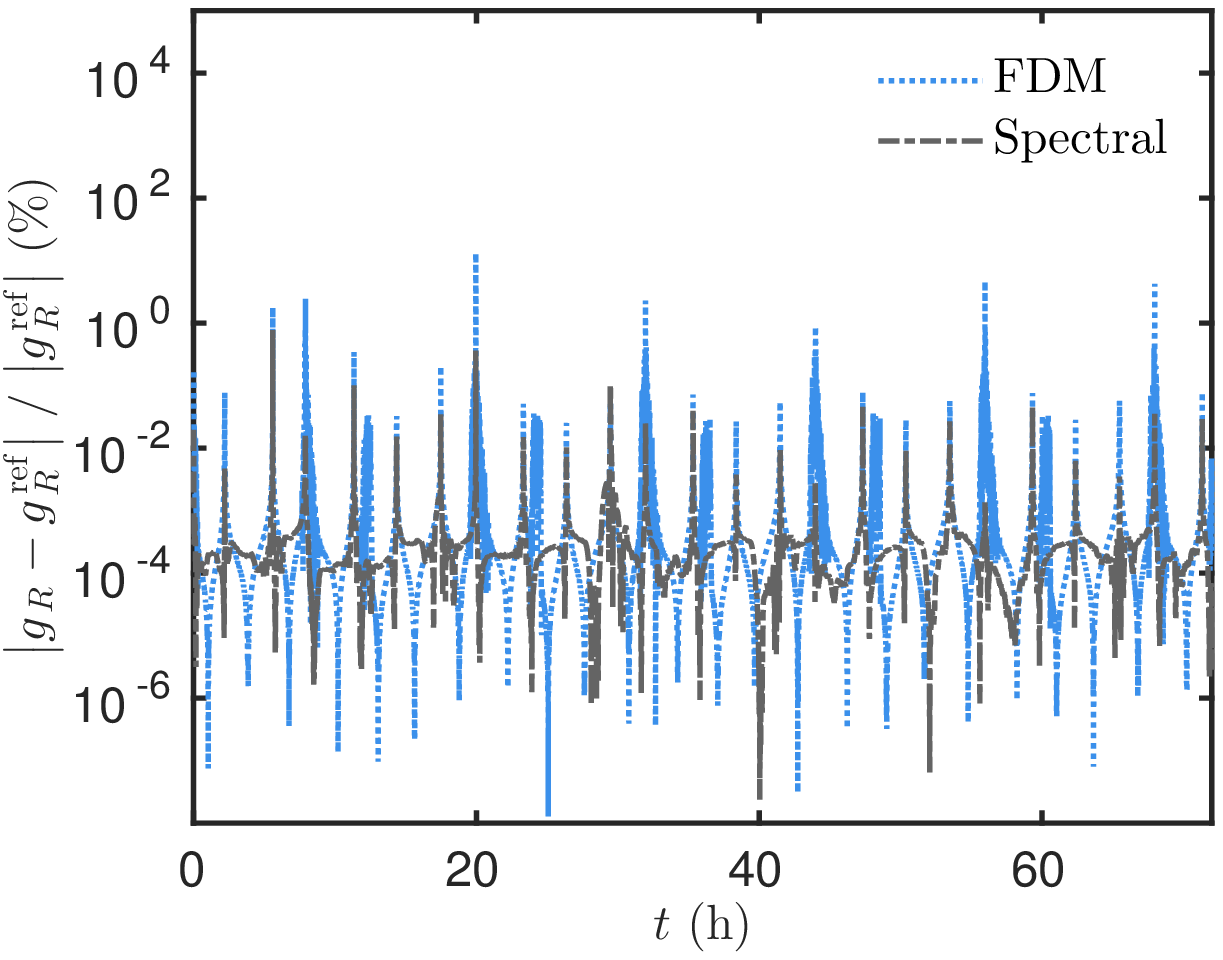}} 
\caption{\emph{(c,d)} absolute and \emph{(e,f)} relative differences of the moisture flux density to the reference solution}
\end{figure}

\begin{figure}
\centering\includegraphics[width=.45\textwidth]{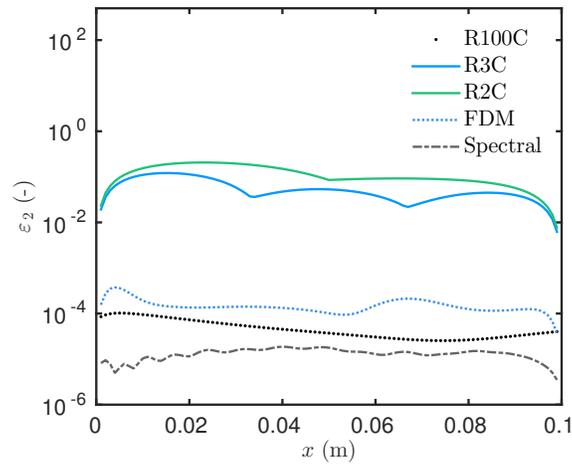}
\caption{Variation of the error $\varepsilon_{\,2}$.}
\label{fig_AN2:e_fx}
\end{figure}

\begin{figure}
\centering\includegraphics[width=.45\textwidth]{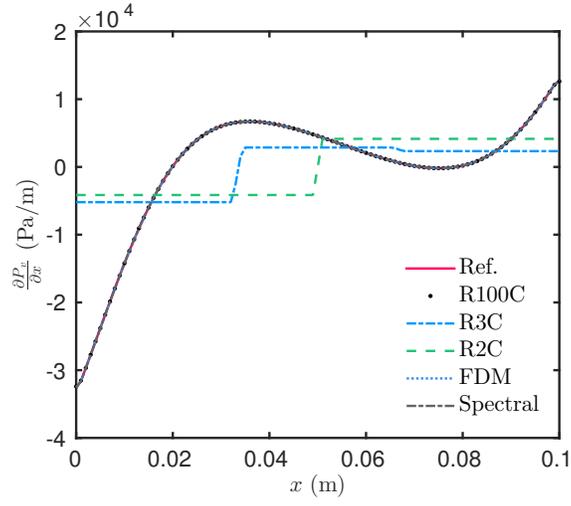}
\caption{Profile of the space derivative of the vapor pressure at $t \egal 72 \unit{h}$.}
\label{fig_AN2:dP24h_fx}
\end{figure}

The solution to problem \eqref{eq:moisture_1d_dim} is computed using the RC approaches for $r \egal \bigl\{\,  2 \,,\, 3 \,,\, 100 \,\bigr\}$, the standard finite-differences one and the spectral one with $N \egal 10$ modes. For the last two methods, the spatial discretisation is $\dx \egal 10^{\,-3} \ \mathsf{m}\,$. The profile of vapor pressure at $t \egal 72 \unit{h}$ is shown in Figure~\ref{fig_AN2:P24h_fx}~\textit{(a,b)}. The time evolution of the vapor pressure in the middle of the layer is given in Figures~\ref{fig_AN2:RC_Pmid_ft} and \ref{fig_AN2:SP_EU_Pmid_ft}. The standard finite-differences and spectral approaches represent accurately the field evolution. On the contrary, an important discrepancy is noted for the approach using two or three resistances (at the order of $\ \simeq \ 200 \unit{Pa}$). The differences become more important when looking at the moisture flux density, illustrated in Figures~ \ref{fig_AN2:RC_gR_ft} and \ref{fig_AN2:SP_EU_gR_ft}. The flux is underestimated by the RC approach based on a few resistances. If we look at Figures~\ref{fig_AN2:RC_EagR_ft} and \ref{fig_AN2:SP_EU_EagR_ft}, the absolute differences of the flux reach $\ \simeq 10^{\,-2}$ for the RC approach. However, as shown in Figure~\ref{fig_AN2:RC_EgR_ft} and \ref{fig_AN2:SP_EU_EgR_ft}, for these approaches the relative difference on the flux scales with $\ \simeq \ 100 \%\,$ for a field of the same order of magnitude. The global error for each approach is given in Figure~\ref{fig_AN2:e_fx}. The most accurate approach is the spectral one, with an error at the order $\mathcal{O}(\,10^{\,-2}\,)\,$. Due to the nonlinearity of the problem, the error of the RC approach, with two or three resistances, have increased when compared to the previous case. As mentioned in the previous case, the error is more important for the \RC{2} and \RC{3} approaches since the space derivative of the field is computed with a low accuracy as illustrated in Figure~\ref{fig_AN2:dP24h_fx}.

The accuracy digits of each method are reported in Table~\ref{tab_AN2:efficiency_num_models}. The digit accuracy of the RC approach with a few number of resistance is much lower compared to the previous case. This lack of accuracy may also be due to the computation of the solution in its physical dimension. Where the temperature scales with $\mathcal{O}(\,10\,)$ in the previous case, here the vapor pressure scales with $\mathcal{O}(\,10^{\,3}\,)\,$. It may introduce additional computational rounding errors. For the spectral approach, it remains stable with four digits of accuracy in the computed vapor pressure. 
In terms of computational time, the \RC{2} and \RC{3} have low computational ratio. However, these methods lack of accuracy to compute the fields, particularly the moisture flux density. The Spectral method has the best efficiency providing an accurate solution at a reasonable computational cost.

\begin{figure}
\centering\includegraphics[width=.45\textwidth]{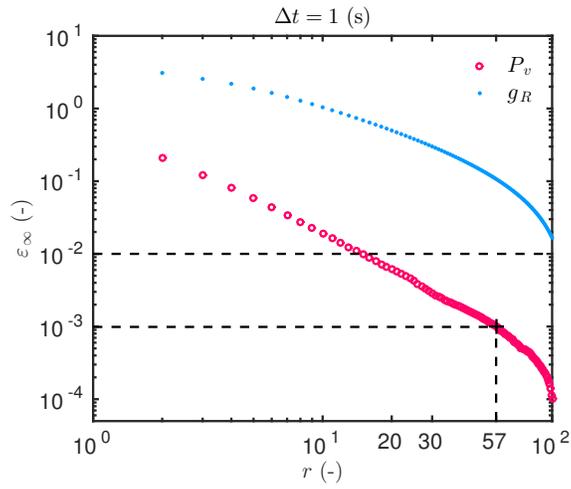}
\caption{Variation of the error $\varepsilon_{\,\infty}$ with the number of resistances $r$ for the vapor pressure and the flux density.}
\label{fig_AN2:err_fN}
\end{figure}

A parametric study on the number of resistances have been carried out and the error $\varepsilon_{\,\infty}$ is shown in Figure~\ref{fig_AN2:err_fN}. Compared to the previous case of linear diffusion, more resistances are required to reach a satisfying accuracy. A minimal number of $r \egal 20$ and $ r \egal 90$ resistances are necessary to represent accurately the vapor pressure or moisture flux evolution, respectively.

\begin{table}
\centering
\caption{Efficiency of the numerical models for nonlinear moisture diffusion.}
\label{tab_AN2:efficiency_num_models}
\begin{tabular}{c |c c c| c}
\hline
\hline
\textit{Numerical Model} 
& $\mathrm{scd} \ \unit{-}$ 
& $\varepsilon_{\,\infty} \ \unit{-}$ for $\Pv$
& $\varepsilon_{\,\infty} \ \unit{-}$ for $g$
& $R_{\,\mathrm{cpu}} \ \unit{ms/h}$\\ \hline \hline 
\RC{2}  
& $0.69$ 
& $0.20$
& $10.7$
&  $0.02$  \\ 	
\RC{3}  
& $0.72$ 
& $0.12$
& $10.6$ 
& $0.11$  \\ 
\RC{100}  
& $3.5$ 
& $2 \e{-4}$ 
& $5 \e{-2}$ 
&  $38$  \\ 
FDM  
& $3.4$ 
& $3.7 \e{-4}$ 
& $8 \e{-2}$ 
& $37$  \\ 
Spectral 
& $4.2$ 
& $2 \e{-5}$ 
& $4 \e{-3}$ 
& $12.3$  \\ 
\hline
\hline
\end{tabular}
\end{table}

\section{Evaluating the methods efficiencies for a real case study}
\label{sec:real_case}

\subsection{Description}

\begin{figure}
\centering
\subfigure[\label{fig_AN3:coupe_mur}]{\includegraphics[width=.45\textwidth]{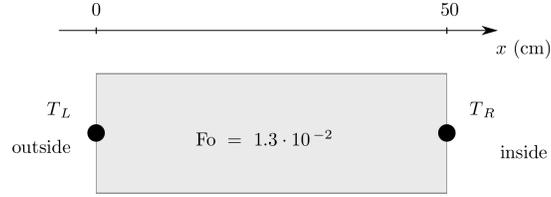}} \\
\vspace{0.5cm}
\subfigure[\label{fig_AN3:TBC_ft}]{\includegraphics[width=.45\textwidth]{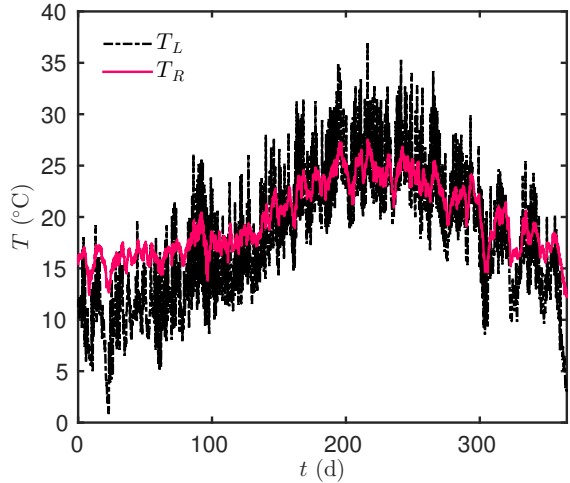}}
\caption{\emph{(a)} Illustration of the wall and the sensor positions. \emph{(b)} Variation of the temperature boundary conditions.}
\end{figure}

The purpose is to evaluate now the efficiencies of the numerical methods considering a more realistic case study of heat transfer. The fields were assessed on a building built in Bayonne, France, at the end of the $19^{\,\mathrm{th}}$ Century. With three basements, a west-oriented wall, located at the first floor in the living room, was monitored. Two calibrated monitoring sensors \texttt{HOBO} \texttt{TMC}--$6$--\texttt{HA} were placed at the surface of the wall, as shown in Figure~\ref{fig_AN3:coupe_mur}. A thermal conductive paste was added on the sensor to reduce the contact resistance. An insulated protection has also been placed to avoid incident radiation on the sensors. The location of the sensors was chosen on a part of the wall where heat transfer can be supposed as uni-dimensional between inside and outside ambient. The data measurements were stored with a period of $1 \unit{h}$ during one year. Interested readers are invited to consult \cite{Cantin_2010,Berger_2016} for complementary information. 

In \cite{Berger_2016}, complementary measures were used to estimate the thermal conductivity of the wall. Here, the purpose is to use the surface one year measurements to provide the boundary conditions. As shown in Figure~\ref{fig_AN3:TBC_ft}, there are distinguished daily variations with a raising between $150 \unit{days}$ and $300 \unit{days}\,$, corresponding to the summer season. 

The wall is considered as homogeneous using the following thermal properties: $k \egal 2.48 \unit{W/m/K}$ for the thermal conductivity, $\rho \cdot c \egal 2.8 \cdot 10^{\,6} \unit{J/(m^{\,3}.K)}$ for the volumetric heat capacity. Considering the length of the wall $L \egal 0.5 \unit{m}$ and a reference time of $1 \unit{h}$, the \textsc{Fourier} dimensionless number equals $Fo \egal 1.3 \e{-2} \,$.

For this real and occupied building the initial condition is not known. In addition, the installation of the sensors may cause perturbations on the thermal behavior of the wall. Thus, the first seven days of measurements are discarded and a linear reconstruction between the measured temperatures at given locations was considered as the initial condition at $t \egal 0 \,$.

\subsection{Results and discussion}

\begin{figure}
\centering
\subfigure[\label{fig_AN3:R2C_diffTx2_ft}]{\includegraphics[width=.45\textwidth]{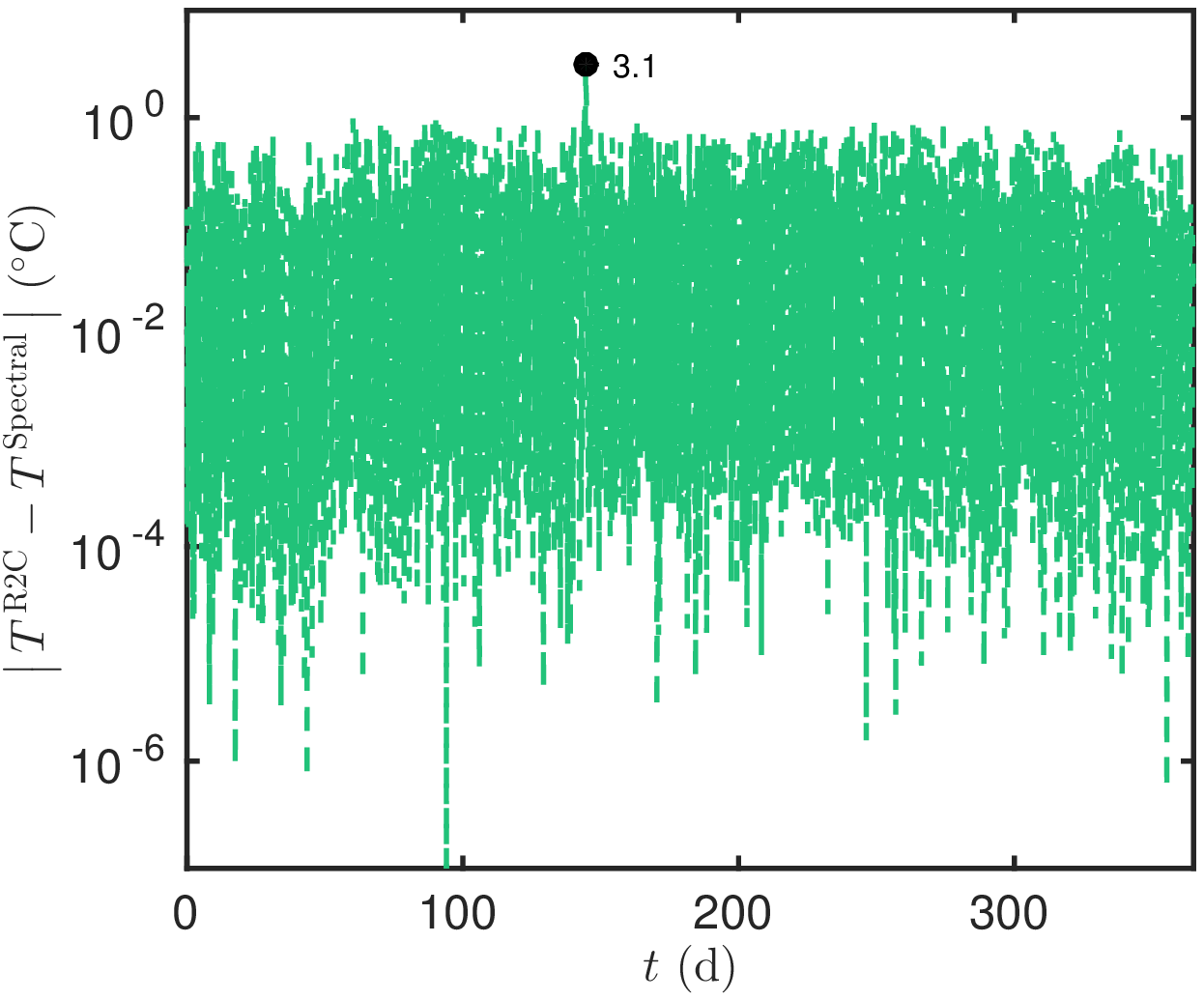}} 
\subfigure[\label{fig_AN3:zoomRC_Tx2_ft}]{\includegraphics[width=.45\textwidth]{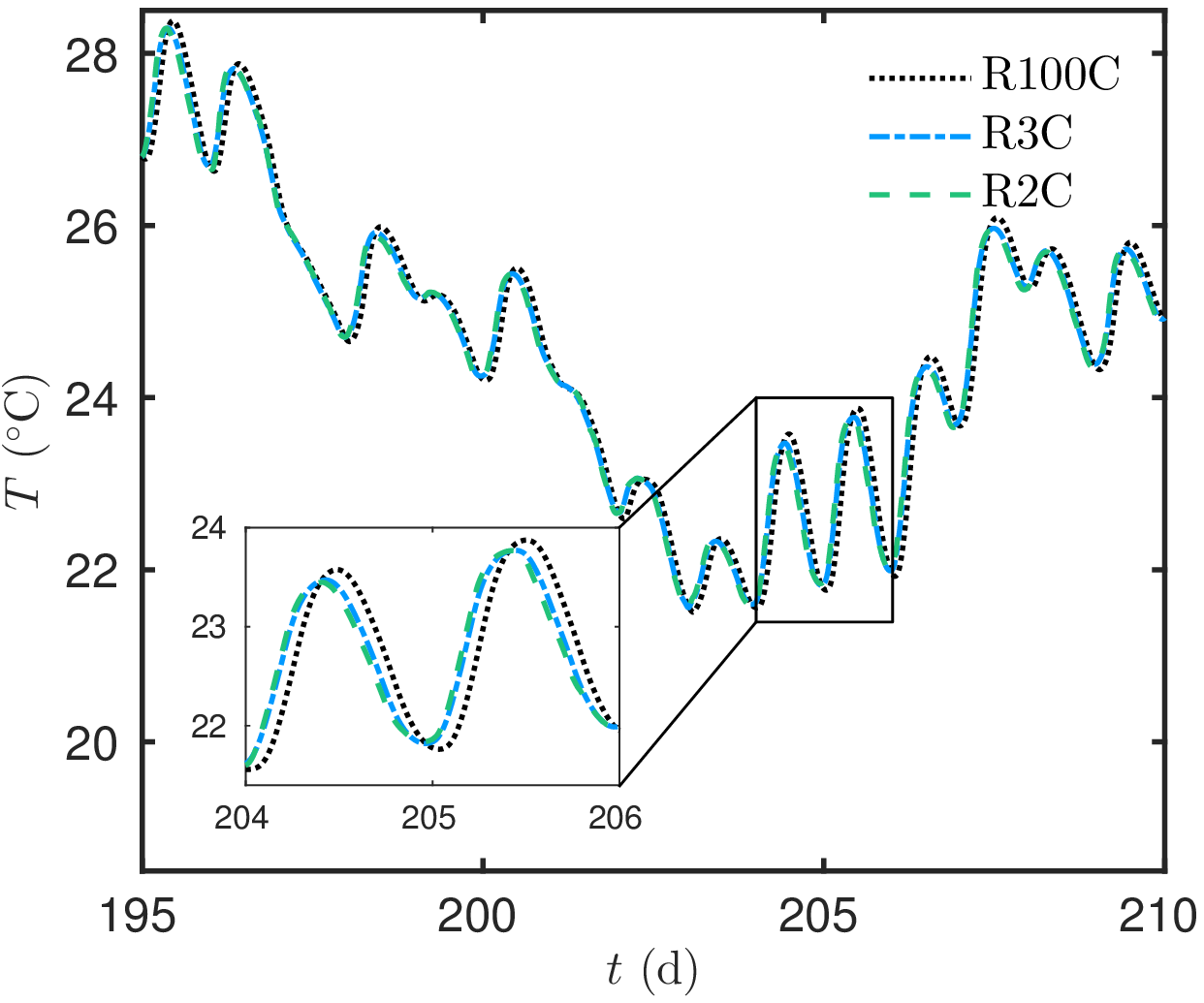}}  \\
\subfigure[\label{fig_AN3:Ex_diffTx2_ft}]{\includegraphics[width=.45\textwidth]{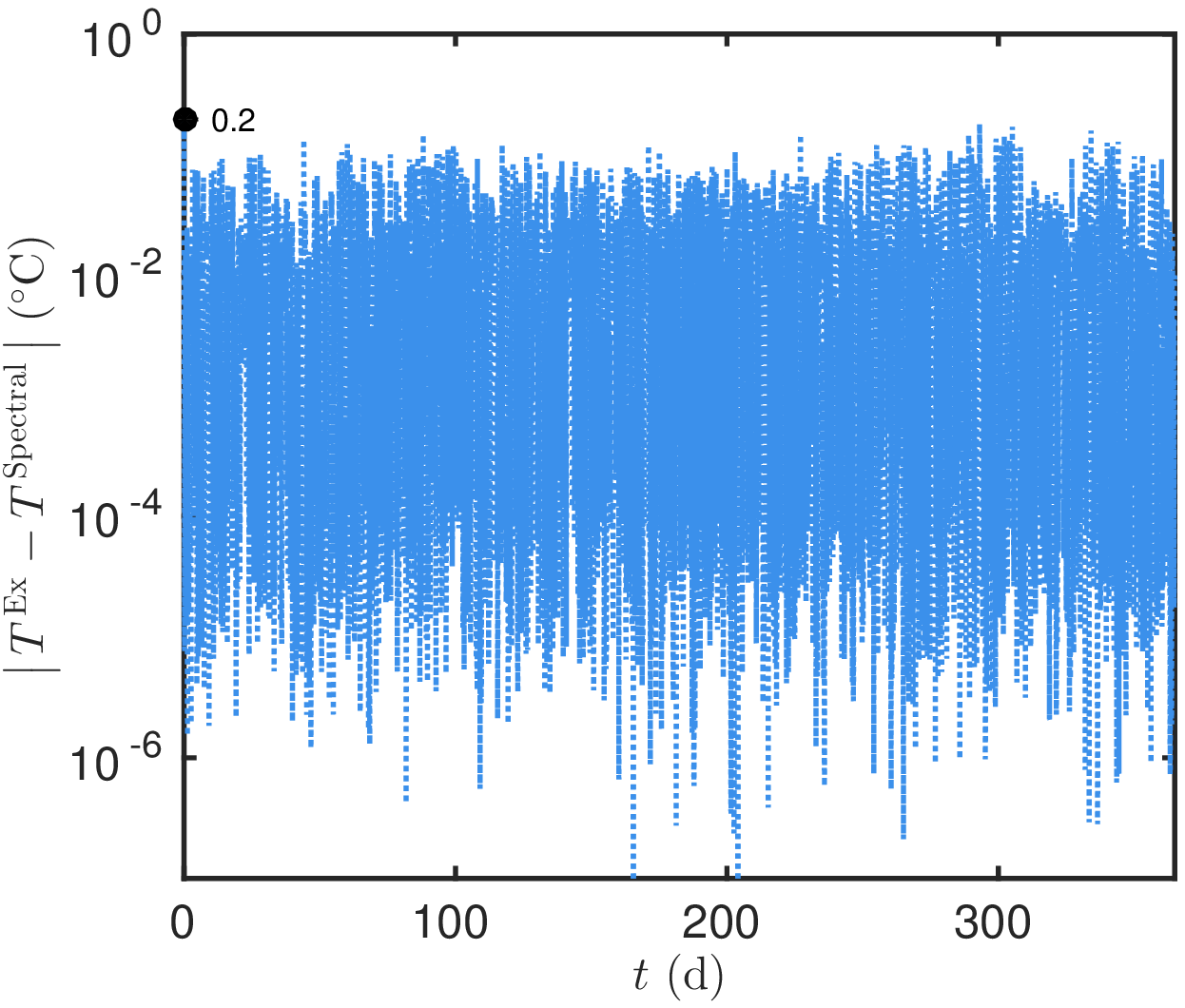}} 
\subfigure[\label{fig_AN3:zoomSP_EU_Tx2_ft}]{\includegraphics[width=.45\textwidth]{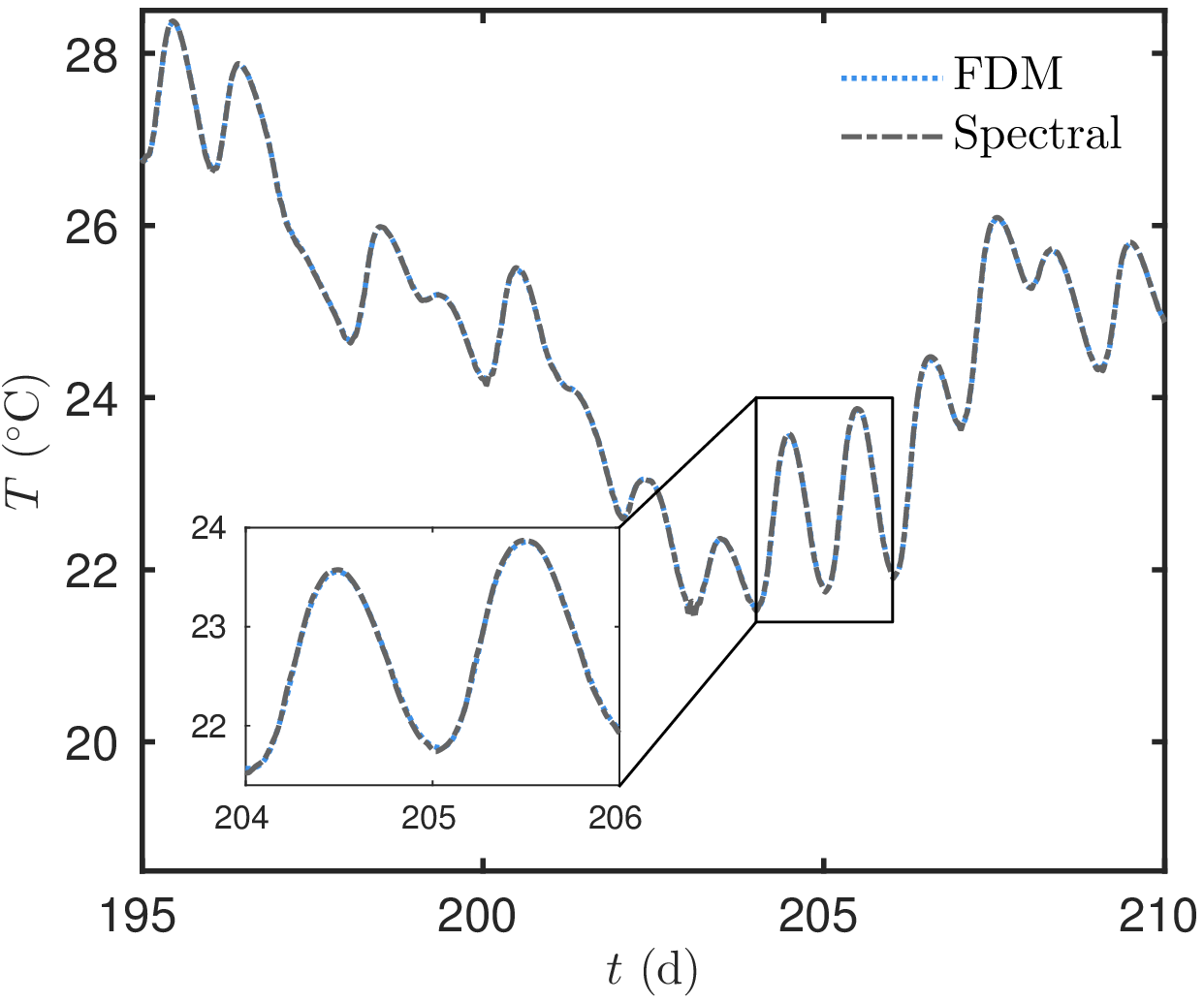}}
\caption{\emph{(a,c)} Temperature absolute difference with the spectral solution for the R$2$C and the Explicit approaches. \emph{(b,d)} Temperature evolution in the wall at $x \egal 0.1 \unit{cm}\,$ during the hottest days.}
\end{figure}

\begin{figure}
\centering
\subfigure[\label{fig_AN3:R2C_DiffqR_ft}]{\includegraphics[width=.45\textwidth]{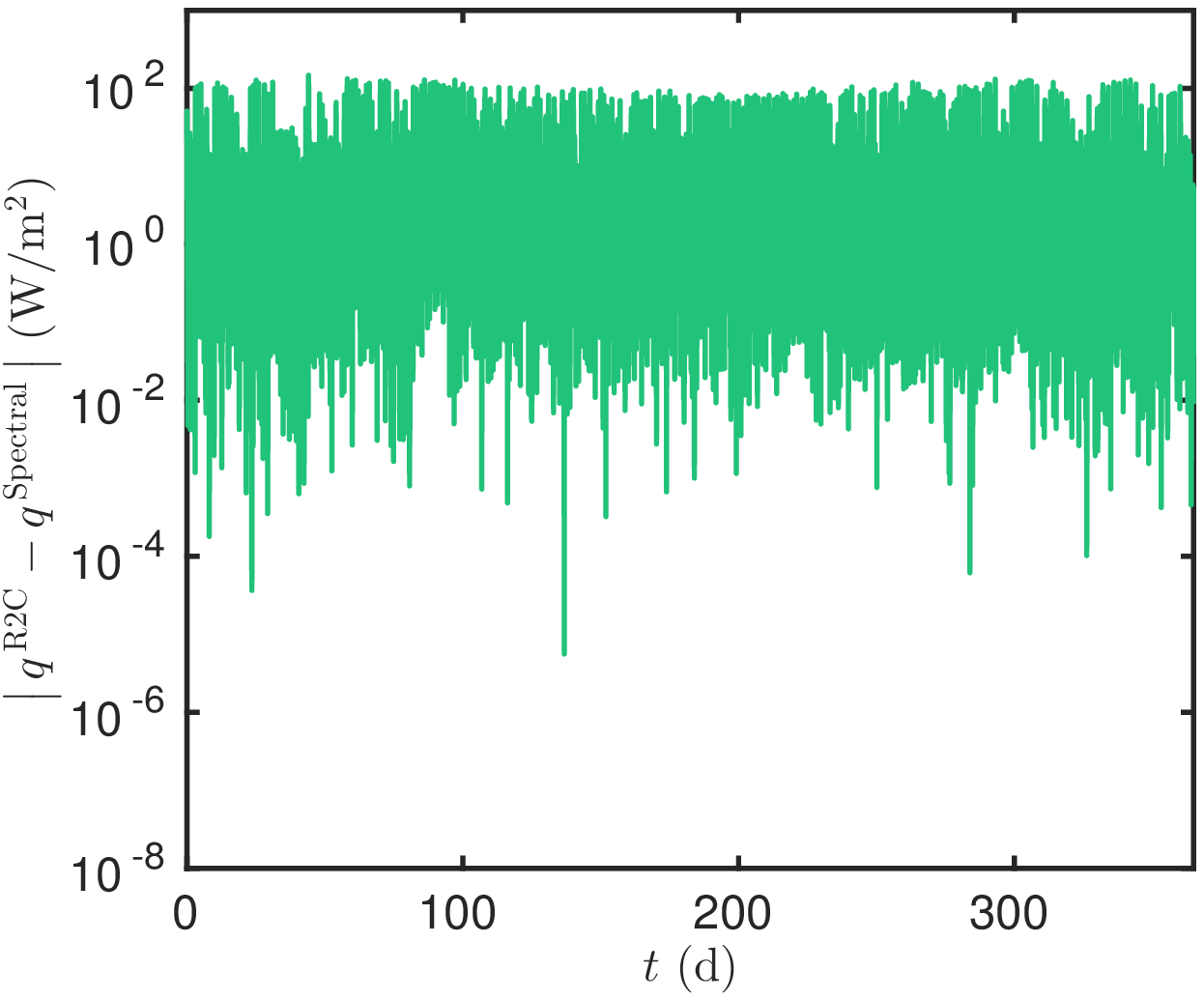}} 
\subfigure[\label{fig_AN3:zoomRC_qR_ft}]{\includegraphics[width=.45\textwidth]{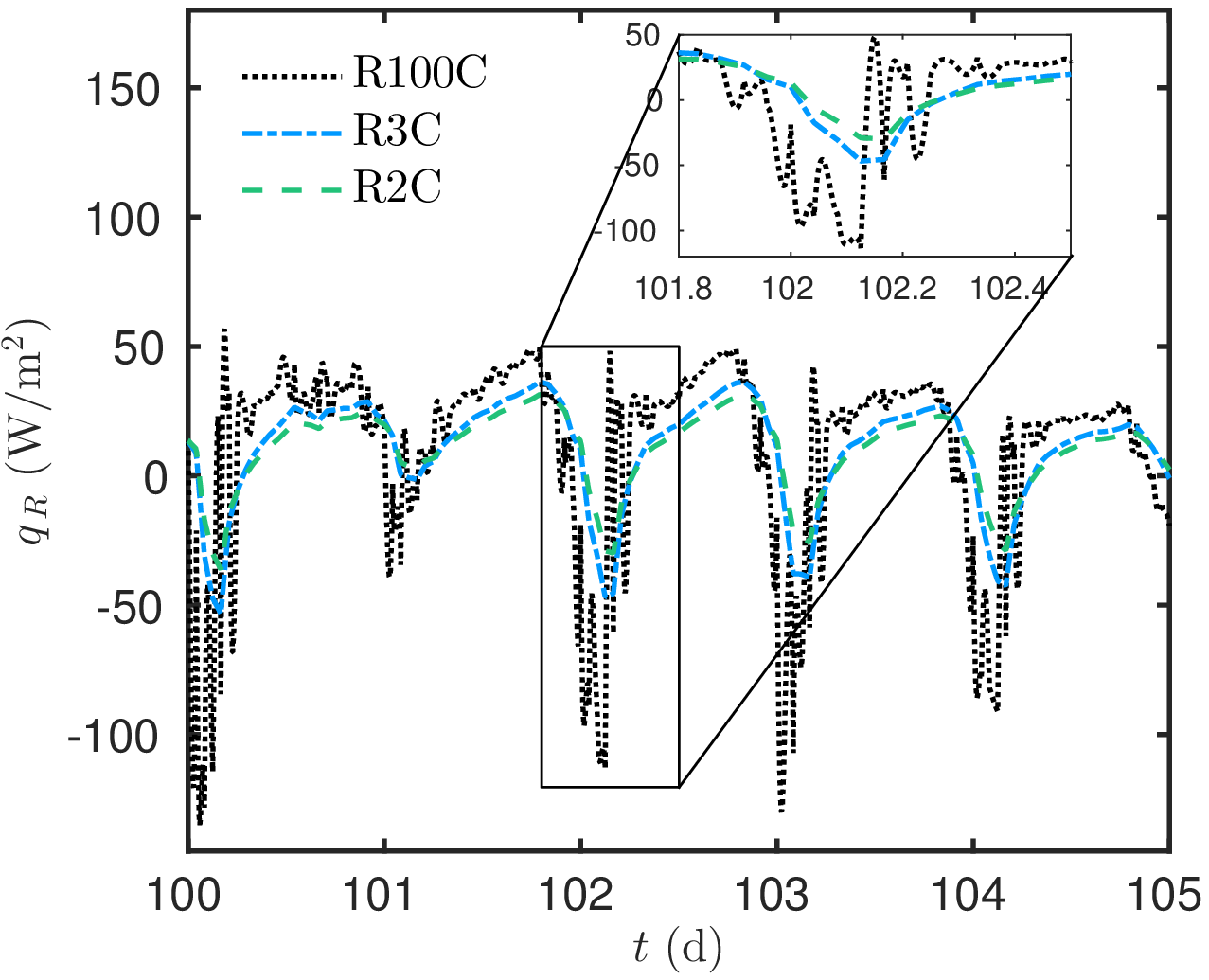}}  \\
\subfigure[\label{fig_AN3:Ex_DiffqR_ft}]{\includegraphics[width=.45\textwidth]{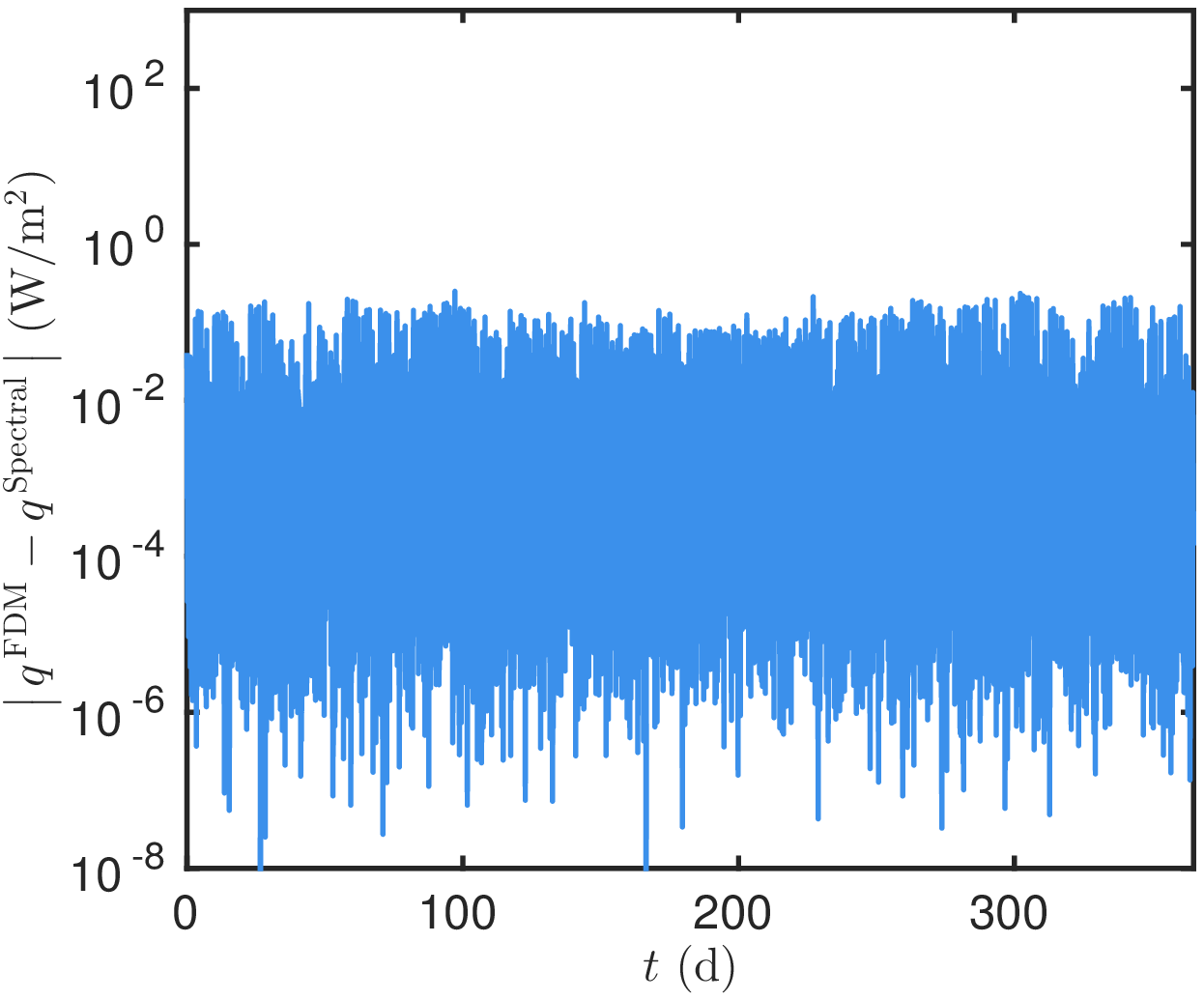}} 
\subfigure[\label{fig_AN3:zoomSP_EU_qR_ft}]{\includegraphics[width=.45\textwidth]{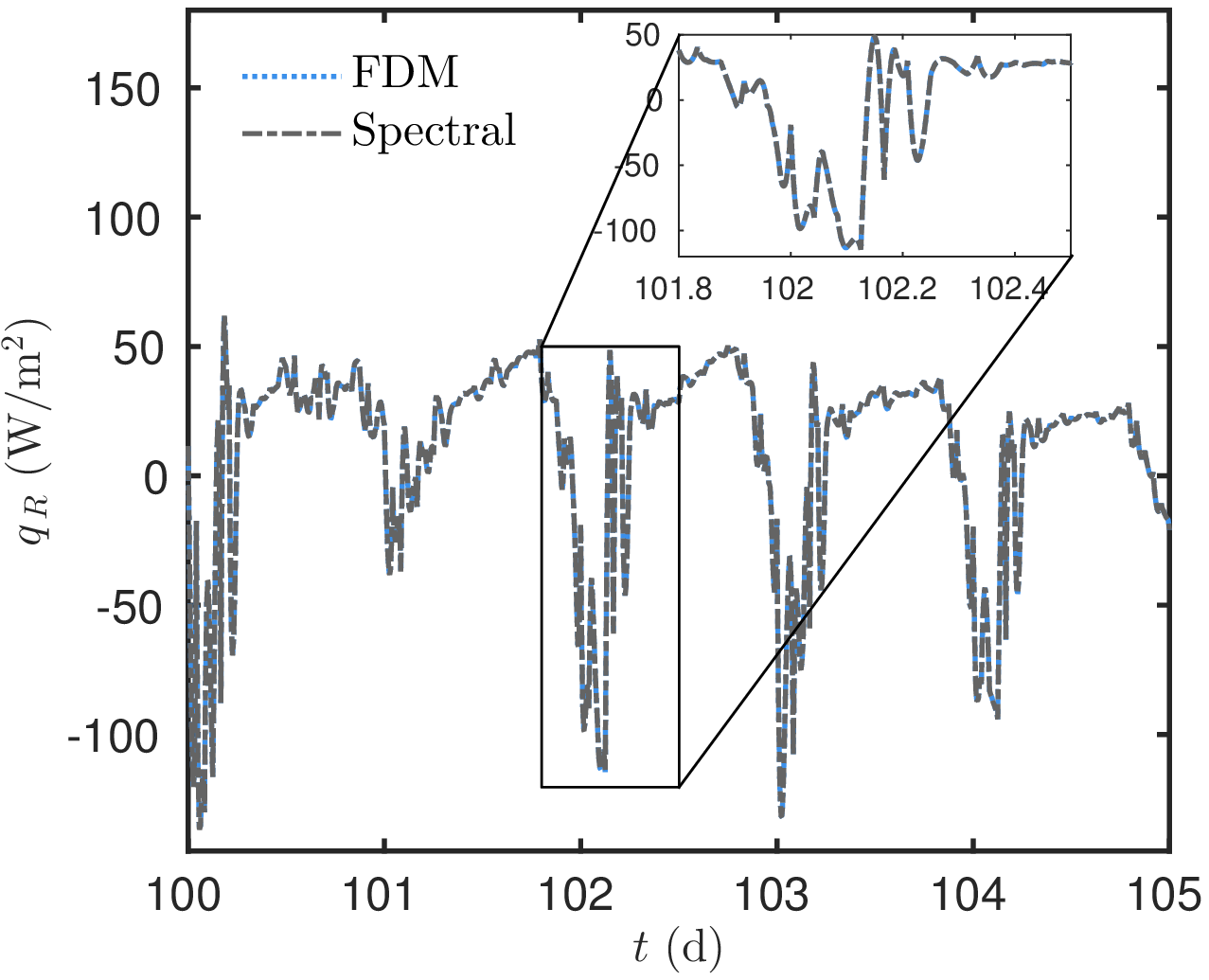}}  
\caption{\emph{(a,c)} Heat flux absolute difference with the spectral solution for the R$2$C and the Explicit approaches. \emph{(b,d)}  Heat flux density evolution in the wall at the right boundary $x \egal 0.5 \unit{cm}\,$ during the winter.}
\end{figure}

The temperature is computed within the wall using the RC with $r \egal \bigl\{\, 2\,,\, 3 \,,\, 100 \,\bigr\}$ resistances, the standard finite-differences and the spectral approaches. For the last two methods, the spatial discretisation is $\dx \egal 10^{\,-3} \ \mathsf{m}\,$. For the last two methods, the spatial discretisation is $\dx \egal 10^{\,-2} \ \mathsf{m}\,$. No reference solution is computed for this case. Thus, as the spectral method was shown as the most precise method in the previous cases, it was used as a reference to compute the absolute temperature difference for the standard finite-differences ~and \RC{2} approaches, illustrated in Figures~\ref{fig_AN3:R2C_diffTx2_ft} and \ref{fig_AN3:Ex_diffTx2_ft}. The standard finite-differences ~approach has less discrepancies with the spectral solution than the \RC{2} one. The maximum difference for the \RC{2} reaches $3.1 \unit{^{\,\circ}C}\,$. As shown in Figures~\ref{fig_AN3:zoomRC_Tx2_ft} and \ref{fig_AN3:zoomSP_EU_Tx2_ft}, there is good agreement between the \RC{100}, standard finite-differences and spectral solutions. As expected, the RC approach with a sufficient high number of resistances enables to accurately represent the physical phenomenon. The heat flux at the right boundary is shown in Figures~\ref{fig_AN3:zoomRC_qR_ft} and \ref{fig_AN3:zoomSP_EU_qR_ft}. More important discrepancies are noted between the RC approaches with two or three resistances than for the other solutions. In addition, the absolute difference with the spectral solution reaches $\simeq \ 10^{\,2} \unit{W/m^{\,2}}$ for the \RC{2} approach. For the standard finite-differences solution, the difference remains lower than $10^{\,-1} \unit{W/m^{\,2}}\,$.

In terms of computational cost, the ratio of CPU time $R_{\,\mathrm{cpu}}$ is $0.17 \ \mathsf{s/h}$ for the \RC{2}, $0.48 \ \mathsf{s/h}$ for the \RC{3}, $1.9 \ \mathsf{s/h}$ for the \RC{100}, $1.7 \ \mathsf{s/h}$ by using the finite-difference approach and $1.0 \ \mathsf{s/h}$ by the spectral model. The latter presents the best efficiency to compute the solution of the problem.

\begin{figure}
\centering
\subfigure[\label{fig_AN3:Tm_ft}]{\includegraphics[width=.45\textwidth]{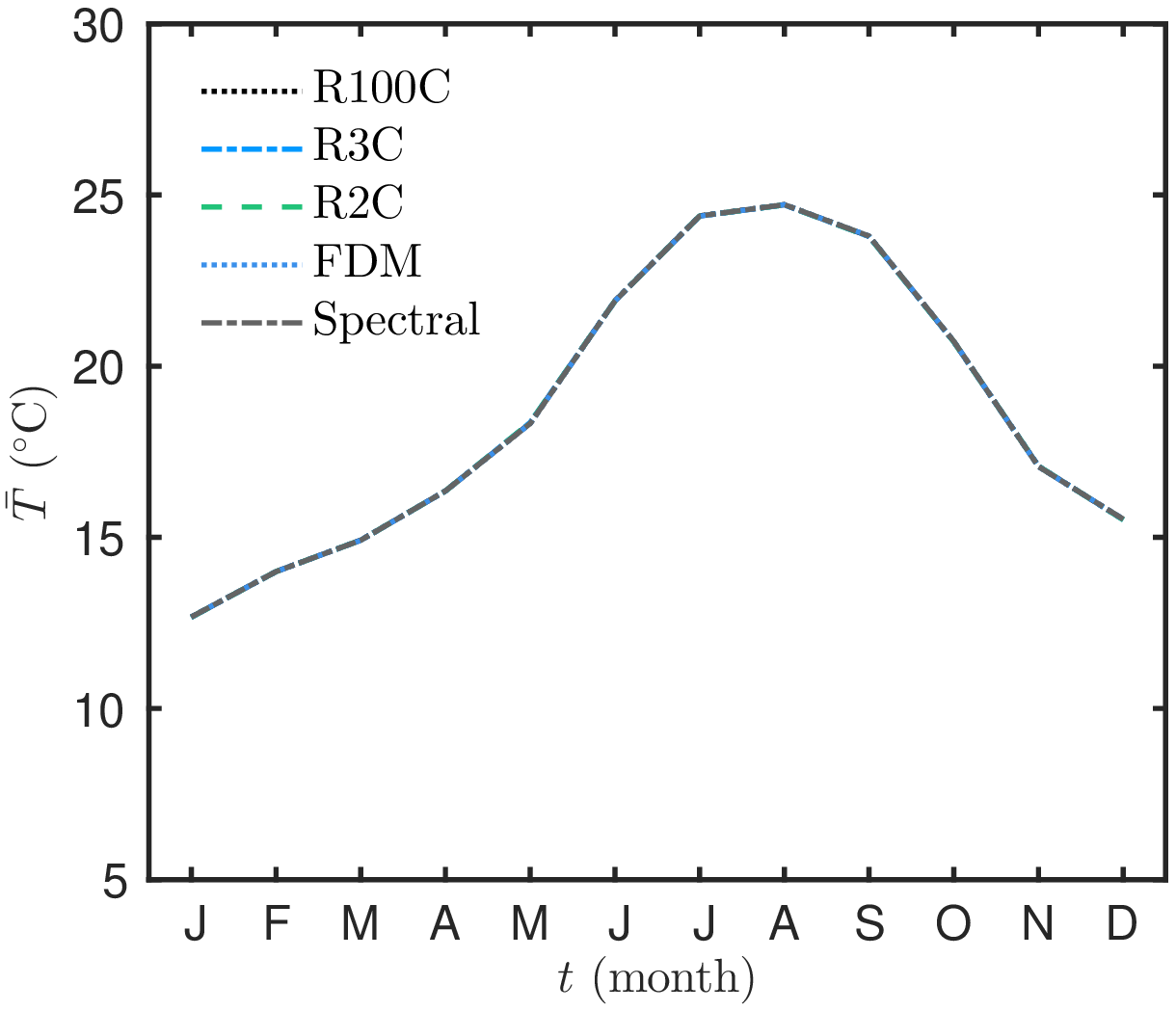}} 
\subfigure[\label{fig_AN3:Tj_ft}]{\includegraphics[width=.45\textwidth]{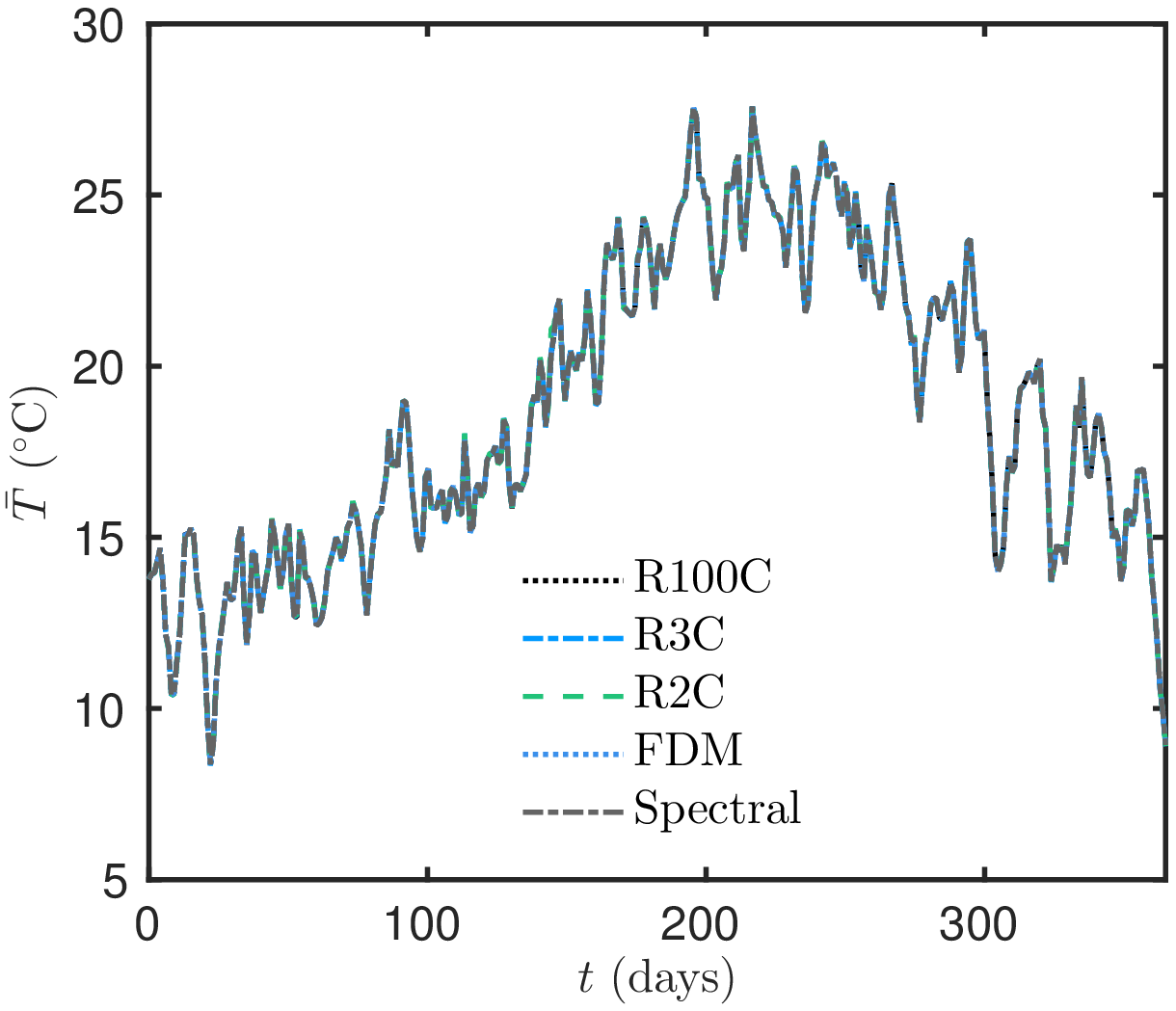}}  \\
\caption{Evolution of the monthly \emph{(a)} and daily \emph{(b)} mean temperature.}
\end{figure}

\begin{figure}
\centering
\subfigure[\label{fig_AN3:E_fmois}]{\includegraphics[width=.45\textwidth]{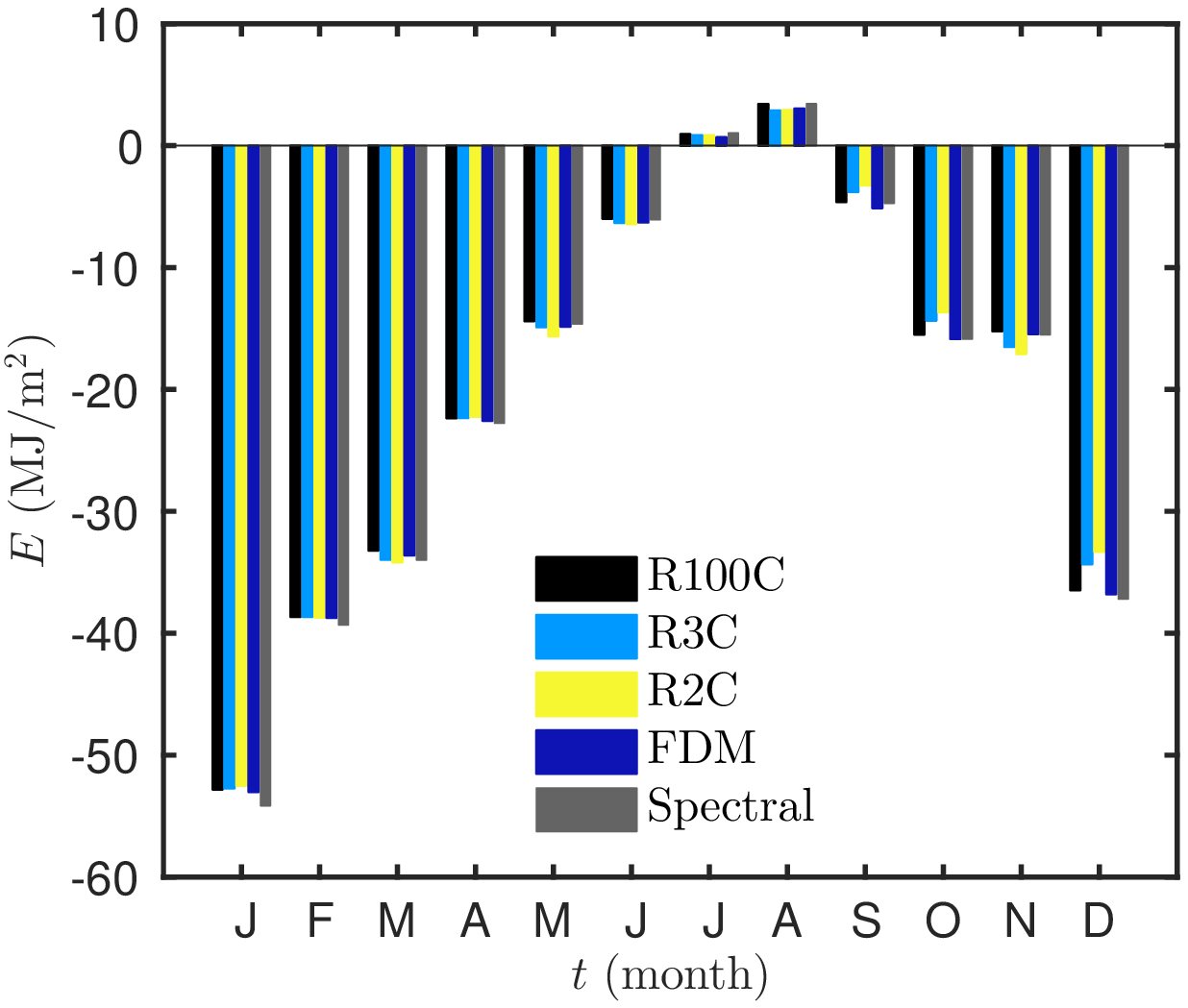}} 
\subfigure[\label{fig_AN3:E_fjour}]{\includegraphics[width=.45\textwidth]{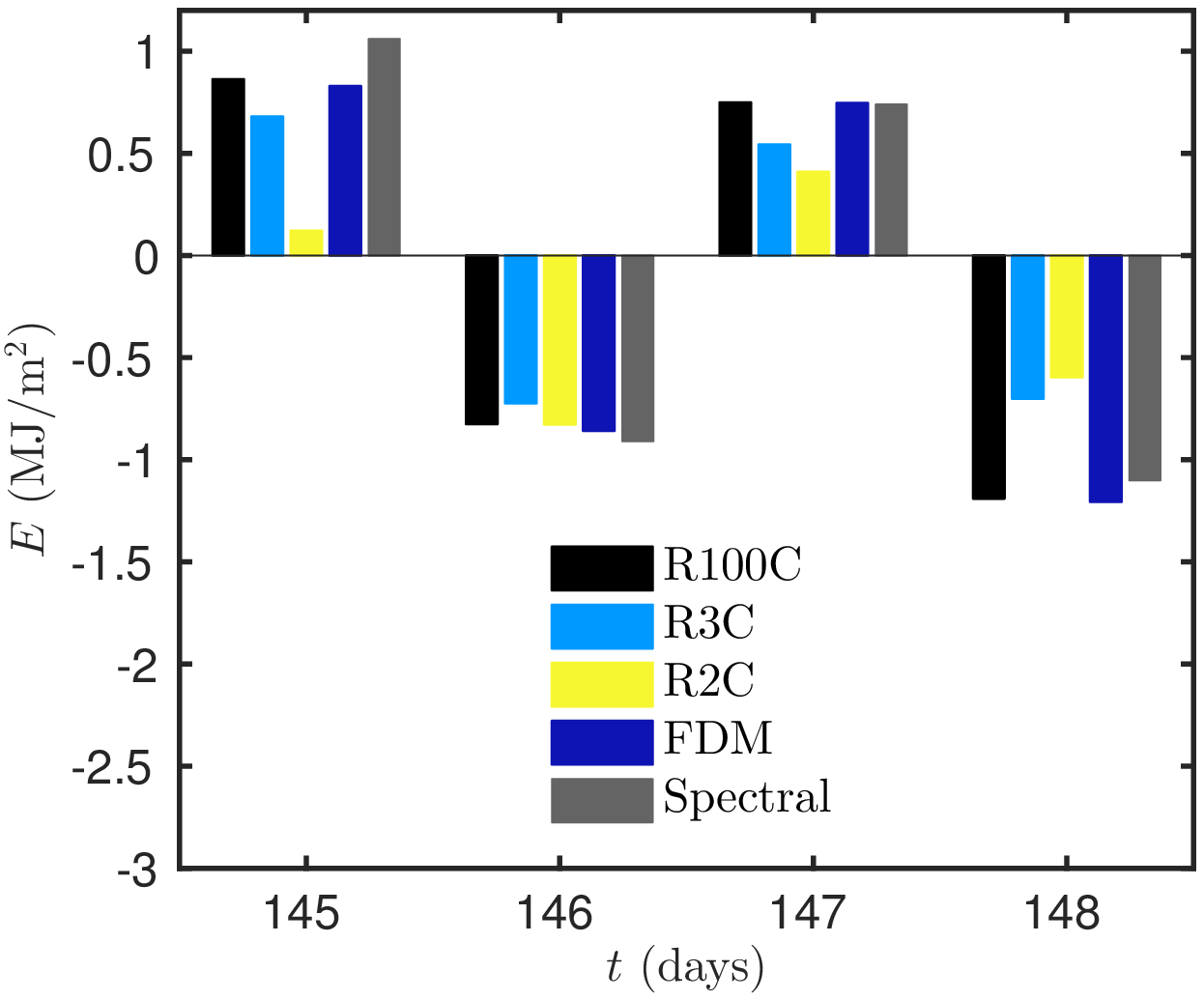}}  \\
\caption{Evolution of the monthly \emph{(a)} and daily \emph{(b)} conduction loads.}
\end{figure}

It can be noted that the daily and monthly averaged temperatures are well represented by all methods, as shown in Figures~\ref{fig_AN3:Tm_ft} and \ref{fig_AN3:Tj_ft}. There is a perfect agreement among all solutions. The conduction loads are presented in Figures~\ref{fig_AN3:E_fmois} and \ref{fig_AN3:E_fjour}. For the analysis on the monthly period, the conduction loads are negative for a ten-month period corresponding to winter time. During the summer, the loads are positive, indicating an inward heat flux density. Globally, all the numerical methods enable to estimate the conduction loads with the same order of accuracy. No important discrepancies are noted. This is due to the definition of the heat loads from Eq.~\eqref{eq:heat_load}. The integration is performed for a monthly period. As noticed in Figure~\ref{fig_AN3:zoomRC_qR_ft}, the \RC{2} approximates well the mean heat flux, compared to the \RC{100} approach. Therefore, when integrating during a monthly period, the error on the conduction loads is smoothed. On the contrary, when looking at the daily loads of a winter period in Figure~\ref{fig_AN3:E_fjour}, the discrepancies among the three methods are higher. Particularly, the \RC{2} and \RC{3} solutions lack of accuracy.  

These results lead to the conclusion that the choice of the methods depends on the characteristic timescale of the selected output. If the issue is to accurately analyze the physical phenomenon at a short time scale, then the methods (i) standard finite-differences, (ii) spectral and (iii) RC with a sufficient enough number of resistances, provide a sufficient accuracy. If the output evolves on a relatively large time scale, the RC approach with a few number of resistances may provide a satisfactory approximation of the physical phenomenon. 

\section{Synopsis of the discussions}

It is of major importance to have efficient numerical methods to investigate physical phenomena associated to heat or moisture diffusion through building envelopes. Building simulation programs detail the governing equations but the efficiency of the numerical methods is rarely discussed. This article proposed to investigate the efficiency of three methods by evaluating their accuracy and calculation speed. The first one is the standard one based on central finite-differences and on the \RK ~explicit approach. The second method is the so-called RC model. This approach was initially used to build analogous electric devices to simulate the heat conduction in the $1940$'s--$1950$'s. It was then implemented in the computer algorithm and is still used for educational, research and regulation purposes in building physics. The last method is the spectral reduced order method. It has been recently proposed for building simulation purposes in \cite{Gasparin_2017a,Gasparin_2017b}. 

Three case studies have been presented in Section~\ref{sec:linear_heat_diff}, a case of linear heat diffusion was considered. The solutions were computed using the three approaches and compared to a reference solution. The results highlighted that accuracy is satisfactory for the standard finite-differences, the \RC{100} and the spectral approaches. Among the three methods, the spectral is the most accurate with the higher number of digits accuracy. It actually gives a good compromise between accuracy and calculation speed. When using a low number of thermal resistances, the accuracy of the RC approach decreases with an absolute difference of the order of $\mathcal{O}(\,10^{\,-1}\,) \unit{^{\,\circ}C}$ with the reference solution. However, the accuracy is lower when looking at the flux density. Indeed, the \RC{2} and \RC{3} approximate the flux with an absolute difference of the order $\mathcal{O}(\,10^{\,2}\,) \unit{W/m^{\,2}}\,$. A parametric study was performed to evaluate the minimal number of thermal resistances to reach a satisfactory precision. For this case study, the minimal number is around $10$ and $30$ resistances for the temperature and the heat flux, respectively. 

A similar numerical investigation was then accomplished for a case of nonlinear moisture diffusion, in Section~\ref{sec:NL_moisture_diff}. The results are very similar and highlight the spectral method is again the most accurate approach with a reasonable computational run time. The speed of computation of the \RC{2} and \RC{3} models is very low. However, the speed of computation decreases with the accuracy of the solution. It was noted that the number of significant digits of the \RC{2} and \RC{3} solutions was much lower than for the previous linear case study due mainly to two reasons. First, this case considered a non-linear problem inducing more errors for coarse approaches. Then, the RC approach computes the solution directly in its physical dimension where the others consider a dimensionless formulation of the problem. The vapor pressure scales with $\mathcal{O}(\,10^{\,3}\,)$ where the temperature with $\mathcal{O}(\,10\,)\,$. Therefore, additional computational rounding errors may be introduced in the second case study since the errors of the floating point arithmetic of the computers are minimal for quantities of the magnitude $\mathcal{O}(\,1\,)\,$. The minimal number of resistances to reach a sufficient accuracy with the RC approach was $10$ and $90$ for the temperature and the heat flux outputs of interest, respectively. 

The third case study was more realistic case with real measured temperature as boundary condition for the study of heat transfer in a $19$ Century building wall. The simulation was performed for $1 \unit{year}$. No reference solution was available for this case. The standard finite-differences, \RC{100} and spectral methods had similar tendencies to represent the temperature evolution in the wall. The \RC{2} approach showed a maximum absolute difference of $3.1 \unit{^{\,\circ}C}$ with the spectral solution. For the heat flux, the absolute differences were of the order of $\mathcal{O}(\,10^{\,2} \unit{W/m^{\,2}}\,)$ between both approaches. Within the context of energy efficiency, a relevant output is the time-integrated conduction loads. The latter were calculated for both monthly and daily periods. All the methods have very similar tendencies to estimate the monthly conduction loads. Nevertheless, the \RC{2} and \RC{3} approaches reveal significant discrepancies with the others when looking at the daily loads. It is due to the integration periods and the fact that the RC approaches with a few temperature approximates well the time mean of the heat flux. 

\section{Conclusion}
\label{sec:conclusion}

With the issue of elaborating reliable models to predict some physical phenomena involved in building energy efficiency, the selection of the numerical methods has to be done considering the characteristic time scale of the aimed output, among other criteria. For the analysis of physical phenomena in a short time scale it is better to use the methods (i) standard finite differences, (ii) spectral or (iii) RC with a sufficient number of resistances. The RC model with a few number of resistances decreases the order of the model. With three resistances for instance, the number of degrees of freedom is only two, reducing consequently the computational effort to solve the problem. However, the fidelity of the model to represent the physical phenomena of heat or moisture transfer is strongly impacted and lacks of accuracy. These results are corroborated with the ones presented in \cite{Kircher_2015}. With the issue of reducing the computational effort, the spectral method is the most efficient among the three approaches. Indeed, a model with less than $10$ degrees of freedom enables to compute the most accurate solution among the other approaches. Another important point concerns the computation of the surface density of fluxes at the boundaries. The numerical approximation to compute the fluxes introduces errors so that the accuracy is generally decreased compared to the one of the governing fields (temperature or vapor pressure). Since the fluxes are of capital importance on the design of energy efficient buildings, the accuracy of the numerical method to compute them should always been carefully verified.

All these results were obtained on rather a simple case study. The physical phenomena are more complex, involving coupled heat and mass transfer and \textsc{Robin}--type boundary conditions with radiation heat transfer and wind-driven rain. Therefore, before analyzing the physical phenomena in building physics, it is of capital importance to compare the accuracy of the numerical model, including the mathematical formulation of the physical problem together with the numerical methods and the discretisation mesh, with a reference solution. This procedure has to be carried out keeping in mind the output of interest. 

\section*{Nomenclature}

\begin{tabular*}{0.7\textwidth}{@{\extracolsep{\fill}} |@{} >{\scriptsize} c >{\scriptsize} l >{\scriptsize} l| }
\hline
\multicolumn{3}{|c|}{\emph{Latin letters}} \\
$E$ & conduction loads &  $[\mathsf{J/m^{\,2})}]$ \\
$c$ & specific heat capacity & $[\mathsf{W/(kg.K)}]$ \\
$g$ & vapor flux density &  $[\mathsf{kg/(s.m^{\,2})}]$ \\
$k$ & thermal conductivity & $[\mathsf{W/(m.K)}]$ \\
$P_{\,c}$ & capillary pressure & $[\mathsf{Pa}]$ \\
$\Pv$ & vapor pressure & $[\mathsf{Pa}]$ \\
$q$ & heat flux density &  $[\mathsf{W/.m^{\,2}}]$ \\
$t$ & time coordinate & $[\mathsf{s}]$ \\
$T$ & temperature & $[\mathsf{K}]$ \\
$x$ & space coordinate & $[\mathsf{m}]$ \\
\hline
\end{tabular*}

\bigskip

\begin{tabular*}{0.7\textwidth}{@{\extracolsep{\fill}} |@{} >{\scriptsize} c >{\scriptsize} l >{\scriptsize} l| }
\hline
\multicolumn{3}{|c|}{\emph{Greek letters}} \\
$\kappa$ & moisture permeability & $[\mathsf{s}]$ \\
$\rho$ & density  & $[\mathsf{kg/(m^3)}]$ \\
$xi$ & moisture capacity & $[\mathsf{kg/m^3}]$ \\
\hline
\end{tabular*}

\section*{Acknowledgments}

The authors acknowledge the Brazilian Agencies CAPES of the Ministry of Education, the CNPQ of the Ministry of Science, Technology and Innovation, for the financial support for the project CAPES-COFECUB Ref. 774/2013 as well as the support of CNRS/INSIS (Cellule \'energie) under the program ``Projets Exploratoires --- 2017''. The authors also acknowledge the Junior Chair Research program ``Building performance assessment, evaluation and enhancement'' from the University of Savoie Mont Blanc in collaboration with The French Atomic and Alternative Energy Center (CEA) and Scientific and Technical Center for Buildings (CSTB). 

\bibliographystyle{apalike} 
\bibliography{references}

\end{document}